\documentclass[aps,prd,floatfix,eqsecnum]{revtex4}
\usepackage{graphicx}
\usepackage{dcolumn}
\usepackage{bm}
\usepackage{amsfonts}
\usepackage{latexsym}
\usepackage{amssymb}
\usepackage{verbatim}
\usepackage{amsthm}
\usepackage{amsmath}

\def\fun#1#2{\lower3.6pt\vbox{\baselineskip0pt\lineskip.9pt
  \ialign{$\mathsurround=0pt#1\hfil##\hfil$\crcr#2\crcr\sim\crcr}}}
\newcommand{\MR}{{{\mathbb R}}}

\newcommand{\RF}{{{\mathbb R}}}

\newtheorem{theorem}{Theorem}[section]

\newtheorem{conjecture}{Conjecture}[section]
\begin{document}


\title{Second-order Gauge-invariant
  Cosmological Perturbation Theory:\\ Current Status updated in 2019}

\author{Kouji Nakamura}
\email{dr.kouji.nakamura@gmail.com}
\affiliation{%
  Gravitational-Wave Science Project,
  National Astronomical Observatory of Japan,
  Osawa, Mitaka, Tokyo 181-8588, Japan.
}%


\date{\today}

\begin{abstract}
  The current status of the recent developments of the second-order
  gauge-invariant cosmological perturbation theory is reviewed.
  To show the essence of this perturbation theory, we concentrate only
  on the universe filled with a single scalar field.
  Through this review, we point out the problems which should be
  clarified for the further theoretical sophistication of this
  perturbation theory.
  This review is an extension of the review paper [K.~Nakamura,
  ``Second-Order Gauge-Invariant Cosmological Perturbation Theory:
  Current Status'', Advances in Astronomy, {\bf 2010} (2010),
  576273.].
  We also expect that this theoretical sophistication will be
  also useful to discuss the future developments in cosmology as a
  precise science.
\end{abstract}

\maketitle

\section{Introduction}
\label{sec:intro}


The general relativistic cosmological {\it linear} perturbation theory
has been developed to a high degree of sophistication during the last
40
years~\cite{Bardeen-1980,Kodama-Sasaki-1984,Mukhanov-Feldman-Brandenberger-1992}.
One of the motivations of this development was to clarify the relation
between the scenarios of the early universe and cosmological data,
such as the cosmic microwave background (CMB) anisotropies.
Recently, the first-order approximation of our universe from a
homogeneous isotropic one was revealed through the observation of the
CMB by the Wilkinson Microwave Anisotropy Probe (WMAP)~\cite{WMAP} and
by the Planck mission~\cite{Planck}, the cosmological parameters are
accurately measured, we have obtained the standard cosmological model,
and the so-called ``precision cosmology'' is developing.
These developments in observations were also supported by the
theoretical sophistication of the linear order cosmological
perturbation theory.


The observational results of CMB also suggest that the fluctuations of
our universe are adiabatic and Gaussian at least in the first-order
approximation.
We are now on the stage to discuss the deviation from this first-order
approximation from the observational~\cite{WMAP,Planck} and
theoretical sides~\cite{V.Acquaviva-N.Bartolo-S.Matarrese-A.Riotto-2003,J.Maldacena-2003,K.A.Malik-D.Wands-2004,N.Bartolo-S.Matarrese-A.Riotto-2004a,N.Bartolo-S.Matarrese-A.Riotto-2004b,D.H.Lyth-Y.Rodriguez-2005,F.Vernizzi-2005,N.Bartolo-S.Matarrese-A.Riotto-2004c,N.Bartolo-S.Matarrese-A.Riotto-2004d,N.Bartolo-E.Komatsu-S.Matarrese-A.Riotto-2004,N.Bartolo-S.Matarrese-A.Riotto-2006,Bartolo:2006cu,Bartolo:2006fj,Nitta:2009jp,Pitrou:2008ak,Senatore:2008vi}
through the non-Gaussianity, the non-adiabaticity, and so on.
These will be goals of future missions of observations.
With the increase of precision of the CMB data, the study of
relativistic cosmological perturbations beyond linear order is a
topical subject.
The {\it second-order} cosmological perturbation theory is one of such
perturbation theories beyond linear order.


Although the second-order perturbation theory in general relativity is
an old topic, a general framework of the gauge-invariant formulation
of the general relativistic second-order perturbation has been
proposed~\cite{kouchan-gauge-inv,kouchan-second}.
This general formulation is an extension of the works of Bruni et
al.~\cite{M.Bruni-S.Matarrese-S.Mollerach-S.Soonego-CQG1997} and has
also been applied to cosmological perturbations:
The derivation of the second-order Einstein equation in a
gauge-invariant manner without any gauge
fixing~\cite{kouchan-cosmo-second-letter,kouchan-cosmo-second-full-paper};
Applicability in more generic
situations~\cite{kouchan-second-cosmo-matter}; Confirmation of the
consistency between all components of the second-order Einstein
equations and equations of
motions~\cite{kouchan-second-cosmo-consistency}; A comparison with a
different formulations~\cite{A.J.Christopherson-et.al-2011}.
We also note that the radiation was discussed by treating the
Boltzmann equation up to second order\cite{Pitro-2007,Pitro-2009}
along the gauge-invariant manner of the above series of papers by the
present author.


On the other hand, more basic issues on the general-relativistic
gauge-invariant higher-order perturbation theory are also developed.
Our general framework is based on the assumption that the linear-order
metric perturbation is decomposed into its gauge-invariant and
gauge-variant parts
(Conjecture~\ref{conjecture:decomp_conjecture_for_hab}, in
Sec.~\ref{sec:gauge-invariant-variables}, below).
In
Refs.~\cite{K.Nakamura-CQG-Letter-2011,K.Nakamura-Progress-Construction-2013},
we proposed a scenario of a proof of
Conjecture~\ref{conjecture:decomp_conjecture_for_hab} and showed that
Conjecture~\ref{conjecture:decomp_conjecture_for_hab} are almost
proved except for the special modes of perturbations due to the
non-local nature in the statement of
Conjecture~\ref{conjecture:decomp_conjecture_for_hab}.
In
Refs.~\cite{K.Nakamura-Progress-Construction-2013,K.Nakamura-IJMP-globalization-2012},
we also pointed out the physical importance of these special modes
which are excluded in our proof proposed in Refs.~\cite{K.Nakamura-CQG-Letter-2011,K.Nakamura-Progress-Construction-2013}.
We also examine the extendibility of our formulation to an arbitrary
higher-order perturbations and concluded that we can extend our
general-formulation of higher-order gauge-invariant perturbation
theory to an arbitrary higher-order, though the arguments of this
examination is still incomplete~\cite{K.Nakamura-CQG-Recursive-2014}.


In this article, we summarize the current status of our development of
the second-order gauge-invariant cosmological perturbation theory
through the simple system of the universe filled with a scalar field.
This review is an updating version of our previous
review~\cite{K.Nakamura-AIA-Review-2010} in 2010.
Through this review, we point out the problems which should be
clarified and directions of the further development of the theoretical
sophistication of the general relativistic higher-order perturbation
theory, especially in cosmological perturbations.
We expect that this theoretical sophistication will be also useful to
discuss the future developments to cosmology as a precise science.


The organization of this paper is as follows.
In
Sec.~\ref{sec:General-framework-of-GR-GI-perturbation-theory},
we review the general framework of the second-order gauge-invariant
perturbation theory developed in
Refs.~\cite{kouchan-gauge-inv,kouchan-cosmo-second-letter,kouchan-cosmo-second-full-paper,kouchan-second,kouchan-LTVII}.
This review also includes additional explanations which were not given
in those papers.
In Sec.~\ref{sec:Perturbation-of-the-field-equations}, we also the
derivations of the second-order perturbation of the Einstein equation
and the energy-momentum tensor from general point of view.
For simplicity, in this review, we only consider a single scalar field
as a matter content.
The ingredients of
Sec.~\ref{sec:General-framework-of-GR-GI-perturbation-theory}
and \ref{sec:Perturbation-of-the-field-equations} are applicable to
perturbation theory in any theory with general covariance, if
Conjecture~\ref{conjecture:decomp_conjecture_for_hab} in
Sec.~\ref{sec:gauge-invariant-variables} is correct.
In Sec.~\ref{sec:Cosmological-Background-spacetime-equations},
we summarize the Einstein equations in the case of a background
homogeneous isotropic universe, which are used in the derivation
of the first- and second-order Einstein equations.
In
Sec.~\ref{sec:Equations-for-the-first-order-cosmological-perturbations},
the first-order perturbation of the Einstein equations and the
Klein-Gordon equations are summarized.
The derivation of the second-order perturbations of the Einstein
equations and the Klein-Gordon equations, and their consistency are
reviewed in
Sec.~\ref{sec:Equations-for-the-second-order-cosmological-perturbations}.
The final section, Sec.~\ref{sec:summary}, is devoted to a summary and
discussions.
In addition to these main text, we briefly explain the derivation of
the general Taylor expansion in
Appendix~\ref{sec:derivation-of-Taylor-expansion}.
Derivation of the formulae for perturbative curvatures in
Appendix~\ref{sec:derivation-of-pert-Einstein-tensors}.
In
Appendix~\ref{sec:Outline-of-the-proof-of-the-decomposition-conjecture},
we briefly show a scenario of the proof for
Conjecture~\ref{conjecture:decomp_conjecture_for_hab} based on the
ingredient in Ref.~\cite{K.Nakamura-Progress-Construction-2013},
though this scenario is still incomplete.


We have to note that this is a review of our own works on general relativistic
higher-order perturbations and is not a survey of a huge number of
papers of this topic.
We hope this review is helpful for the future development of
perturbation theories in general relativity not only for cosmology but
also for any other situations of gravitational fields.


\section{General framework of the general relativistic gauge-invariant
  perturbation theory}
\label{sec:General-framework-of-GR-GI-perturbation-theory}


In this section, we review the general framework of the
gauge-invariant perturbation theory developed in Refs.~\cite{kouchan-gauge-inv,kouchan-second,M.Bruni-S.Matarrese-S.Mollerach-S.Soonego-CQG1997,kouchan-cosmo-second-letter,kouchan-cosmo-second-full-paper,kouchan-LTVII,R.K.Sachs-1964,J.M.Stewart-M.Walker11974,J.M.Stewart-M.Walker11990,J.M.Stewart-1991,S.Sonego-M.Bruni-CMP1998,Matarrese-Mollerach-Bruni-1998,Bruni-Gualtieri-Sopuerta-2003,Sopuerta-Bruni-Gualtieri-2004}.
To develop the general relativistic gauge-invariant perturbation
theory, we first explain the general arguments of the Taylor expansion
on a manifold without introducing an explicit coordinate system in Sec.\ref{sec:Taylor-expansion-of-tensors-on-a-manifold}.
Further, we also have to clarify the notion of ``gauge'' in general
relativity to develop the gauge-invariant perturbation theory from
general point of view, which is explained in
Sec.~\ref{sec:Gauge-degree-of-freedom-in-general-relativity}.
After clarifying the notion of ``gauge'' in general relativistic
perturbations, in Sec.~\ref{sec:Formulation-of-perturbation-theory},
we explain the formulation of the general relativistic gauge-invariant
perturbation theory from general point of view.
Although our understanding of ``gauge'' in general relativistic
perturbations is essentially different from ``degree of freedom of
coordinates'' in many literature, ``a coordinate transformation''
is induced by our understanding of ``gauge,'' as a result.
This situation is explained in
Sec.~\ref{sec:Induced-coordiante-transformations}.
Sec.~\ref{sec:Induced-coordiante-transformations} also includes
explanations of the conceptual relation between general covariance and
gauge invariance.
To exclude ``gauge degree of freedom'' which is unphysical degree of
freedom in perturbations, we construct ``gauge-invariant variables''
of perturbations as reviewed in
Sec.~\ref{sec:gauge-invariant-variables}.
These ``gauge-invariant variables'' are regarded as physical
quantities of perturbations in theories with general covariance.


\subsection{Taylor expansion of tensors on a manifold}
\label{sec:Taylor-expansion-of-tensors-on-a-manifold}


First, we briefly review the issues on the general form of the Taylor
expansion of tensors on a manifold ${\cal M}$.
The gauge issue of general relativistic perturbation theories which we
will discuss is related to the coordinate transformation as the result.
Therefore, we first have to discuss the general form of the Taylor
expansion without the explicit introduction of coordinate systems.
Although we only consider the Taylor expansion of a scalar function
$f:{\cal M}\mapsto\MR$, here, the resulting formula is extended to
that for any tensor field on a manifold as in Appendix
\ref{sec:derivation-of-Taylor-expansion}.
We have to emphasize that the general formula of the Taylor expansion
shown here is the starting point of our gauge-invariant formulation of
the second-order general relativistic perturbation theory.


The Taylor expansion of a function $f$ is an approximated form of
$f(q)$ at $q\in{\cal M}$ in terms of the variables at $p\in{\cal M}$,
where $q$ is in the neighborhood of $p$.
To derive the formula for the Taylor expansion of $f$, we have to
compare the values of $f$ at the different points on the manifold.
To accomplish this, we introduce a one-parameter {\it family} of
diffeomorphisms $\Phi_{\lambda}:{\cal M}\mapsto{\cal M}$, where
$\Phi_{\lambda}(p)=q$ and $\Phi_{\lambda=0}(p)=p$.
One example of a diffeomorphisms $\Phi_{\lambda}$ is an exponential
map with a generator.
However, we consider a more general class of diffeomorphisms, as seen
below.


The diffeomorphism $\Phi_{\lambda}$ induces the pull-back
$\Phi_{\lambda}^{*}$ of the function $f$ and this pull-back enable us
to compare the values of the function $f$ at different points.
Further, the Taylor expansion of the function $f(q)$ is given by
\begin{eqnarray}
  f(q)
  &=& f(\Phi_{\lambda}(p))
  =: (\Phi^{*}_{\lambda}f)(p)
  \nonumber\\
  &=&
  f(p)
  +
  \left.\frac{\partial}{\partial\lambda}(\Phi^{*}_{\lambda}f)\right|_{p}
  \lambda
  +
  \frac{1}{2}
  \left.\frac{\partial^{2}}{\partial\lambda^{2}}(\Phi^{*}_{\lambda}f)\right|_{p}
  \lambda^{2}
  + O(\lambda^{3}).
  \label{eq:symbolic-Taylor-expansion-of-f}
\end{eqnarray}
Since this expression hold for an arbitrary smooth function $f$, the
function $f$ in Eq.~(\ref{eq:symbolic-Taylor-expansion-of-f}) can be
regarded as a dummy.
Therefore, we should regard the Taylor expansion
(\ref{eq:symbolic-Taylor-expansion-of-f}) to be the expansion of the
pull-back $\Phi_{\lambda}^{*}$ of the diffeomorphism $\Phi_{\lambda}$,
rather than the expansion of the function $f$.


According to this point of view, Sonego and
Bruni~\cite{S.Sonego-M.Bruni-CMP1998} showed that there exist vector
fields $\xi_{1}^{a}$ and $\xi_{2}^{a}$ such that the expansion
(\ref{eq:symbolic-Taylor-expansion-of-f}) is given by
\begin{eqnarray}
  f(q)
  &=& (\Phi^{*}_{\lambda}f)(p)
  \nonumber\\
  &=& f(p)
  + \left.\left({\pounds}_{\xi_{1}}f\right)\right|_{p} \lambda
  + \frac{1}{2}
  \left.\left({\pounds}_{\xi_{2}}+{\pounds}_{\xi_{1}}^{2}\right)f\right|_{p}
  \lambda^{2}
  + O(\lambda^{3}),
  \label{eq:Taylor-expansion-of-f}
\end{eqnarray}
without loss of generality (see Appendix
\ref{sec:derivation-of-Taylor-expansion}).
Equation (\ref{eq:Taylor-expansion-of-f}) is not only the
representation of the Taylor expansion of the function $f$, but also
the definitions of the generators $\xi_{1}^{a}$ and $\xi_{2}^{a}$.
These generators of the one-parameter family of diffeomorphisms
$\Phi_{\lambda}$ represent the direction along which the Taylor
expansion is carried out.
The generator $\xi_{1}^{a}$ is the first-order approximation of the
flow of the diffeomorphism $\Phi_{\lambda}$, and the generator
$\xi_{2}^{a}$ is the second-order correction to this flow.
We should regard the generators $\xi_{1}^{a}$ and $\xi_{2}^{a}$ to be
independent.
Further, as shown in Appendix
\ref{sec:derivation-of-Taylor-expansion}, the representation of the
Taylor expansion of an arbitrary scalar function $f$ is extended to
that for an arbitrary tensor field $Q$ just through the replacement
$f\rightarrow Q$.


We must note that, in general, the representation
(\ref{eq:Taylor-expansion-of-f}) of the Taylor expansion is different
from an usual exponential map which is generated by a vector field.
In general,
\begin{eqnarray}
  \label{eq:Phi-is-not-one-parameter-group-of-diffeomorphism}
  \Phi_{\sigma}\circ\Phi_{\lambda}\neq\Phi_{\sigma+\lambda}, \quad
  \Phi_{\lambda}^{-1}\neq\Phi_{-\lambda}.
\end{eqnarray}
As noted in
Ref.~\cite{M.Bruni-S.Matarrese-S.Mollerach-S.Soonego-CQG1997},
if the second-order generator $\xi_{2}^{a}$ in
Eq.~(\ref{eq:Taylor-expansion-of-f}) is proportional to the
first-order generator $\xi_{1}^{a}$ in
Eq.~(\ref{eq:Taylor-expansion-of-f}), the diffeomorphism
$\Phi_{\lambda}$ is reduced to an exponential map.
Therefore, one may reasonably doubt that $\Phi_{\lambda}$ forms a
group except under very special conditions.
However, we have to note that the properties
(\ref{eq:Phi-is-not-one-parameter-group-of-diffeomorphism}) does
not directly mean that $\Phi_{\lambda}$ does not form a group.
There will be possibilities that $\Phi_{\lambda}$ form a group in a
different sense from exponential maps, in which the properties
(\ref{eq:Phi-is-not-one-parameter-group-of-diffeomorphism}) will be
maintained.


\begin{figure}
  \begin{center}
    \includegraphics[width=0.5\textwidth]{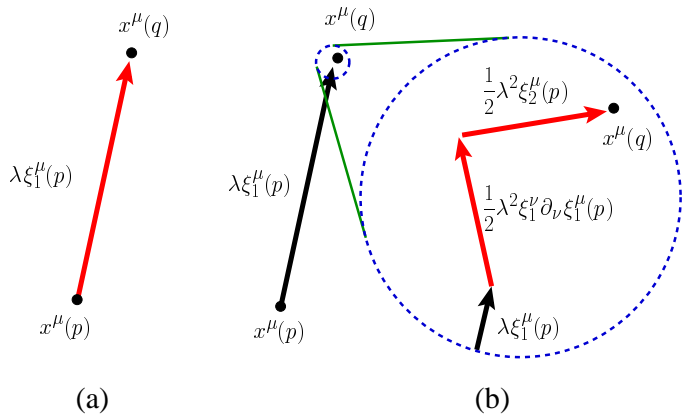}
  \end{center}
  \caption{
    (a) The second term $\lambda \xi_{1}(p)$ in
    Eq.~(\ref{eq:Taylor-expansion-of-xmu}) is the vector which
    points from the point $x^{\mu}(p)$ to the point $x^{\mu}(q)$
    in the sense of the first-order correction. (b) If we look
    at the neighborhood of the point $x^{\mu}(q)$ in detail, the
    vector $\lambda \xi_{1}(p)$ may fail to point to
    $x^{\mu}(q)$ in the sense of the second order.
    Therefore, it is necessary to add the second-order
    correction
    $\frac{1}{2}\lambda^{2}(\xi_{2}^{\mu}+\xi^{\nu}_{1}(p)\partial_{\nu}\xi^{\mu}_{1}(p))$.
  }
  \label{fig:Taylor-expansion-of-coordinate-function}
\end{figure}


Now, we give an intuitive explanation of the representation
(\ref{eq:Taylor-expansion-of-f}) of the Taylor expansion through the
case where the scalar function $f$ in
Eq.~(\ref{eq:Taylor-expansion-of-f}) is a coordinate function.
When two points $p,q\in{\cal M}$ in
Eq.~(\ref{eq:Taylor-expansion-of-f}) are in the neighborhood of each
other, we can apply a coordinate system ${\cal M}\mapsto\RF^{n}$
($n=\dim{\cal M}$), which denoted by $\{x^{\mu}\}$, to an open set
which includes these two points.
Then, we can measure the relative position of these two points $p$ and
$q$ in ${\cal M}$ in terms of this coordinate system in $\RF^{n}$
through the Taylor expansion (\ref{eq:Taylor-expansion-of-f}).
In this case, we may regard that the scalar function $f$ in
Eq.~(\ref{eq:Taylor-expansion-of-f}) is a coordinate function
$x^{\mu}$ and Eq.~(\ref{eq:Taylor-expansion-of-f}) yields
\begin{eqnarray}
  x^{\mu}(q)
  &=& (\Phi^{*}_{\lambda}x^{\mu})(p)
  \nonumber\\
  &=& x^{\mu}(p)
  + \lambda \xi_{1}^{\mu}(p)
  + \frac{1}{2} \lambda^{2}
  \left.\left(\xi_{2}^{\mu}+\xi^{\nu}_{1}\partial_{\nu}\xi^{\mu}_{1}\right)\right|_{p}
  + O(\lambda^{3}).
  \label{eq:Taylor-expansion-of-xmu}
\end{eqnarray}
The second term $\lambda \xi_{1}^{\mu}(p)$ in the right hand side of
Eq.~(\ref{eq:Taylor-expansion-of-xmu}) is familiar.
This is regarded as the vector which points from the point
$x^{\mu}(p)$ to the point $x^{\mu}(q)$ in the sense of the first-order
correction as shown in
Fig.\ref{fig:Taylor-expansion-of-coordinate-function}(a).
However, in the sense of the second order, this vector
$\lambda \xi_{1}^{\mu}(p)$ may fail to point to $x^{\mu}(q)$.
Therefore, it is necessary to add the second-order correction as shown
in Fig.\ref{fig:Taylor-expansion-of-coordinate-function}(b).
As a correction of the second order, we may add the term
$\frac{1}{2} \lambda^{2} \xi^{\nu}_{1}(p)\partial_{\nu}\xi^{\mu}_{1}(p)$.
This second-order correction corresponds to that coming from the
exponential map which is generated by the vector field
$\xi_{1}^{\mu}$.
However, this correction completely determined by the vector field
$\xi_{1}^{\mu}$.
Even if we add this correction comes from the exponential map, there
is no guarantee that the corrected vector $\lambda\xi_{1}^{\mu}(p)+\frac{1}{2}\lambda^{2}\xi^{\nu}_{1}(p)\partial_{\nu}\xi^{\mu}_{1}(p)$
does point to $x^{\mu}(q)$ in the sense of the second order.
Thus, we have to add the new correction
$\frac{1}{2}\lambda^{2}\xi^{\nu}_{2}(p)$ of the second order, in
general.


Of course, without this correction
$\frac{1}{2}\lambda^{2}\xi^{\nu}_{2}(p)$, the vector which comes only
from the exponential map generated by the vector field $\xi_{1}^{\mu}$
might point to the point $x^{\mu}(q)$.
Actually, this is possible if we carefully choose the vector field
$\xi^{\mu}_{1}$ taking into account of the deviations at the second
order.
However, this means that we have to take care of the second-order
correction when we determine the first-order correction.
This contradicts to the philosophy of the Taylor expansion as a
perturbative expansion, in which we can determine everything order by
order.
Therefore, we should regard that the correction
$\frac{1}{2}\lambda^{2}\xi^{\nu}_{2}(p)$ is necessary in general
situations.


\subsection{Gauge degree of freedom in general relativity}
\label{sec:Gauge-degree-of-freedom-in-general-relativity}


Since we want to explain the gauge-invariant perturbation theory in
general relativity, first of all, we have to explain the notion of
``gauge'' in general relativity\cite{kouchan-LTVII}.
General relativity is a theory with general covariance, which
intuitively states that there is no preferred coordinate system in
nature.
This general covariance also introduce the notion of ``gauge'' in the
theory.
In the theory with general covariance, these ``gauges'' give rise to
the unphysical degree of freedom and we have to fix the ``gauges'' or
to extract some invariant quantities to obtain physical results.
Therefore, treatments of ``gauges'' are crucial in general relativity
and this situation becomes more delicate in general relativistic
perturbation theory as explained below.


In 1964, Sachs\cite{R.K.Sachs-1964} pointed out that there are two
kinds of ``gauges'' in general relativity.
Sachs called these two ``gauges'' as the first- and the second-kind of
gauges, respectively.
Here, we review these concepts of ``gauge,'' which are different from
each other.


\subsubsection{First kind gauge}
\label{sec:first-kind-gauge}


{\it The first kind gauge} is a coordinate system on a single manifold
${\cal M}$.
Although this first kind gauge is not important in this paper, we
explain this to emphasize the ``gauge'' discussing in this review is
different from this first kind gauge.


In the standard text book of manifolds (for example, see
\cite{Kobayashi-Nomizu-I-1996}), the following property of a manifold
is written,
``On a manifold, we can always introduce a coordinate system as a
diffeomorphism $\psi_{\alpha}$ from an open set
$O_{\alpha}\subset{\cal M}$ to an open set
$\psi_{\alpha}(O_{\alpha})\subset\RF^{n}$ ($n=\dim{\cal M}$).''
This diffeomorphism $\psi_{\alpha}$, i.e., coordinate system of the
open set $O_{\alpha}$, is called {\it gauge choice} (of the first
kind).
If we consider another open set in $O_{\beta}\subset{\cal M}$, we have
another gauge choice
$\psi_{\beta}:O_{\beta}\mapsto\psi_{\beta}(O_{\beta})\subset\RF^{n}$
for $O_{\beta}$.
If these two open sets $O_{\alpha}$ and $O_{\beta}$ have the
intersection $O_{\alpha}\cap O_{\beta}\neq\emptyset$, we can consider
the diffeomorphism $\psi_{\beta}\circ\psi_{\alpha}^{-1}$.
This diffeomorphism $\psi_{\beta}\circ\psi_{\alpha}^{-1}$ is just a
coordinate transformation: $\psi_{\alpha}(O_{\alpha}\cap
O_{\beta})\subset\RF^{n}\mapsto \psi_{\beta}(O_{\alpha}\cap
O_{\beta})\subset\RF^{n}$, which is called {\it gauge transformation}
(of the first kind) in general relativity.


According to the theory of a manifold, coordinate system are not on a
manifold itself but we can always introduce a coordinate system
through a map from an open set in the manifold ${\cal M}$ to an open
set of $\RF^{n}$.
For this reason, general covariance in general relativity is
automatically included in the premise that our spacetime is regarded
as a single manifold.
The first kind gauge does arise due to this general covariance.
The gauge issue of the first kind is usually represented by the question,
``Which coordinate system is convenient?''
The answer to this question depends on the problem which we are
addressing, i.e., what we want to clarify.
In some case, this gauge issue of the first kind is an important.
However, in many case, it becomes harmless if we apply a covariant
theory on the manifold.


\subsubsection{Second kind gauge}
\label{sec:second-kind-gauge}


{\it The second kind gauge} appears in perturbation theories in a
theory with general covariance.
This notion of the second kind ``gauge'' is the main issue of this
article.
To explain this, we have to remind what we are doing in perturbation
theories.


First, in any perturbation theories, we always treat two spacetime
manifolds.
One is the {\it physical spacetime} ${\cal M}$.
We want to describe the properties of this physical spacetime
${\cal M}$ through perturbative analyses.
This physical spacetime ${\cal M}$ is usually identified with our
nature itself.
The other is the {\it background spacetime} ${\cal M}_{0}$.
This background spacetime have nothing to do with our nature and
is a fictitious manifold which is introduced as a reference to carry
out perturbative analyses by us.
We emphasize that these two spacetime manifolds ${\cal M}$ and
${\cal M}_{0}$ are distinct.
Let us denote the physical spacetime by $({\cal M},\bar{g}_{ab})$ and
the background spacetime by $({\cal M}_{0},g_{ab})$, where
$\bar{g}_{ab}$ is the metric on the physical spacetime manifold,
${\cal M}$, and $g_{ab}$ is the metric on the background spacetime
manifold, ${\cal M}_{0}$.
Further, we formally denote the spacetime metric and the other
physical tensor fields on ${\cal M}$ by $Q$ and its background value
on ${\cal M}_{0}$ by $Q_{0}$.


Second, in any perturbation theories, we always write equations for
the perturbation of the physical variable $Q$ in the form
\begin{equation}
  \label{eq:variable-symbolic-perturbation}
  Q(``p\mbox{''}) = Q_{0}(p) + \delta Q(p).
\end{equation}
Usually, this equation is simply regarded as a relation between the
physical variable $Q$ and its background value $Q_{0}$, or as the
definition of the deviation $\delta Q$ of the physical variable $Q$
from its background value $Q_{0}$.
However, Eq.~(\ref{eq:variable-symbolic-perturbation}) has deeper
implications.
Keeping in our mind that we always treat two different spacetimes,
${\cal M}$ and ${\cal M}_{0}$, in perturbation theory,
Eq.~(\ref{eq:variable-symbolic-perturbation}) is a rather curious
equation in the following sense:
The variable on the left-hand side of
Eq.~(\ref{eq:variable-symbolic-perturbation}) is a variable on
${\cal M}$, while the variables on the right-hand side of
Eq.~(\ref{eq:variable-symbolic-perturbation}) are variables on
${\cal M}_{0}$.
Hence, Eq.~(\ref{eq:variable-symbolic-perturbation}) gives a relation
between variables on two different manifolds.


Furthermore, through Eq.~(\ref{eq:variable-symbolic-perturbation}), we
have implicitly identified points in these two different manifolds.
More specifically, $Q(``p\mbox{''})$ on the left-hand side of
Eq.~(\ref{eq:variable-symbolic-perturbation}) is a field on
${\cal M}$, and $``p\mbox{''}\in{\cal M}$.
Similarly, we should regard the background value $Q_{0}(p)$ of
$Q(``p\mbox{''})$ and its deviation $\delta Q(p)$ of $Q(``p\mbox{''})$
from $Q_{0}(p)$, which are on the right-hand side of
Eq.~(\ref{eq:variable-symbolic-perturbation}), as fields on
${\cal M}_{0}$, and $p\in{\cal M}_{0}$.
Because Eq.~(\ref{eq:variable-symbolic-perturbation}) is regarded as
an equation for a field variable, it implicitly states that the points
$``p\mbox{''}\in{\cal M}$ and $p\in{\cal M}_{0}$ are same.
This represents the implicit assumption of the existence of a map
${\cal M}_{0}\rightarrow{\cal M}$ $:$ $p\in{\cal M}_{0}\mapsto
``p\mbox{''}\in{\cal M}$, which is usually called a {\it gauge choice}
(of the second kind) in perturbation
theory\cite{J.M.Stewart-M.Walker11974,J.M.Stewart-M.Walker11990,J.M.Stewart-1991}.


It is important to note that the second kind gauge choice between
points on ${\cal M}_{0}$ and ${\cal M}$, which is established by such
a relation as Eq.~(\ref{eq:variable-symbolic-perturbation}), is not
unique in theories with general covariance.
Rather, Eq.~(\ref{eq:variable-symbolic-perturbation}) involves the
degree of freedom corresponding to the choice of the map ${\cal X}$
$:$ ${\cal M}_{0}\mapsto{\cal M}$.
This is called the {\it gauge degree of freedom} (of the second kind).
Such a degree of freedom always exists in perturbations of a theory
with general covariance.
General covariance intuitively means that there is no preferred
coordinate system in the theory as mentioned above.
If general covariance is not imposed on the theory, there is a
preferred coordinate system in the theory, and we naturally introduce
this preferred coordinate system onto both ${\cal M}_{0}$ and
${\cal M}$.
Then, we can choose the identification map ${\cal X}$ using this
preferred coordinate system.
However, there is no such coordinate system in general relativity due
to general covariance, and we have no guiding principle to choose the
identification map ${\cal X}$.
Indeed, we may identify $``p\mbox{''}\in{\cal M}$ with
$q\in{\cal M}_{0}$ ($q\neq p$) instead of $p\in{\cal M}_{0}$.
In the above understanding of the concept of ``gauge'' (of the second
kind) in general relativistic perturbation theory, a gauge
transformation is simply a change of the map ${\cal X}$.


These are the basic ideas of gauge degree of freedom (of the second
kind) in the general relativistic perturbation theory which are
pointed out by Sacks\cite{R.K.Sachs-1964} and mathematically clarified
by Stewart and Walker\cite{J.M.Stewart-M.Walker11974,J.M.Stewart-M.Walker11990,J.M.Stewart-1991}.
Based on these ideas, higher-order perturbation theory has been
developed in Refs.~\cite{kouchan-gauge-inv,kouchan-second,M.Bruni-S.Matarrese-S.Mollerach-S.Soonego-CQG1997,kouchan-cosmo-second-letter,kouchan-cosmo-second-full-paper,kouchan-second-cosmo-matter,kouchan-second-cosmo-consistency,A.J.Christopherson-et.al-2011,K.Nakamura-CQG-Letter-2011,K.Nakamura-IJMP-globalization-2012,K.Nakamura-Progress-Construction-2013,K.Nakamura-CQG-Recursive-2014,kouchan-LTVII,Bruni-Gualtieri-Sopuerta-2003,Sopuerta-Bruni-Gualtieri-2004,M.Bruni-S.Sonego-CQG1999}.


\subsection{Formulation of perturbation theory}
\label{sec:Formulation-of-perturbation-theory}


To formulate the above understanding in more detail, we introduce an
infinitesimal parameter $\lambda$ for the perturbation.
Further, we consider the $4+1$-dimensional manifold
${\cal N}={\cal M}\times\RF$, where $4=\dim{\cal M}$ and
$\lambda\in\MR$.
The background spacetime
${\cal M}_{0}=\left.{\cal N}\right|_{\lambda=0}$ and the physical
spacetime
${\cal M}={\cal M}_{\lambda}=\left.{\cal N}\right|_{\MR=\lambda}$ are
also submanifolds embedded in the extended manifold ${\cal N}$.
Each point on ${\cal N}$ is identified by a pair $(p,\lambda)$, where
$p\in{\cal M}_{\lambda}$, and each point in
${\cal M}_{0}\subset{\cal N}$ is identified by $\lambda=0$.


Through this construction, the manifold ${\cal N}$ is foliated by
four-dimensional submanifolds ${\cal M}_{\lambda}$ of each $\lambda$,
and these are diffeomorphic to ${\cal M}$ and ${\cal M}_{0}$.
The manifold ${\cal N}$ has a natural differentiable structure
consisting of the direct product of ${\cal M}$ and $\RF$.
Further, the perturbed spacetimes ${\cal M}_{\lambda}$ for each
$\lambda$ must have the same differential structure with this
construction.
In other words, we require that perturbations be continuous in the
sense that ${\cal M}$ and ${\cal M}_{0}$ are connected by a continuous
curve within the extended manifold ${\cal N}$.
Hence, the changes of the differential structure resulting from the
perturbation, for example the formation of singularities and singular
perturbations in the sense of fluid mechanics, are excluded from
consideration.


Let us consider the set of field equations
\begin{equation}
  \label{eq:field-eq-for-Q}
  {\cal E}[Q_{\lambda}] = 0
\end{equation}
on the physical spacetime ${\cal M}_{\lambda}$ for the physical
variables $Q_{\lambda}$ on ${\cal M}_{\lambda}$.
The field equation (\ref{eq:field-eq-for-Q}) formally represents the
Einstein equation for the metric on ${\cal M}_{\lambda}$ and the
equations for matter fields on ${\cal M}_{\lambda}$.
If a tensor field $Q_{\lambda}$ is given on each ${\cal M}_{\lambda}$,
$Q_{\lambda}$ is automatically extended to a tensor field on
${\cal N}$ by $Q(p,\lambda):=Q_{\lambda}(p)$, where
$p\in{\cal M}_{\lambda}$.
In this extension, the field equation (\ref{eq:field-eq-for-Q}) is
regarded as an equation on the extended manifold ${\cal N}$.
Thus, we have extended an arbitrary tensor field and the field
equations (\ref{eq:field-eq-for-Q}) on each ${\cal M}_{\lambda}$ to
those on the extended manifold ${\cal N}$.


Tensor fields on ${\cal N}$ obtained through the above construction
are necessarily ``tangent'' to each ${\cal M}_{\lambda}$.
To consider the basis of the tangent space of ${\cal N}$, we introduce
the normal form and its dual, which are normal to each
${\cal M}_{\lambda}$ in ${\cal N}$.
These are denoted by $(d\lambda)_{a}$ and
$(\partial/\partial\lambda)^{a}$, respectively, and they satisfy
$(d\lambda)_{a}(\partial/\partial\lambda)^{a}=1$.
The form $(d\lambda)_{a}$ and its dual,
$(\partial/\partial\lambda)^{a}$, are normal to any tensor field
extended from the tangent space on each ${\cal M}_{\lambda}$ through
the above construction.
The set consisting of $(d\lambda)_{a}$,
$(\partial/\partial\lambda)^{a}$ and the basis of the tangent space on
each ${\cal M}_{\lambda}$ is regarded as the basis of the tangent
space of ${\cal N}$.


Now, we define the perturbation of an arbitrary tensor field $Q$.
We compare $Q$ on ${\cal M}_{\lambda}$ with $Q_{0}$ on ${\cal M}_{0}$,
and it is necessary to identify the points of ${\cal M}_{\lambda}$
with those of ${\cal M}_{0}$ as mentioned above.
This point identification map is the gauge choice of the second kind
as mentioned above.
The gauge choice is made by assigning a diffeomorphism
${\cal X}_{\lambda}$ $:$ ${\cal N}$ $\rightarrow$ ${\cal N}$ such that
${\cal X}_{\lambda}$ $:$ ${\cal M}_{0}$ $\rightarrow$
${\cal M}_{\lambda}$.
Following the paper of Bruni et
al.~\cite{M.Bruni-S.Matarrese-S.Mollerach-S.Soonego-CQG1997}, we
introduce a gauge choice ${\cal X}_{\lambda}$ as an one-parameter
groups of diffeomorphisms, i.e., an exponential map, for simplicity.
We denote the generator of this exponential map by
${}_{{\cal X}}\!\eta^{a}$.
This generator ${}_{{\cal X}}\!\eta^{a}$ is decomposed by the basis on
${\cal N}$ which are constructed above.
Although the generator ${}_{{\cal X}}\!\eta^{a}$ should satisfy some
appropriate properties\cite{kouchan-gauge-inv}, the arbitrariness of
the gauge choice ${\cal X}_{\lambda}$ is represented by the tangential
component of the generator ${}_{{\cal X}}\!\eta^{a}$ to
${\cal M}_{\lambda}$.


\begin{figure}
  \begin{center}
    \includegraphics[width=0.5\textwidth]{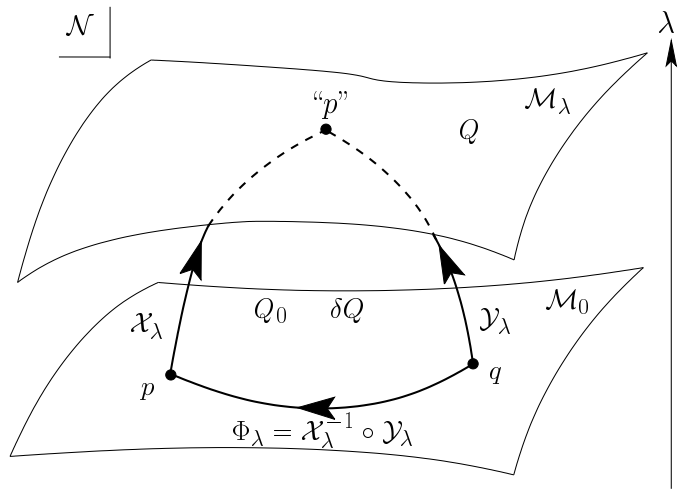}
  \end{center}
  \caption{
    The second kind gauge is a point-identification between the
    physical spacetime ${\cal M}_{\lambda}$ and the background
    spacetime ${\cal M}_{0}$ on the extended manifold ${\cal N}$.
    Through Eq.~(\ref{eq:variable-symbolic-perturbation}), we
    implicitly assume the existence of a point-identification map
    between ${\cal M}_{\lambda}$ and ${\cal M}_{0}$.
    However, this point-identification is not unique by virtue of the
    general covariance in the theory.
    We may chose the gauge of the second kind so that
    $p\in{\cal M}_{0}$ and ``$p$''$\in{\cal M}_{\lambda}$ is same
    (${\cal X}_{\lambda}$).
    We may also choose the gauge so that $q\in{\cal M}_{0}$ and
    ``$p$''$\in{\cal M}_{\lambda}$ is same (${\cal Y}_{\lambda}$).
    These are different gauge choices.
    The gauge transformation
    ${\cal X}_{\lambda}\rightarrow{\cal Y}_{\lambda}$ is given
    by the diffeomorphism
    $\Phi={\cal X}_{\lambda}^{-1}\circ{\cal Y}_{\lambda}$.
  }
  \label{fig:gauge-choice-is-point-identification}
\end{figure}


The pull-back ${\cal X}_{\lambda}^{*}Q$, which is induced by the
exponential map ${\cal X}_{\lambda}$, maps a tensor field $Q$ on the
physical manifold ${\cal M}_{\lambda}$ to a tensor field
${\cal X}_{\lambda}^{*}Q$ on the background spacetime.
In terms of this generator ${}_{{\cal X}}\!\eta^{a}$, the pull-back
${\cal X}_{\lambda}^{*}Q$ is represented by the Taylor expansion
\begin{eqnarray}
  Q(r)
  &=&
  Q({\cal X}_{\lambda}(p))
  =
  {\cal X}_{\lambda}^{*}Q(p)
  \nonumber\\
  &=&
  Q(p)
  + \lambda \left.{\pounds}_{{}_{{\cal X}}\!\eta}Q \right|_{p}
  + \frac{1}{2} \lambda^{2}
  \left.{\pounds}_{{}_{{\cal X}}\!\eta}^{2}Q\right|_{p}
  + O(\lambda^{3}),
  \label{eq:Taylor-expansion-of-calX-org}
\end{eqnarray}
where $r={\cal X}_{\lambda}(p)\in{\cal M}_{\lambda}$.
Because $p\in{\cal M}_{0}$, we may regard the equation
\begin{eqnarray}
  {\cal X}_{\lambda}^{*}Q(p)
  &=&
  Q_{0}(p)
  + \lambda \left.{\pounds}_{{}_{{\cal X}}\!\eta}Q\right|_{{\cal M}_{0}}(p)
  + \frac{1}{2} \lambda^{2}
  \left.{\pounds}_{{}_{{\cal X}}\!\eta}^{2}Q\right|_{{\cal M}_{0}}(p)
  + O(\lambda^{3})
  \label{eq:Taylor-expansion-of-calX}
\end{eqnarray}
as an equation on the background spacetime ${\cal M}_{0}$, where
$Q_{0}=\left.Q\right|_{{\cal M}_{0}}$ is the background value of the
physical variable of $Q$.
Once the definition of the pull-back of the gauge choice
${\cal X}_{\lambda}$ is given, the first- and the second-order
perturbations ${}^{(1)}_{\;\cal X}\!Q$ and ${}^{(2)}_{\;\cal X}\!Q$ of
a tensor field $Q$ under the gauge choice ${\cal X}_{\lambda}$ are
simply given by the expansion
\begin{equation}
  \label{eq:Bruni-35}
  \left.{\cal X}^{*}_{\lambda}Q_{\lambda}\right|_{{\cal M}_{0}}
  =
  Q_{0}
  + \lambda {}^{(1)}_{\;\cal X}\!Q
  + \frac{1}{2} \lambda^{2} {}^{(2)}_{\;\cal X}\!Q
  + O(\lambda^{3})
\end{equation}
with respect to the infinitesimal parameter $\lambda$.
Comparing Eqs.~(\ref{eq:Taylor-expansion-of-calX}) and
(\ref{eq:Bruni-35}), we define the first- and the second-order
perturbations of a physical variable $Q_{\lambda}$ under the gauge
choice ${\cal X}_{\lambda}$ by
\begin{eqnarray}
  {}^{(1)}_{\;\cal X}\!Q :=
  \left.{\pounds}_{{}_{\cal X}\!\eta} Q\right|_{{\cal M}_{0}},
  \quad
  {}^{(2)}_{\;\cal X}\!Q :=
  \left.{\pounds}_{{}_{\cal X}\!\eta}^{2} Q\right|_{{\cal M}_{0}}.
  \label{eq:representation-of-each-order-perturbation}
\end{eqnarray}
We note that all variables in Eq.~(\ref{eq:Bruni-35}) are defined on
${\cal M}_{0}$.


Now, we consider two {\it different gauge choices} based on the above
understanding of the second kind gauge choice.
Suppose that ${\cal X}_{\lambda}$ and ${\cal Y}_{\lambda}$ are two
exponential maps with the generators ${}_{\cal X}\!\eta^{a}$ and
${}_{\cal Y}\!\eta^{a}$ on ${\cal N}$, respectively.
In other words, ${\cal X}_{\lambda}$ and ${\cal Y}_{\lambda}$ are two
gauge choices (see
Fig.~\ref{fig:gauge-choice-is-point-identification}).
Then, the integral curves of each ${}_{\cal X}\!\eta^{a}$ and
${}_{\cal Y}\!\eta^{a}$ in ${\cal N}$ are the orbits of the actions of
the gauge choices ${\cal X}_{\lambda}$ and ${\cal Y}_{\lambda}$,
respectively.
Since we choose the generators ${}_{\cal X}\!\eta^{a}$ and
${}_{\cal Y}\!\eta^{a}$ so that these are transverse to each
${\cal M}_{\lambda}$ everywhere on ${\cal N}$, the integral curves of
these vector fields intersect with each ${\cal M}_{\lambda}$.
Therefore, points lying on the same integral curve of either of the
two are to be regarded as {\it the same point} within the respective
gauges.
When these curves are not identical, i.e., the tangential components
to each ${\cal M}_{\lambda}$ of ${}_{\cal X}\!\eta^{a}$ and
${}_{\cal Y}\!\eta^{a}$ are different, these point identification maps
${\cal X}_{\lambda}$ and ${\cal Y}_{\lambda}$ are regarded as
{\it two different gauge choices}.


We next introduce the concept of {\it gauge invariance}.
In particular, we consider the concept of {\it order by order gauge
  invariance}~\cite{kouchan-second-cosmo-matter}, in this article.
Suppose that ${\cal X}_{\lambda}$ and ${\cal Y}_{\lambda}$ are two
different gauge choices which are generated by the vector fields
${}_{\cal X}\!\eta^{a}$ and ${}_{\cal Y}\!\eta^{a}$, respectively.
These gauge choices also pull back a generic tensor field $Q$ on
${\cal N}$ to two other tensor fields, ${\cal X}_{\lambda}^{*}Q$ and
${\cal Y}_{\lambda}^{*}Q$, for any given value of $\lambda$.
In particular, on ${\cal M}_{0}$, we now have three tensor fields
associated with a tensor field $Q$; one is the background value
$Q_{0}$ of $Q$, and the other two are the pulled-back variables of $Q$
from ${\cal M}_{\lambda}$ to ${\cal M}_{0}$ by the two different gauge
choices,
\begin{eqnarray}
  {}_{\cal X}\!Q_{\lambda} &:=&
  \left.{\cal X}^{*}_{\lambda}Q\right|_{{\cal M}_{0}}
  \nonumber\\
  &=&
  Q_{0}
  + \lambda {}^{(1)}_{\;{\cal X}}\!Q
  + \frac{1}{2} \lambda^{2} {}^{(2)}_{\;{\cal X}}\!Q
  + O(\lambda^{3})
  \label{eq:Bruni-39-one}
  \\
  {}_{\cal Y}\!Q_{\lambda} &:=&
  \left.{\cal Y}^{*}_{\lambda}Q\right|_{{\cal M}_{0}}
  \nonumber\\
  &=&
  Q_{0}
  + \lambda {}^{(1)}_{\;{\cal Y}}\!Q
  + \frac{1}{2} \lambda^{2} {}^{(2)}_{\;{\cal Y}}\!Q
  + O(\lambda^{3})
  \label{eq:Bruni-40-one}
\end{eqnarray}
Here, we have used Eq.~(\ref{eq:Bruni-35}).
Because ${\cal X}_{\lambda}$ and ${\cal Y}_{\lambda}$ are gauge
choices which map from ${\cal M}_{0}$ to ${\cal M}_{\lambda}$,
${}_{\cal X}Q_{\lambda}$ and ${}_{\cal Y}Q_{\lambda}$ are the
different representations on ${\cal M}_{0}$ in the two different
gauges of the same perturbed tensor field $Q$ on
${\cal M}_{\lambda}$.
The quantities ${}^{(k)}_{\;\cal X}\!Q$ and ${}^{(k)}_{\;\cal Y}\!Q$
in Eqs.~(\ref{eq:Bruni-39-one}) and (\ref{eq:Bruni-40-one}) are the
perturbations of $O(k)$ in the gauges ${\cal X}_{\lambda}$ and
${\cal Y}_{\lambda}$, respectively.
We say that the $k$th-order perturbation ${}^{(k)}_{\;\cal X}\!Q$ of
$Q$ is {\it order by order gauge invariant} if and only if for any two
gauges ${\cal X}_{\lambda}$ and ${\cal Y}_{\lambda}$ the following
holds:
\begin{equation}
  {}^{(k)}_{\;\cal X}\!Q = {}^{(k)}_{\;\cal Y}\!Q.
\end{equation}


Now, we consider the {\it gauge transformation rules} between
different gauge choices.
In general, the representation ${}^{\cal X}Q_{\lambda}$ on
${\cal M}_{0}$ of the perturbed variable $Q$ on ${\cal M}_{\lambda}$
depends on the gauge choice ${\cal X}_{\lambda}$.
If we employ a different gauge choice, the representation of
$Q_{\lambda}$ on ${\cal M}_{0}$ may change.
Suppose that ${\cal X}_{\lambda}$ and ${\cal Y}_{\lambda}$ are
different gauge choices, which are the point identification maps from
${\cal M}_{0}$ to ${\cal M}_{\lambda}$, and the generators of these
gauge choices are given by ${}_{\cal X}\!\eta^{a}$ and
${}_{\cal Y}\!\eta^{a}$, respectively.
Then, the change of the gauge choice from ${\cal X}_{\lambda}$ to
${\cal Y}_{\lambda}$ is represented by the diffeomorphism
\begin{equation}
  \label{eq:diffeo-def-from-Xinv-Y}
  \Phi_{\lambda} :=
  ({\cal X}_{\lambda})^{-1}\circ{\cal Y}_{\lambda}.
\end{equation}
This diffeomorphism $\Phi_{\lambda}$ is the map $\Phi_{\lambda}$ $:$
${\cal M}_{0}$ $\rightarrow$ ${\cal M}_{0}$ for each value of
$\lambda\in\MR$.
The diffeomorphism $\Phi_{\lambda}$ does change the point
identification, as expected from the understanding of the gauge choice
discussed above.
Therefore, the diffeomorphism $\Phi_{\lambda}$ is regarded as the
gauge transformation $\Phi_{\lambda}$ $:$ ${\cal X}_{\lambda}$
$\rightarrow$ ${\cal Y}_{\lambda}$.


The gauge transformation $\Phi_{\lambda}$ induces a pull-back from the
representation ${}_{\cal X}\!Q_{\lambda}$ of the perturbed tensor
field $Q$ in the gauge choice ${\cal X}_{\lambda}$ to the
representation ${}_{\cal Y}\!Q_{\lambda}$ in the gauge choice
${\cal Y}_{\lambda}$.
Actually, the tensor fields ${}_{\cal X}\!Q_{\lambda}$ and
${}_{\cal Y}\!Q_{\lambda}$, which are defined on ${\cal M}_{0}$, are
connected by the linear map $\Phi^{*}_{\lambda}$ as
\begin{eqnarray}
  {}_{\cal Y}\!Q_{\lambda}
  &=&
  \left.{\cal Y}^{*}_{\lambda}Q\right|_{{\cal M}_{0}}
  =
  \left.\left(
      {\cal Y}^{*}_{\lambda}
      \left({\cal X}_{\lambda}
      {\cal X}_{\lambda}^{-1}\right)^{*}Q\right)
  \right|_{{\cal M}_{0}}
  \nonumber\\
  &=&
  \left.
    \left(
      {\cal X}^{-1}_{\lambda}
      {\cal Y}_{\lambda}
    \right)^{*}
    \left(
      {\cal X}^{*}_{\lambda}Q
    \right)
  \right|_{{\cal M}_{0}}
  =  \Phi^{*}_{\lambda} {}_{\cal X}\!Q_{\lambda}.
  \label{eq:Bruni-45-one}
\end{eqnarray}
According to generic arguments concerning the Taylor expansion of the
pull-back of a tensor field on the same manifold, given in
\S\ref{sec:Taylor-expansion-of-tensors-on-a-manifold}, it should be
possible to express the gauge transformation
$\Phi^{*}_{\lambda} {}_{\cal X}\!Q_{\lambda}$ in the form
\begin{eqnarray}
  \Phi^{*}_{\lambda} {}_{\cal X}\!Q = {}_{\cal X}\!Q
  + \lambda {\pounds}_{\xi_{1}} {}_{\cal X}\!Q
  + \frac{\lambda^{2}}{2} \left\{
    {\pounds}_{\xi_{2}} + {\pounds}_{\xi_{1}}^{2}
  \right\} {}_{\cal X}\!Q
  + O(\lambda^{3}),
  \label{eq:Bruni-46-one}
\end{eqnarray}
where the vector fields $\xi_{1}^{a}$ and $\xi_{2}^{a}$ are the
generators of the gauge transformation $\Phi_{\lambda}$ (see
Eq.~(\ref{eq:Taylor-expansion-of-f})).


Comparing the representation (\ref{eq:Bruni-46-one}) of the Taylor
expansion in terms of the generators $\xi_{1}^{a}$ and $\xi_{2}^{a}$
of the pull-back $\Phi_{\lambda}^{*}{}_{\cal X}\!Q$ and that in terms
of the generators ${}_{\cal X}\!\eta^{a}$ and ${}_{\cal Y}\!\eta^{a}$
of the pull-back
${\cal Y}^{*}_{\lambda}\circ\left({\cal X}_{\lambda}^{-1}\right)^{*}\;{}_{{\cal X}}\!Q$
($=\Phi_{\lambda}^{*}{}_{\cal X}\!Q$), we readily obtain explicit
expressions for the generators $\xi_{1}^{a}$ and $\xi_{2}^{a}$ of the
gauge transformation
$\Phi={\cal X}^{-1}_{\lambda}\circ{\cal Y}_{\lambda}$ in terms of the
generators ${}_{\cal X}\!\eta^{a}$ and ${}_{\cal Y}\!\eta^{a}$ of each
gauge choices as follows:
\begin{eqnarray}
  \xi_{1}^{a}
  =
  {}_{\cal Y}\!\eta^{a}
  -
  {}_{\cal X}\!\eta^{a},
  \quad
  \xi_{2}^{a}
  =
  \left[
    {}_{\cal Y}\!\eta
    ,
    {}_{\cal X}\!\eta
  \right]^{a}.
  \label{eq:relation-between-xi-eta}
\end{eqnarray}
Further, because the gauge transformation $\Phi_{\lambda}$ is a map
within the background spacetime ${\cal M}_{0}$, the generator should
consist of vector fields on ${\cal M}_{0}$.
This can be satisfied by imposing some appropriate conditions on the
generators ${}_{\cal Y}\!\eta^{a}$ and ${}_{\cal X}\!\eta^{a}$.


We can now derive the relation between the perturbations in the two
different gauges.
Up to second order, these relations are derived by substituting
(\ref{eq:Bruni-39-one}) and (\ref{eq:Bruni-40-one}) into
(\ref{eq:Bruni-46-one}):
\begin{eqnarray}
  \label{eq:Bruni-47-one}
  {}^{(1)}_{\;{\cal Y}}\!Q - {}^{(1)}_{\;{\cal X}}\!Q &=&
  {\pounds}_{\xi_{1}}Q_{0}, \\
  \label{eq:Bruni-49-one}
  {}^{(2)}_{\;\cal Y}\!Q - {}^{(2)}_{\;\cal X}\!Q &=&
  2 {\pounds}_{\xi_{1}} {}^{(1)}_{\;\cal X}\!Q
  +\left\{{\pounds}_{\xi_{2}}+{\pounds}_{\xi_{1}}^{2}\right\} Q_{0}.
\end{eqnarray}


Here, we should comment on the gauge choice in the above explanation.
We have introduced an exponential map ${\cal X}_{\lambda}$ (or
${\cal Y}_{\lambda}$) as the gauge choice, for simplicity.
However, this simplified introduction of ${\cal X}_{\lambda}$ as an
exponential map is not essential to the gauge transformation rules
(\ref{eq:Bruni-47-one}) and (\ref{eq:Bruni-49-one}).
Actually, we can generalize the diffeomorphism ${\cal X}_{\lambda}$
from an exponential map.
For example, the diffeomorphism whose pull-back is represented by the
Taylor expansion (\ref{eq:Taylor-expansion-of-f}) is a candidate of
the generalization.
If we generalize the diffeomorphism ${\cal X}_{\lambda}$, the
representation (\ref{eq:Taylor-expansion-of-calX}) of the pulled-back
variable ${\cal X}_{\lambda}^{*}Q(p)$, the representations of the
perturbations (\ref{eq:representation-of-each-order-perturbation}),
and the relations (\ref{eq:relation-between-xi-eta}) between
generators of $\Phi_{\lambda}$, ${\cal X}_{\lambda}$, and
${\cal Y}_{\lambda}$ will be changed.
However, the gauge transformation rules (\ref{eq:Bruni-47-one}) and
(\ref{eq:Bruni-49-one}) are direct consequences of the generic Taylor
expansion (\ref{eq:Bruni-46-one}) of $\Phi_{\lambda}$.
Generality of the representation of the Taylor expansion
(\ref{eq:Bruni-46-one}) of $\Phi_{\lambda}$ implies that the gauge
transformation rules (\ref{eq:Bruni-47-one}) and
(\ref{eq:Bruni-49-one}) will not be changed, even if we generalize the
each gauge choice ${\cal X}_{\lambda}$.
Further, the relations (\ref{eq:relation-between-xi-eta}) between
generators also imply that, even if we employ simple exponential maps
as gauge choices, both of the generators $\xi_{1}^{a}$ and
$\xi_{2}^{a}$ are naturally induced by the generators of the original
gauge choices.
Hence, we conclude that the gauge transformation rules
(\ref{eq:Bruni-47-one}) and (\ref{eq:Bruni-49-one}) are quite general
and irreducible.
In this article, we review the development of a second-order
gauge-invariant cosmological perturbation theory based on the above
understanding of the gauge degree of freedom only through the gauge
transformation rules (\ref{eq:Bruni-47-one}) and
(\ref{eq:Bruni-49-one}).
Hence, the developments of the cosmological perturbation theory
presented below will not be changed even if we generalize the gauge
choice ${\cal X}_{\lambda}$ from a simple exponential map.


We also have to emphasize the physical implication of the gauge
transformation rules (\ref{eq:Bruni-47-one}) and
(\ref{eq:Bruni-49-one}).
According to the above construction of the perturbation theory, gauge
degree of freedom, which induces the transformation rules
(\ref{eq:Bruni-47-one}) and (\ref{eq:Bruni-49-one}), is unphysical
degree of freedom.
As emphasized above, the physical spacetime ${\cal M}_{\lambda}$ is
identified with our nature itself, while there is no background
spacetime ${\cal M}_{0}$ in our nature.
The background spacetime ${\cal M}_{0}$ is a fictitious spacetime and
it have nothing to do with our nature.
Since the gauge choice ${\cal X}_{\lambda}$ just gives a relation
between ${\cal M}_{\lambda}$ and ${\cal M}_{0}$, the gauge choice
${\cal X}_{\lambda}$ also have nothing to do with our nature.
On the other hand, any observations and experiments are carried out
only on the physical spacetime ${\cal M}_{\lambda}$ through the
physical processes within the physical spacetime ${\cal M}_{\lambda}$.
Therefore, any direct observables in any observations and experiments
should be independent of the gauge choice ${\cal X}_{\lambda}$, i.e.,
should be gauge invariant.
Keeping this fact in our mind, the gauge transformation rules
(\ref{eq:Bruni-47-one}) and (\ref{eq:Bruni-49-one}) imply that the
perturbations ${}^{(1)}_{\;{\cal X}}\!Q$ and ${}^{(2)}_{\;\cal X}\!Q$
include unphysical degree of freedom, i.e., gauge degree of freedom,
if these perturbations are transformed as (\ref{eq:Bruni-47-one}) or
(\ref{eq:Bruni-49-one}) under the gauge transformation ${\cal
  X}_{\lambda}\rightarrow{\cal Y}_{\lambda}$.
If the perturbations ${}^{(1)}_{\;{\cal X}}\!Q$ and
${}^{(2)}_{\;\cal X}\!Q$ are independent of the gauge choice, these
variables are order by order gauge invariant.
Therefore, order by order gauge-invariant variables does not include
unphysical degree of freedom and should be related to the physics on
the physical spacetime ${\cal M}_{\lambda}$.


\subsection{Coordinate transformations induced by the second
  kind gauge transformation}
\label{sec:Induced-coordiante-transformations}


In many literature, gauge degree of freedom is regarded as the degree
of freedom of the coordinate transformation.
In the linear-order perturbation theory, these two degree of freedom
are equivalent with each other.
However, in the higher order perturbations, we should regard that
these two degree of freedom are different.
Although the essential understanding of the gauge degree of freedom
(of the second kind) is as that explained above, the gauge
transformation (of the second kind) also induces the infinitesimal
coordinate transformation on the physical spacetime
${\cal M}_{\lambda}$ as a result.
In many case, the understanding of ``gauges'' in perturbations based
on coordinate transformations leads mistakes.
Therefore, we did not use any ingredient of this subsection in our
series of
papers~\cite{kouchan-gauge-inv,kouchan-second,kouchan-cosmo-second-letter,kouchan-cosmo-second-full-paper,kouchan-second-cosmo-matter,kouchan-second-cosmo-consistency}
concerning about higher-order general relativistic gauge-invariant
perturbation theory.
However, we comment on the relations between the coordinate
transformation, briefly.
Details can be seen in
Refs.~\cite{Matarrese-Mollerach-Bruni-1998,Bruni-Gualtieri-Sopuerta-2003,kouchan-gauge-inv}.


To see that the gauge transformation of the second kind induces the
coordinate transformation, we introduce the coordinate system
$\{O_{\alpha},\psi_{\alpha}\}$ on the ``background spacetime''
${\cal M}_{0}$, where $O_{\alpha}$ are open sets on the background
spacetime and $\psi_{\alpha}$ are diffeomorphisms from $O_{\alpha}$ to
$\RF^{4}$ ($4=\dim{{\cal M}_{0}}$).
The coordinate system $\{O_{\alpha},\psi_{\alpha}\}$ is the set of the
collection of the pair of open sets $O_{\alpha}$ and diffeomorphism
$\psi_{\alpha}:$ $O_{\alpha}\mapsto\RF^{4}$.
If we employ a gauge choice ${\cal X}_{\lambda}$, we have the
correspondence of ${\cal M}_{\lambda}$ and ${\cal M}_{0}$.
Together with the coordinate system $\psi_{\alpha}$ on ${\cal M}_{0}$,
this correspondence between ${\cal M}_{\lambda}$ and ${\cal M}_{0}$
induces the coordinate system on ${\cal M}_{\lambda}$.
Actually, ${\cal X}_{\lambda}(O_{\alpha})$ for each $\alpha$ is an
open set of ${\cal M}_{\lambda}$.
Then, $\psi_{\alpha}\circ{\cal X}_{\lambda}^{-1}$ becomes a
diffeomorphism from an open set
${\cal X}_{\lambda}(O_{\alpha})\subset{\cal M}_{\lambda}$ to
$\RF^{4}$.
This diffeomorphism $\psi_{\alpha}\circ{\cal X}_{\lambda}^{-1}$
induces a coordinate system of an open set on ${\cal M}_{\lambda}$.


When we have two different gauge choices ${\cal X}_{\lambda}$ and
${\cal Y}_{\lambda}$, $\psi_{\alpha}\circ{\cal X}_{\lambda}^{-1}$ and
$\psi_{\alpha}\circ{\cal Y}_{\lambda}^{-1}$ become different
coordinate systems on ${\cal M}_{\lambda}$.
We can also consider the coordinate transformation from the coordinate
system $\psi_{\alpha}\circ{\cal X}_{\lambda}^{-1}$ to another
coordinate system $\psi_{\alpha}\circ{\cal Y}_{\lambda}^{-1}$.
Since the gauge transformation
${\cal X}_{\lambda}$ $\rightarrow$ ${\cal Y}_{\lambda}$ is induced by
the diffeomorphism $\Phi_{\lambda}$ defined by
Eq.~(\ref{eq:diffeo-def-from-Xinv-Y}), the induced coordinate
transformation is given by
\begin{eqnarray}
  y^{\mu}(q) := x^{\mu}(p) = \left(\left(\Phi^{-1}\right)^{*}x^{\mu}\right)(q)
\end{eqnarray}
in the {\it passive} point of
view~\cite{Matarrese-Mollerach-Bruni-1998,Bruni-Gualtieri-Sopuerta-2003,kouchan-gauge-inv}.
If we represent this coordinate transformation in terms of the Taylor
expansion in Sec.~\ref{sec:Taylor-expansion-of-tensors-on-a-manifold},
up to third order, we have the coordinate transformation
\begin{eqnarray}
  y^{\mu}(q) &=& x^{\mu}(q) - \lambda \xi^{\mu}_{1}(q)
  + \frac{\lambda^{2}}{2} \left\{
    - \xi^{\mu}_{2}(q)
    + \xi^{\nu}_{1}(q)\partial_{\nu}\xi^{\mu}_{1}(q)
  \right\}
  + O(\lambda^{3})
  .
\end{eqnarray}


Here again, we note that we have coordinate system
$\{{\cal X}_{\lambda}(O_{\alpha})$ $,$
$\psi_{\alpha}\circ{\cal X}_{\lambda}^{-1}\}$ ``on the physical
spacetime ${\cal M}_{\lambda}$'' if we introduce the coordinate system
$\psi_{\alpha}$ ``on the background spacetime ${\cal M}_{0}$'' and if
we introduce a diffeomorphism ${\cal X}_{\lambda}$ $:$ ${\cal M}_{0}$
$\mapsto$ ${\cal M}_{\lambda}$ as  the point identification map
between the background spacetime ${\cal M}_{0}$ and the physical
spacetime ${\cal M}_{\lambda}$.
If we apply the notion of the above (order-by-order) gauge-invariance
in the system, this application states that the system which we want
to describe is independent of the gauge-choice ${\cal X}_{\lambda}$.
On the other hand, if we apply the general covariance to the system
``on the background spacetime ${\cal M}_{0}$'', this application
implies that the system on the background spacetime ${\cal M}_{0}$ is
independent of the choice of the coordinate system
$\{O_{\alpha},\psi_{\alpha}\}$.
The general covariance ``on the background spacetime ${\cal M}_{0}$''
is accomplished by the introduction of a covariant theory on the
background spacetime.
In addition to this covariant theory ``on the background spacetime
${\cal M}_{0}$'', if we impose on the (order-by-order)
gauge-invariance for the perturbations, this implies that the system
``on the physical spacetime ${\cal M}_{\lambda}$'' is independent of
the choice of the coordinate system $\{{\cal X}_{\lambda}(O_{\alpha})$
$,$ $\psi_{\alpha}\circ{\cal X}_{\lambda}^{-1}\}$ ``on the physical
spacetime ${\cal M}_{\lambda}$''.
This is the statement of the general covariance ``on the physical
spacetime ${\cal M}_{\lambda}$''.
Thus, if we apply the gauge-invariance to ``perturbations'' together
with the covariant theory ``on the background spacetime
${\cal M}_{0}$'', this application corresponds to the general
covariance ``on the physical spacetime ${\cal M}_{\lambda}$''.
Therefore, the general covariance on the physical spacetime in
perturbation theory is guaranteed by the imposition of the
gauge-invariance to ``perturbations'' and a covariant theory on the
background spacetime.
This is the physical meaning of gauge-invariance for perturbations.


\subsection{Gauge-invariant variables}
\label{sec:gauge-invariant-variables}


Here, inspecting the gauge transformation rules
(\ref{eq:Bruni-47-one}) and (\ref{eq:Bruni-49-one}), we define the
gauge-invariant variables for the metric perturbations and for arbitrary
matter fields (tensor fields).
Employing the idea of order by order gauge invariance for
perturbations~\cite{kouchan-second-cosmo-matter} introduced in
Sec.~\ref{sec:Formulation-of-perturbation-theory}, we  proposed a
procedure to construct gauge-invariant variables of higher-order
perturbations~\cite{kouchan-gauge-inv}.
Our proposal is as follows.
First, we decompose a linear-order metric perturbation into its
gauge invariant and variant parts.
The procedure for decomposing linear-order metric perturbations is
easily extended to second-order metric perturbations, and we can
decompose the second-order metric perturbation into gauge invariant
and variant parts.
By using the gauge-variant parts of the first- and the second-order
metric perturbations, we can define the gauge-invariant variables for
the first- and second-order perturbations of an arbitrary field other
than the metric.


Now, we review the above strategy to construct gauge-invariant
variables.
To consider a metric perturbation, we expand the metric on the
physical spacetime ${\cal M}_{\lambda}$, which is pulled back to the
background spacetime ${\cal M}_{0}$ using a gauge choice in the form
given in (\ref{eq:Bruni-35}):
\begin{eqnarray}
  {\cal X}^{*}_{\lambda}\bar{g}_{ab}
  &=&
  g_{ab} + \lambda {}_{{\cal X}}\!h_{ab}
  + \frac{\lambda^{2}}{2} {}_{{\cal X}}\!l_{ab}
  + O^{3}(\lambda),
  \label{eq:metric-expansion}
\end{eqnarray}
where $g_{ab}$ is the metric on ${\cal M}_{0}$.
Of course, the expansion (\ref{eq:metric-expansion}) of the
metric depends entirely on the gauge choice ${\cal X}_{\lambda}$.
Nevertheless, henceforth, we do not explicitly express the index of
the gauge choice ${\cal X}_{\lambda}$ in an expression if there is no
possibility of confusion.


Our starting point to construct gauge-invariant variables is the
following conjecture:
\begin{conjecture}
  \label{conjecture:decomp_conjecture_for_hab}
  If there is a symmetric tensor field $h_{ab}$ of the second rank,
  whose gauge-transformation rule with the generator $\xi$ is given by
  \begin{eqnarray}
    \label{eq:hab-gauge-trans}
    {}_{\;{\cal Y}}\!h_{ab} - {}_{\;{\cal X}}\!h_{ab} = {\pounds}_{\xi}g_{ab},
  \end{eqnarray}
  Then there exist a tensor field ${\cal H}_{ab}$ and a vector field
  $X^{a}$ such that $h_{ab}$ is decomposed as
  \begin{eqnarray}
    \label{eq:hab-calHab+LieXgab}
    h_{ab} = : {\cal H}_{ab} + {\pounds}_{X}g_{ab},
  \end{eqnarray}
  where ${\cal H}_{ab}$ and $X^{a}$ are transformed as
  \begin{eqnarray}
    \label{eq:gauge-trans-calHab-Xa}
    {}_{\;{\cal Y}}\!{\cal H}_{ab} - {}_{\;{\cal X}}\!{\cal H}_{ab} = 0
    , \quad
    {}_{\;{\cal Y}}\!X^{a} - {}_{\;{\cal X}}\!X^{a} = \xi^{a}
  \end{eqnarray}
  under the gauge-transformation (\ref{eq:hab-gauge-trans}),
  respectively.
\end{conjecture}


In this conjecture, ${\cal H}_{ab}$ is gauge invariant and call
${\cal H}_{ab}$ as the {\it gauge-invariant part} of the tensor field
$h_{ab}$.
On the other hand, the vector field $X^{a}$ in
Eq.~(\ref{eq:hab-calHab+LieXgab}) is gauge dependent, and we call
$X^{a}$ as the {\it gauge-variant part} of the tensor field $h_{ab}$.


Since Conjecture~\ref{conjecture:decomp_conjecture_for_hab} can be
directly applied to the linear metric perturbation $h_{ab}$, a
linear metric perturbation $h_{ab}$ is decomposed as
\begin{eqnarray}
  h_{ab} =: {\cal H}_{ab} + {\pounds}_{X}g_{ab},
  \label{eq:linear-metric-decomp}
\end{eqnarray}
due to the Conjecture~\ref{conjecture:decomp_conjecture_for_hab},
where ${\cal H}_{ab}$ and $X^{a}$ are the gauge invariant and variant
parts of the linear-order metric perturbations $h_{ab}$, i.e., under
the gauge transformation (\ref{eq:Bruni-47-one}), these are
transformed as
\begin{equation}
    \label{eq:linear-metric-decomp-gauge-trans}
  {}_{{\cal Y}}\!{\cal H}_{ab} - {}_{{\cal X}}\!{\cal H}_{ab} =  0,
  \quad
  {}_{\quad{\cal Y}}\!X^{a} - {}_{{\cal X}}\!X^{a} = \xi^{a}_{1}
\end{equation}
due to Eqs.~(\ref{eq:gauge-trans-calHab-Xa}) in
Conjecture~\ref{conjecture:decomp_conjecture_for_hab}.


As emphasized in our series of papers
\cite{kouchan-gauge-inv,kouchan-second,kouchan-cosmo-second-letter,kouchan-cosmo-second-full-paper,kouchan-second-cosmo-matter,kouchan-second-cosmo-consistency,K.Nakamura-CQG-Letter-2011,K.Nakamura-IJMP-globalization-2012,K.Nakamura-Progress-Construction-2013,K.Nakamura-CQG-Recursive-2014},
Conjecture~\ref{conjecture:decomp_conjecture_for_hab} is still quite
non-trivial and it is not simple to carry out the systematic
decomposition (\ref{eq:linear-metric-decomp}) on an arbitrary
background spacetime, since this procedure depends completely on the
background spacetime $({\cal M}_{0},g_{ab})$.
Actually, in Ref.~\cite{K.Nakamura-Progress-Construction-2013}, we
showed an scenario of the proof of
Conjecture~\ref{conjecture:decomp_conjecture_for_hab}.
This scenario of the proof of
Conjecture~\ref{conjecture:decomp_conjecture_for_hab} and remaining
problems in this proof are briefly explained in
Appendix~\ref{sec:Outline-of-the-proof-of-the-decomposition-conjecture}.
This scenario is incomplete due to the non-local nature in the
definition of the gauge-invariant part ${\cal H}_{ab}$ and the
gauge-variant part $X^{a}$.
Furthermore,  as we will show below,
Conjecture~\ref{conjecture:decomp_conjecture_for_hab} is almost
correct at least in the case of cosmological perturbations of a
homogeneous and isotropic universe in
Sec.~\ref{sec:Gauge-invariant-metric-perturbations} except for some
special modes of perturbations which is ignore in this review.


Once we accept Conjecture~\ref{conjecture:decomp_conjecture_for_hab},
we can always find gauge-invariant variables for higher-order
perturbations\cite{kouchan-gauge-inv,K.Nakamura-CQG-Recursive-2014}.
According to the gauge transformation rule (\ref{eq:Bruni-49-one}),
the second-order metric perturbation $l_{ab}$ is transformed as
\begin{eqnarray}
  \label{eq:second-order-gauge-trans-of-metric}
  {}^{(2)}_{\;\cal Y}\!l_{ab} - {}^{(2)}_{\;\cal X}\!l_{ab}
  =
  2 {\pounds}_{\xi_{1}} {}_{\;\cal X}\!h_{ab}
  +\left\{{\pounds}_{\xi_{2}}+{\pounds}_{\xi_{1}}^{2}\right\} g_{ab}
\end{eqnarray}
under the gauge transformation
$\Phi_{\lambda}$ $=$
$({\cal X}_{\lambda})^{-1}\circ{\cal Y}_{\lambda}$ $:$
${\cal X}_{\lambda}\rightarrow{\cal Y}_{\lambda}$.
Although this gauge transformation rule is slightly complicated,
inspecting this gauge transformation rule, we first introduce the
variable $\hat{L}_{ab}$ defined by
\begin{equation}
  \label{eq:Lhatab-def}
  \hat{L}_{ab}
  :=
  l_{ab}
  - 2 {\pounds}_{X} h_{ab}
  + {\pounds}_{X}^{2} g_{ab},
\end{equation}
where the vector $X^{a}$ is that introduced by
Eq.~(\ref{eq:linear-metric-decomp}).
Under the gauge transformation
$\Phi_{\lambda}$ $=$
$({\cal X}_{\lambda})^{-1}\circ{\cal Y}_{\lambda}$ $:$
${\cal X}_{\lambda}\rightarrow{\cal Y}_{\lambda}$, the variable
$\hat{L}_{ab}$ is transformed as
\begin{eqnarray}
  {}_{\;\cal Y}\!\hat{L}_{ab} - {}_{\;\cal X}\!\hat{L}_{ab}
  &=&
  {\pounds}_{\sigma} g_{ab},
  \label{eq:kouchan-4.67}
  \\
  \sigma^{a} &:=& \xi_{2}^{a} + [\xi_{1},X]^{a}.
  \label{eq:sigma-def}
\end{eqnarray}
The gauge transformation rule (\ref{eq:kouchan-4.67}) is identical to
Eq.~(\ref{eq:hab-gauge-trans}) in
Conjecture~\ref{conjecture:decomp_conjecture_for_hab}.
Therefore, we may apply
Conjecture~\ref{conjecture:decomp_conjecture_for_hab} not only to
the linear-order metric perturbation  $h_{ab}$ but also to the
variable $\hat{L}_{ab}$ associated with the second-order metric
perturbation.
Then, $\hat{L}_{ab}$ can be decomposed as
\begin{eqnarray}
  \label{eq:Lhatab-decomposition}
  \hat{L}_{ab} = {\cal L}_{ab} + {\pounds}_{Y}g_{ab},
\end{eqnarray}
where ${\cal L}_{ab}$ is the gauge-invariant part of the variable
$\hat{L}_{ab}$, or equivalently, of the second-order metric
perturbation $l_{ab}$, and $Y^{a}$ is the gauge-variant part of
$\hat{L}_{ab}$, i.e., of the second-order metric perturbation
$l_{ab}$.
Under the gauge transformation $\Phi_{\lambda}$ $=$
$({\cal X}_{\lambda})^{-1}\circ{\cal Y}_{\lambda}$, the variables
${\cal L}_{ab}$ and $Y^{a}$ are transformed as
\begin{equation}
  {}_{\;\cal Y}\!{\cal L}_{ab} - {}_{\;\cal X}\!{\cal L}_{ab} = 0,
  \quad
  {}_{\;\cal Y}\!Y_{a} - {}_{\;\cal Y}\!Y_{a} = \sigma_{a},
\end{equation}
respectively.
Thus, once we accept
Conjecture~\ref{conjecture:decomp_conjecture_for_hab},
the second-order metric perturbations are decomposed as
\begin{eqnarray}
  \label{eq:H-ab-in-gauge-X-def-second-1}
  l_{ab}
  &=:&
  {\cal L}_{ab} + 2 {\pounds}_{X} h_{ab}
  + \left(
      {\pounds}_{Y}
    - {\pounds}_{X}^{2}
  \right)
  g_{ab},
\end{eqnarray}
where ${\cal L}_{ab}$ and $Y^{a}$ are the gauge invariant and variant
parts of the second order metric perturbations, i.e.,
\begin{eqnarray}
  {}_{{\cal Y}}\!{\cal L}_{ab} - {}_{{\cal X}}\!{\cal L}_{ab} = 0,
  \quad
  {}_{{\cal Y}}\!Y^{a} - {}_{{\cal X}}\!Y^{a}
  = \xi_{2}^{a} + [\xi_{1},X]^{a}.
\end{eqnarray}


Furthermore, as shown in Ref.~\cite{kouchan-gauge-inv}, using the
first- and second-order gauge variant parts, $X^{a}$ and $Y^{a}$, of
the metric perturbations, the gauge-invariant variables for an
arbitrary field $Q$ other than the metric are given by
\begin{eqnarray}
  \label{eq:matter-gauge-inv-def-1.0}
  {}^{(1)}\!{\cal Q} &:=& {}^{(1)}\!Q - {\pounds}_{X}Q_{0}
  , \\
  \label{eq:matter-gauge-inv-def-2.0}
  {}^{(2)}\!{\cal Q} &:=& {}^{(2)}\!Q - 2 {\pounds}_{X} {}^{(1)}Q
  - \left\{ {\pounds}_{Y} - {\pounds}_{X}^{2} \right\} Q_{0}
  .
\end{eqnarray}
It is straightforward to confirm that the variables
${}^{(p)}\!{\cal Q}$ defined by (\ref{eq:matter-gauge-inv-def-1.0})
and (\ref{eq:matter-gauge-inv-def-2.0}) are gauge invariant under the
gauge transformation rules (\ref{eq:Bruni-47-one}) and
(\ref{eq:Bruni-49-one}), respectively.


Equations (\ref{eq:matter-gauge-inv-def-1.0}) and
(\ref{eq:matter-gauge-inv-def-2.0}) have very important implications.
To see this, we represent these equations as
\begin{eqnarray}
  \label{eq:matter-gauge-inv-decomp-1.0}
  {}^{(1)}\!Q &=& {}^{(1)}\!{\cal Q} + {\pounds}_{X}Q_{0}
  , \\
  \label{eq:matter-gauge-inv-decomp-2.0}
  {}^{(2)}\!Q  &=& {}^{(2)}\!{\cal Q} + 2 {\pounds}_{X} {}^{(1)}Q
  + \left\{ {\pounds}_{Y} - {\pounds}_{X}^{2} \right\} Q_{0}
  .
\end{eqnarray}
These equations imply that any perturbation of first- and second-order
can always be decomposed into gauge-invariant and gauge-variant parts
as Eqs.~(\ref{eq:matter-gauge-inv-decomp-1.0}) and
(\ref{eq:matter-gauge-inv-decomp-2.0}), respectively.
These decomposition formulae (\ref{eq:matter-gauge-inv-decomp-1.0})
and (\ref{eq:matter-gauge-inv-decomp-2.0}) are important ingredients
in the general framework of the second-order general relativistic
gauge-invariant perturbation theory.


\section{Perturbations of the field equations}
\label{sec:Perturbation-of-the-field-equations}


In terms of the gauge-invariant variables defined last section, we
derive the field equations, i.e., Einstein equations and the equation
for a matter field.
To derive the perturbation of the Einstein equations and the equation
for a matter field (Klein-Gordon equation), first of all, we have to
derive the perturbative expressions of the Einstein
tensor~\cite{kouchan-second}.
This is reviewed in
Sec.~\ref{sec:Perturbation-of-the-Einstein-tensor}.
We also derive the first- and the second-order perturbations of the
energy momentum tensor for a scalar field and the Klein-Gordon
equation~\cite{kouchan-second-cosmo-matter} in Sec.~\ref{sec:Perturbation-of-the-energy-momentum-tensor-and-KG-eq}.
Finally, we consider the first- and the second-order the Einstein
equations in Sec.~\ref{sec:Perturbation-of-the-Einstein-equation}.


\subsection{Perturbations of the Einstein curvature}
\label{sec:Perturbation-of-the-Einstein-tensor}


The relation between the curvatures associated with the metrics on the
physical spacetime ${\cal M}_{\lambda}$ and the background spacetime
${\cal M}_{0}$ is given by the relation between the pulled-back
operator
${\cal X}_{\lambda}^{*}\bar{\nabla}_{a}\left({\cal X}^{-1}_{\lambda}\right)^{*}$
of the covariant derivative $\bar{\nabla}_{a}$ associated with the
metric $\bar{g}_{ab}$ on ${\cal M}_{\lambda}$ and the covariant
derivative $\nabla_{a}$ associated with the metric $g_{ab}$ on
${\cal M}_{0}$.
The pulled-back covariant derivative
${\cal X}_{\lambda}^{*}\bar{\nabla}_{a}\left({\cal X}^{-1}_{\lambda}\right)^{*}$
depends entirely on the gauge choice ${\cal X}_{\lambda}$.
The property of the derivative operator
${\cal X}^{*}_{\lambda}\bar{\nabla}_{a}\left({\cal X}^{-1}_{\lambda}\right)^{*}$
as the covariant derivative on ${\cal M}_{\lambda}$ is given by
\begin{equation}
  {\cal X}^{*}_{\lambda}\bar{\nabla}_{a}
  \left(
    \left(
      {\cal X}^{-1}_{\lambda}\right)^{*}{\cal X}^{*}_{\lambda}\bar{g}_{ab}
  \right) = 0,
  \label{eq:property-as-covariant-derivative-on-phys-sp}
\end{equation}
where ${\cal X}^{*}_{\lambda}\bar{g}_{ab}$ is the pull-back of the
metric on ${\cal M}_{\lambda}$, which is expanded as
Eq.~(\ref{eq:metric-expansion}).
In spite of the gauge dependence of the operator
${\cal X}^{*}_{\lambda}\bar{\nabla}_{a}\left({\cal X}^{-1}_{\lambda}\right)^{*}$,
we simply denote this operator by $\bar{\nabla}_{a}$, because our
calculations are carried out only on ${\cal M}_{0}$ in the same gauge
choice ${\cal X}_{\lambda}$.
Further, we denote the pulled-back metric
${\cal X}^{*}_{\lambda}\bar{g}_{ab}$ on ${\cal M}_{\lambda}$ by
$\bar{g}_{ab}$, as mentioned above.


Since the derivative operator $\bar{\nabla}_{a}$
($={\cal X}^{*}\bar{\nabla}_{a}\left({\cal X}^{-1}\right)^{*}$)
may be regarded as a derivative operator on ${\cal M}_{0}$ that
satisfies the property
(\ref{eq:property-as-covariant-derivative-on-phys-sp}), there exists a
tensor field $C^{c}_{\;\;ab}$ on ${\cal M}_{0}$ such that
\begin{equation}
  \bar{\nabla}_{a}\omega_{b}
  = \nabla_{a}\omega_{b} - C^{c}_{\;\;ab} \omega_{c},
\end{equation}
where $\omega_{a}$ is an arbitrary one-form on
${\cal M}_{0}$~\cite{Wald-book}.
From the property
(\ref{eq:property-as-covariant-derivative-on-phys-sp}) of the
covariant derivative operator $\bar{\nabla}_{a}$ on
${\cal M}_{\lambda}$, the tensor field $C^{c}_{\;\;ab}$ on
${\cal M}_{0}$ is given by
\begin{equation}
  C^{c}_{\;\;ab} = \frac{1}{2} \bar{g}^{cd}
  \left(
      \nabla_{a}\bar{g}_{db}
    + \nabla_{b}\bar{g}_{da}
    - \nabla_{d}\bar{g}_{ab}
  \right),
  \label{eq:c-connection}
\end{equation}
where $\bar{g}^{ab}$ is the inverse of $\bar{g}_{ab}$ (see Appendix
\ref{sec:derivation-of-pert-Einstein-tensors}).
We note that the gauge dependence of the covariant derivative
$\bar{\nabla}_{a}$ appears only through $C^{c}_{\;\;ab}$.
The Riemann curvature $\bar{R}_{abc}^{\;\;\;\;\;\;d}$ on
${\cal M}_{\lambda}$, which is also pulled back to ${\cal M}_{0}$, is
given by~\cite{Wald-book}:
\begin{equation}
  \bar{R}_{abc}^{\;\;\;\;\;\;d} = R_{abc}^{\;\;\;\;\;\;d}
  - 2 \nabla_{[a}^{} C^{d}_{\;\;b]c}
  + 2 C^{e}_{\;\;c[a} C^{d}_{\;\;b]e},
  \label{eq:phys-riemann-back-riemann-rel}
\end{equation}
where $R_{abc}^{\;\;\;\;\;\;d}$ is the Riemann curvature on
${\cal M}_{0}$.
The perturbative expression for the curvatures are obtained from the
expansion of Eq.~(\ref{eq:phys-riemann-back-riemann-rel}) through the
expansion of $C^{c}_{\;\;ab}$.


The first- and the-second order perturbations of the Riemann, the
Ricci, the scalar, the Weyl curvatures, and the Einstein tensors on
the general background spacetime are summarized in
Ref.~\cite{kouchan-second}.
We also derived the perturbative form of the divergence of an
arbitrary tensor field of second rank to check the perturbative
Bianchi identities in Ref.~\cite{kouchan-second}.
In this article, we only present the perturbative expression for the
Einstein tensor.
The derivations in of the formulae are  given in
Appendix~\ref{sec:derivation-of-pert-Einstein-tensors}.


We expand the Einstein tensor
$\bar{G}_{a}^{\;\;b}:=\bar{R}_{a}^{\;\;b}-\frac{1}{2}\delta_{a}^{\;\;b}\bar{R}$
on ${\cal M}_{\lambda}$ as
\begin{equation}
  \bar{G}_{a}^{\;\;b}
  =
  G_{a}^{\;\;b}
  + \lambda {}^{(1)}\!G_{a}^{\;\;b}
  + \frac{1}{2} \lambda^{2} {}^{(2)}\!G_{a}^{\;\;b}
  + O(\lambda^{3}).
\end{equation}
As shown in Appendix
\ref{sec:derivation-of-pert-Einstein-tensors}, each order perturbation
of the Einstein tensor is given by
\begin{eqnarray}
  \label{eq:linear-Einstein}
  {}^{(1)}\!G_{a}^{\;\;b}
  &=&
  {}^{(1)}{\cal G}_{a}^{\;\;b}\left[{\cal H}\right]
  + {\pounds}_{X} G_{a}^{\;\;b}
  ,\\
  \label{eq:second-Einstein-2,0-0,2}
  {}^{(2)}\!G_{a}^{\;\;b}
  &=&
  {}^{(1)}{\cal G}_{a}^{\;\;b}\left[{\cal L}\right]
  + {}^{(2)}{\cal G}_{a}^{\;\;b} \left[{\cal H}, {\cal H}\right]
  + 2 {\pounds}_{X} {}^{(1)}\!\bar{G}_{a}^{\;\;b}
  + \left\{ {\pounds}_{Y} - {\pounds}_{X}^{2} \right\} G_{a}^{\;\;b},
\end{eqnarray}
where
\begin{eqnarray}
  \label{eq:cal-G-def-linear}
  {}^{(1)}{\cal G}_{a}^{\;\;b}\left[A\right]
  &:=&
       {}^{(1)}\Sigma_{a}^{\;\;b}\left[A\right]
       - \frac{1}{2} \delta_{a}^{\;\;b} {}^{(1)}\Sigma_{c}^{\;\;c}\left[A\right]
       ,
  {}^{(1)}\Sigma_{a}^{\;\;b}\left[A\right]
  :=
       - 2 \nabla_{[a}^{}H_{d]}^{\;\;\;bd}\left[A\right]
       - A^{cb} R_{ac}
       ,
  \\
  \label{eq:cal-G-def-second}
  {}^{(2)}{\cal G}_{a}^{\;\;b}\left[A, B\right]
  &:=&
  {}^{(2)}\Sigma_{a}^{\;\;b}\left[A, B\right]
  - \frac{1}{2} \delta_{a}^{\;\;b} {}^{(2)}\Sigma_{c}^{\;\;c}\left[A, B\right]
  , \\
  {}^{(2)}\Sigma_{a}^{\;\;b}\left[A, B\right]
  &:=&
    2 R_{ad} B_{c}^{\;\;(b}A^{d)c}
  + 2 H_{[a}^{\;\;\;de}\left[A\right] H_{d]\;\;e}^{\;\;\;b}\left[B\right]
  + 2 H_{[a}^{\;\;\;de}\left[B\right] H_{d]\;\;e}^{\;\;\;b}\left[A\right]
  + 2 A_{e}^{\;\;d} \nabla_{[a}H_{d]}^{\;\;\;be}\left[B\right]
  \nonumber\\
  &&
  + 2 B_{e}^{\;\;d} \nabla_{[a}H_{d]}^{\;\;\;be}\left[A\right]
  + 2 A_{c}^{\;\;b} \nabla_{[a}H_{d]}^{\;\;\;cd}\left[B\right]
  + 2 B_{c}^{\;\;b} \nabla_{[a}H_{d]}^{\;\;\;cd}\left[A\right]
  \label{eq:(2)Sigma-def-second}
  ,
\end{eqnarray}
and
\begin{eqnarray}
  H_{ab}^{\;\;\;\;c}\left[A\right]
  &:=&
  \nabla_{(a}^{}A_{b)}^{\;\;\;c}
  - \frac{1}{2} \nabla^{c}_{}A_{ab}
  \label{eq:Habc-def-1}
  , \\
  H_{abc}\left[A\right]
  &:=&
  g_{cd} H_{ab}^{\;\;\;\;d}\left[A\right]
  ,
  \quad
  H_{a}^{\;\;bc}\left[A\right]
  :=
  g^{bd} H_{ad}^{\;\;\;\;c}\left[A\right]
  ,
       \quad
  H_{a\;\;c}^{\;\;b}\left[A\right]
  :=
  g_{cd} H_{a}^{\;\;bd}\left[A\right].
  \label{eq:Habc-def-2}
\end{eqnarray}
We note that ${}^{(1)}{\cal G}_{a}^{\;\;b}\left[*\right]$ and
${}^{(2)}{\cal G}_{a}^{\;\;b}\left[*,*\right]$ in
Eqs.~(\ref{eq:linear-Einstein}) and (\ref{eq:second-Einstein-2,0-0,2})
are the gauge invariant parts of the perturbative Einstein tensors,
and Eqs.~(\ref{eq:linear-Einstein}) and
(\ref{eq:second-Einstein-2,0-0,2}) have the same forms as
Eqs.~(\ref{eq:matter-gauge-inv-def-1.0}) and
(\ref{eq:matter-gauge-inv-decomp-2.0}), respectively.
The expression of ${}^{(2)}{\cal G}_{a}^{\;\;b}\left[A, B\right]$ in
Eq.~(\ref{eq:cal-G-def-second}) with
Eq.~(\ref{eq:(2)Sigma-def-second}) was derived by the consideration of
the general relativistic gauge-invariant perturbation theory with two
infinitesimal parameters in
Refs.~\cite{kouchan-gauge-inv,kouchan-second}.


We also note that ${}^{(1)}{\cal G}_{a}^{\;\;b}\left[*\right]$ and
${}^{(2)}{\cal G}_{a}^{\;\;b}\left[*,*\right]$ defined by
Eqs.~(\ref{eq:cal-G-def-linear})--(\ref{eq:(2)Sigma-def-second})
satisfy the identities
\begin{eqnarray}
  \nabla_{a}
  {}^{(1)}{\cal G}_{b}^{\;\;a}\left[A\right]
  &=&
  - H_{ca}^{\;\;\;\;a}\left[A\right] G_{b}^{\;\;c}
  + H_{ba}^{\;\;\;\;c}\left[A\right] G_{c}^{\;\;a}
  \label{eq:linear-order-divergence-of-calGab}
  , \\
  \nabla_{a}{}^{(2)}{\cal G}_{b}^{\;\;a}\left[A, B\right]
  &=&
      - H_{ca}^{\;\;\;\;a}\left[A\right]
      {}^{(1)}\!{\cal G}_{b}^{\;\;c}\left[B\right]
      - H_{ca}^{\;\;\;\;a}\left[B\right]
      {}^{(1)}\!{\cal G}_{b}^{\;\;c}\left[A\right]
     + H_{ba}^{\;\;\;\;e}\left[A\right]
     {}^{(1)}\!{\cal G}_{e}^{\;\;a}\left[B\right]
     + H_{ba}^{\;\;\;\;e}\left[B\right]
     {}^{(1)}\!{\cal G}_{e}^{\;\;a}\left[A\right]
     \nonumber\\
  &&
     - \left(
     H_{bad}\left[B\right] A^{dc} + H_{bad}\left[A\right] B^{dc}
     \right)
     G_{c}^{\;\;a}
     + \left(
     H_{cad}\left[B\right] A^{ad} + H_{cad}\left[A\right] B^{ad}
     \right)
     G_{b}^{\;\;c}
     ,
     \label{eq:second-div-of-calGab-1,1}
\end{eqnarray}
for arbitrary tensor fields $A_{ab}$ and $B_{ab}$, respectively.
We can directly confirm these identities without specifying arbitrary
tensors $A_{ab}$ and $B_{ab}$ of the second rank,
respectively~\cite{kouchan-second}.
This implies that our general framework of the second-order
gauge-invariant perturbation theory discussed here gives a
self-consistent formulation of the second-order perturbation theory.
These identities (\ref{eq:linear-order-divergence-of-calGab})
and (\ref{eq:second-div-of-calGab-1,1}) guarantee the first- and
second-order perturbations of the Bianchi identity
$\bar{\nabla}_{b}\bar{G}_{a}^{\;\;b}=0$ and are also useful when we
check whether the derived components of
Eqs.~(\ref{eq:cal-G-def-linear}) and (\ref{eq:cal-G-def-second}) are
correct.


\subsection{Perturbations of the energy momentum tensor and
  Klein-Gordon equation}
\label{sec:Perturbation-of-the-energy-momentum-tensor-and-KG-eq}


Here, we consider the perturbations of the energy momentum tensor and
the equation of motion.
As a model of the matter field, we only consider the scalar field, for
simplicity.
Then, equation of motion for a scalar field is the Klein-Gordon
equation.


The energy momentum tensor for a scalar field $\bar{\varphi}$ is given
by
\begin{eqnarray}
  \bar{T}_{a}^{\;\;b} =
  \bar{\nabla}_{a}\bar{\varphi} \bar{\nabla}^{b}\bar{\varphi}
  - \frac{1}{2} \delta_{a}^{\;\;b}
  \left(
    \bar{\nabla}_{c}\bar{\varphi}\bar{\nabla}^{c}\bar{\varphi}
    + 2 V(\bar{\varphi})
  \right),
  \label{eq:MFB-6.2-again}
\end{eqnarray}
where $V(\bar{\varphi})$ is the potential of the scalar field
$\bar{\varphi}$.
We expand the scalar field $\bar{\varphi}$ as
\begin{eqnarray}
  \bar{\varphi}
  =
  \varphi
  + \lambda \hat{\varphi}_{1}
  + \frac{1}{2} \lambda^{2} \hat{\varphi}_{2}
  + O(\lambda^{3}),
  \label{eq:scalar-field-expansion-second-order}
\end{eqnarray}
where $\varphi$ is the background value of the scalar field
$\bar{\varphi}$.
Further, following to the decomposition formulae
(\ref{eq:matter-gauge-inv-def-1.0}) and
(\ref{eq:matter-gauge-inv-def-2.0}), each order perturbation of the
scalar field $\bar{\varphi}$ is decomposed as
\begin{eqnarray}
  \label{eq:varphi-1-def}
  \hat{\varphi}_{1} &=:& \varphi_{1} + {\pounds}_{X}\varphi, \\
  \label{eq:varphi-2-def}
  \hat{\varphi}_{2} &=:& \varphi_{2}
  + 2 {\pounds}_{X}\hat{\varphi}_{1}
  + \left( {\pounds}_{Y} - {\pounds}_{X}^{2} \right) \varphi,
\end{eqnarray}
where $\varphi_{1}$ and $\varphi_{2}$ are the first- and the
second-order gauge-invariant perturbations of the scalar field,
respectively.


Through the perturbative expansions
(\ref{eq:scalar-field-expansion-second-order}) of the scalar field
$\bar{\varphi}$ and Eq.~(\ref{eq:inverse-metric-expansion}) for the
inverse metric, the energy momentum tensor (\ref{eq:MFB-6.2-again}) is
also expanded as
\begin{eqnarray}
  \bar{T}_{a}^{\;\;b} = T_{a}^{\;\;b}
  +
  \lambda {}^{(1)}\!\left(T_{a}^{\;\;b}\right)
  +
  \frac{1}{2} \lambda^{2} {}^{(2)}\!\left(T_{a}^{\;\;b}\right)
  + O(\lambda^{3}).
\end{eqnarray}
The background energy momentum tensor $T_{a}^{\;\;b}$ is given by the
replacement $\bar{\varphi}\rightarrow\varphi$ in
Eq.~(\ref{eq:MFB-6.2-again}).
Further, through the decompositions (\ref{eq:linear-metric-decomp}),
(\ref{eq:H-ab-in-gauge-X-def-second-1}), (\ref{eq:varphi-1-def}), and
(\ref{eq:varphi-2-def}), the perturbations of the energy momentum
tensor ${}^{(1)}\!\left(T_{a}^{\;\;b}\right)$ and
${}^{(2)}\!\left(T_{a}^{\;\;b}\right)$ are also decomposed as
\begin{eqnarray}
  {}^{(1)}\!\left(T_{a}^{\;\;b}\right)
  &=:&
  {}^{(1)}\!{\cal T}_{a}^{\;\;b} + {\pounds}_{X}T_{a}^{\;\;b}
  \label{eq:first-order-energy-momentum-scalar-decomp}
  ,
  \\
  {}^{(2)}\!\left(T_{a}^{\;\;b}\right)
  &=:&
  {}^{(2)}\!{\cal T}_{a}^{\;\;b}
  + 2 {\pounds}_{X}{}^{(1)}\!\left(T_{a}^{\;\;b}\right)
  + \left( {\pounds}_{Y} - {\pounds}_{X}^{2}\right) T_{a}^{\;\;b},
  \label{eq:second-order-energy-momentum-scalar-decomp}
\end{eqnarray}
where the gauge-invariant parts ${}^{(1)}\!{\cal T}_{a}^{\;\;b}$
and ${}^{(2)}\!{\cal T}_{a}^{\;\;b}$ of the first- and the second-order
are given by
\begin{eqnarray}
  {}^{(1)}\!{\cal T}_{a}^{\;\;b}
  &:=&
       \nabla_{a}\varphi \nabla^{b}\varphi_{1}
       - \nabla_{a}\varphi {\cal H}^{bc} \nabla_{c}\varphi
       + \nabla_{a}\varphi_{1} \nabla^{b} \varphi
     - \delta_{a}^{\;\;b}
     \left(
     \nabla_{c}\varphi\nabla^{c}\varphi_{1}
     - \frac{1}{2} \nabla_{c}\varphi {\cal H}^{dc} \nabla_{d} \varphi
     + \varphi_{1} \frac{\partial V}{\partial\varphi}
     \right)
  \label{eq:first-order-energy-momentum-scalar-gauge-inv}
  , \\
  {}^{(2)}\!{\cal T}_{a}^{\;\;b}
  &:=&
       \nabla_{a}\varphi \nabla^{b}\varphi_{2}
       + \nabla_{a}\varphi_{2} \nabla^{b}\varphi
       - \nabla_{a}\varphi g^{bd} {\cal L}_{dc} \nabla^{c}\varphi
       - 2 \nabla_{a}\varphi {\cal H}^{bc} \nabla_{c}\varphi_{1}
       + 2 \nabla_{a}\varphi {\cal H}^{bd}{\cal H}_{dc} \nabla^{c}\varphi
       \nonumber\\
  &&
       + 2 \nabla_{a}\varphi_{1} \nabla^{b}\varphi_{1}
  - 2 \nabla_{a}\varphi_{1} {\cal H}^{bc} \nabla_{c}\varphi
       \nonumber\\
  &&
     - \delta_{a}^{\;\;b}
     \left(
     \nabla_{c}\varphi\nabla^{c}\varphi_{2}
     - \frac{1}{2} \nabla^{c}\varphi {\cal L}_{dc}\nabla^{d}\varphi
    + \nabla^{c}\varphi {\cal H}^{de}{\cal H}_{ec} \nabla_{d}\varphi
    - 2 \nabla_{c}\varphi {\cal H}^{dc} \nabla_{d}\varphi_{1}
    + \nabla_{c}\varphi_{1}\nabla^{c}\varphi_{1}
  \right.
  \nonumber\\
  && \quad\quad\quad\quad
  \left.
    + \varphi_{2}\frac{\partial V}{\partial\varphi}
    + \varphi_{1}^{2}\frac{\partial^{2}V}{\partial\varphi^{2}}
  \right)
  \label{eq:second-order-energy-momentum-scalar-gauge-inv}
  .
\end{eqnarray}
We note that
Eq.~(\ref{eq:first-order-energy-momentum-scalar-decomp}) and
(\ref{eq:second-order-energy-momentum-scalar-decomp}) have the
same form as (\ref{eq:matter-gauge-inv-decomp-1.0}) and
(\ref{eq:matter-gauge-inv-decomp-2.0}), respectively.


Next, we consider the perturbation of the Klein-Gordon equation
\begin{eqnarray}
  \bar{C}_{(K)} := \bar{\nabla}^{a}\bar{\nabla}_{a}\bar{\varphi}
  - \frac{\partial V}{\partial\bar{\varphi}}(\bar{\varphi}) = 0.
  \label{eq:Klein-Gordon-equation}
\end{eqnarray}
Through the perturbative expansions
(\ref{eq:scalar-field-expansion-second-order}) and
(\ref{eq:metric-expansion}), the Klein-Gordon equation
(\ref{eq:Klein-Gordon-equation}) is expanded as
\begin{eqnarray}
  \bar{C}_{(K)} =: C_{(K)} + \lambda \stackrel{(1)}{C_{(K)}} +
  \frac{1}{2} \lambda^{2} \stackrel{(2)}{C_{(K)}} +
  O(\lambda^{3}).
\end{eqnarray}
$C_{(K)}$ is the background Klein-Gordon equation
\begin{eqnarray}
  \label{eq:Klein-Gordon-eq-background}
  C_{(K)}
  &:=&
  \nabla_{a}\nabla^{a}\varphi
  - \frac{\partial V}{\partial\bar{\varphi}}(\varphi)
  = 0
  .
\end{eqnarray}
The first- and the second-order perturbations
$\stackrel{(1)}{C_{(K)}}$ and $\stackrel{(2)}{C_{(K)}}$ are also
decomposed into the gauge-invariant and the gauge-variant parts as
\begin{eqnarray}
  \stackrel{(1)}{C_{(K)}}
  =:
  \stackrel{(1)}{{\cal C}_{(K)}}
  + {\pounds}_{X}C_{(K)},
  \quad
  \stackrel{(2)}{C_{(K)}}
  =:
  \stackrel{(2)}{{\cal C}_{(K)}}
  + 2 {\pounds}_{X}\stackrel{(1)}{C_{(K)}}
  + \left( {\pounds}_{Y} - {\pounds}_{X}^{2} \right) C_{(K)}
  ,
  \label{eq:Klein-Gordon-eq-first-decomp}
\end{eqnarray}
where
\begin{eqnarray}
  \stackrel{(1)}{{\cal C}_{(K)}}
  &:=&
       \nabla^{a}\nabla_{a}\varphi_{1}
       - H_{a}^{\;\;ac}[{\cal H}]\nabla_{c}\varphi
       - {\cal H}^{ab} \nabla_{a}\nabla_{b}\varphi
     - \varphi_{1} \frac{\partial^{2}V}{\partial\bar{\varphi}^{2}}(\varphi)
     \label{eq:Klein-Gordon-eq-first-gauge-inv-def}
     ,
  \\
  \stackrel{(2)}{{\cal C}_{(K)}}
  &:=&
       \nabla^{a}\nabla_{a}\varphi_{2}
       -   H_{a}^{\;\;ac}[{\cal L}] \nabla_{c}\varphi
       + 2 H_{a}^{\;\;ad}[{\cal H}] {\cal H}_{cd} \nabla^{c}\varphi
  - 2 H_{a}^{\;\;ac}[{\cal H}] \nabla_{c}\varphi_{1}
  + 2 {\cal H}^{ab} H_{ab}^{\;\;\;\;c}[{\cal H}] \nabla_{c}\varphi
  \nonumber\\
  &&
  -   {\cal L}^{ab} \nabla_{a}\nabla_{b}\varphi
  + 2 {\cal H}^{a}_{\;\;d} {\cal H}^{db} \nabla_{a}\nabla_{b}\varphi
  - 2 {\cal H}^{ab} \nabla_{a}\nabla_{b}\varphi_{1}
  -   \varphi_{2} \frac{\partial^{2}V}{\partial\bar{\varphi}^{2}}(\varphi)
  -   (\varphi_{1})^{2}\frac{\partial^{3}V}{\partial\bar{\varphi}^{3}}(\varphi)
  \label{eq:Klein-Gordon-eq-second-gauge-inv-def}
  .
\end{eqnarray}
Here, we note that Eqs.~(\ref{eq:Klein-Gordon-eq-first-decomp})
have the same form as
Eqs.~(\ref{eq:matter-gauge-inv-decomp-1.0}) and
(\ref{eq:matter-gauge-inv-decomp-2.0}).


By virtue of the order by order evaluations of the Klein-Gordon
equation, the first- and the second-order perturbation of the
Klein-Gordon equation are necessarily given in gauge-invariant
form as
\begin{eqnarray}
  \label{eq:Klein-Gordon-eq-first-second-gauge-inv}
  \stackrel{(1)}{{\cal C}_{(K)}} = 0, \quad
  \stackrel{(2)}{{\cal C}_{(K)}} = 0.
\end{eqnarray}


We should note that, in Ref.~\cite{kouchan-second-cosmo-matter}, we
summarized the formulae of the energy momentum tensors for an perfect
fluid, an imperfect fluid, and a scalar field.
Further, we also summarized the equations of motion of these three
matter fields: i.e., the energy continuity equation and the Euler
equation for a perfect fluid; the energy continuity equation and the
Navier-Stokes equation for an imperfect fluid; the Klein-Gordon
equation for a scalar field.
All these formulae also have the same form as the decomposition
formulae (\ref{eq:matter-gauge-inv-decomp-1.0}) and
(\ref{eq:matter-gauge-inv-decomp-2.0}).
In this sense, we may say that the decomposition formulae
(\ref{eq:matter-gauge-inv-decomp-1.0}) and
(\ref{eq:matter-gauge-inv-decomp-2.0}) are universal.


\subsection{Perturbations of the Einstein equation}
\label{sec:Perturbation-of-the-Einstein-equation}


Finally, we impose the perturbed Einstein equation of each order,
\begin{equation}
  {}^{(1)}G_{a}^{\;\;b} = 8\pi G \;\; {}^{(1)}T_{a}^{\;\;b},
  \quad
  {}^{(2)}G_{a}^{\;\;b} = 8\pi G \;\; {}^{(2)}T_{a}^{\;\;b}.
\end{equation}
Then, the perturbative Einstein equation is given by
\begin{eqnarray}
  \label{eq:linear-order-Einstein-equation}
  {}^{(1)}\!{\cal G}_{a}^{\;\;b}\left[{\cal H}\right]
  &=&
  8\pi G {}^{(1)}{\cal T}_{a}^{\;\;b}
\end{eqnarray}
at linear order and
\begin{eqnarray}
  \label{eq:second-order-Einstein-equation}
  {}^{(1)}\!{\cal G}_{a}^{\;\;b}\left[{\cal L}\right]
  + {}^{(2)}\!{\cal G}_{a}^{\;\;b}\left[{\cal H}, {\cal H}\right]
  &=&
  8\pi G \;\; {}^{(2)}{\cal T}_{a}^{\;\;b}
\end{eqnarray}
at second order.
These explicitly show that, order by order, the Einstein
equations are necessarily given in terms of gauge-invariant
variables only.


Together with Eqs.~(\ref{eq:Klein-Gordon-eq-first-second-gauge-inv}),
we have seen that the first- and the second-order perturbations
of the Einstein equations and the Klein-Gordon equation are
necessarily given in gauge-invariant form.
This implies that we do not have to consider the gauge degree of
freedom, at least in the level where we concentrate only on the
equations of the system.


We have reviewed the general outline of the second-order
gauge-invariant perturbation theory.
We also note that the ingredients of this section are independent of
the explicit form of the background metric $g_{ab}$, except for
Conjecture~\ref{conjecture:decomp_conjecture_for_hab}.
Therefore, if Conjecture~\ref{conjecture:decomp_conjecture_for_hab} is
correct for the general background spacetime, the ingredients of this
section are also valid not only in cosmological perturbation case but
also the other generic situation.
Since this is the review of cosmological perturbation theory, in next
section, we develop a second-order cosmological perturbation theory in
terms of the gauge-invariant variables within this general framework.


\section{Cosmological background spacetime and equations}
\label{sec:Cosmological-Background-spacetime-equations}


The background spacetime ${\cal M}_{0}$ considered in cosmological
perturbation theory is a homogeneous, isotropic universe that is
foliated by the three-dimensional hypersurface $\Sigma(\eta)$, which
is parametrized by $\eta$.
Each hypersurface of $\Sigma(\eta)$ is a maximally symmetric
three-space~\cite{Weinberg1972}, and the spacetime metric of this
universe is given by
\begin{eqnarray}
  g_{ab} = a^{2}(\eta)\left(
    - (d\eta)_{a}(d\eta)_{b}
    + \gamma_{ij}(dx^{i})_{a}(dx^{j})_{b}
  \right),
  \label{eq:background-metric}
\end{eqnarray}
where $a=a(\eta)$ is the scale factor, $\gamma_{ij}$ is the metric on
the maximally symmetric 3-space with curvature constant $K$, i.e., the
spatial curvature associated with the metric $\gamma_{ij}$ is given by
\begin{eqnarray}
  {}^{(3)}R_{ijkl} = 2 K \gamma_{k[i} \gamma_{j]l}, \quad
  {}^{(3)}R_{ij} = 2 K \gamma_{ij}, \quad
  {}^{(3)} R = 6 K.
  \label{eq:spatial-curvature}
\end{eqnarray}
The indices $i,j,k,...$ for the spatial components run from 1 to 3.


To study the Einstein equation for this background spacetime, we
introduce the energy-momentum tensor for a scalar field, which is
given by
\begin{eqnarray}
  T_{a}^{\;\;b}
  &=&
  \nabla_{a}\varphi\nabla^{b}\varphi -
  \frac{1}{2}\delta_{a}^{\;\;b}\left(\nabla_{c}\varphi\nabla^{c}\varphi +
    2V(\varphi)\right)
  \label{eq:energy-momentum-single-scalar}
  \\
  &=&
  -
  \left(
      \frac{1}{2a^{2}} (\partial_{\eta}\varphi)^{2}
    + V(\varphi)
  \right)
  (d\eta)_{a} \left(\frac{\partial}{\partial\eta}\right)^{b}
  +
  \left(
    \frac{1}{2a^{2}} (\partial_{\eta}\varphi)^{2}
    - V(\varphi)
  \right)
  \gamma_{a}^{\;\;b},
  \label{eq:energy-momentum-single-scalar-homogeneous}
\end{eqnarray}
where we assumed that the scalar field $\varphi$ is homogeneous
\begin{eqnarray}
  \label{eq:background-varphi-is-homogeneous}
  \varphi=\varphi(\eta)
\end{eqnarray}
and $\gamma_{a}^{\;\;b}$ are defined as
\begin{eqnarray}
  \gamma_{ab} := \gamma_{ij}(dx^{i})_{a}(dx^{j})_{b}, \;\;
  \gamma_{a}^{\;\;b}:=\gamma_{i}^{\;\;j}(dx^{i})_{a}(\partial/\partial x^{j})^{b}.
  \label{eq:gammaab-gammaab-def}
\end{eqnarray}


The background Einstein equations $G_{a}^{\;\;b}=8\pi GT_{a}^{\;\;b}$
for this background spacetime filled with the single scalar field are
given by
\begin{eqnarray}
  \label{eq:background-Einstein-equations-scalar-1}
  &&\!\!\!\!\!\!\!\!\!\!\!\!\!\!\!\!
  {\cal H}^{2} + K = \frac{8 \pi G}{3} a^{2} \left(
    \frac{1}{2a^{2}} (\partial_{\eta}\varphi)^{2} + V(\varphi)
  \right)
  ,\\
  \label{eq:background-Einstein-equations-scalar-2}
  &&\!\!\!\!\!\!\!\!\!\!\!\!\!\!\!\!
  2 \partial_{\eta}{\cal H} + {\cal H}^{2} + K = 8 \pi G
  \left(-\frac{1}{2} (\partial_{\eta}\varphi)^{2} + a^{2} V(\varphi)\right)
  .
\end{eqnarray}
We also note that the equations
(\ref{eq:background-Einstein-equations-scalar-1}) and
(\ref{eq:background-Einstein-equations-scalar-2}) lead to
\begin{eqnarray}
  \label{eq:background-Einstein-equations-scalar-3}
  {\cal H}^{2} + K - \partial_{\eta}{\cal H}
  = 4\pi G (\partial_{\eta}\varphi)^{2}.
\end{eqnarray}
Equation (\ref{eq:background-Einstein-equations-scalar-3}) is also
useful when we derive the perturbative Einstein equations.


Next, we consider the background Klein-Gordon equation which is the
equation of motion $\nabla_{a}^{}T_{b}^{\;\;a}=0$ for the scalar field
\begin{eqnarray}
  \label{eq:background-Klein-Gordon-equation}
  \partial_{\eta}^{2}\varphi + 2 {\cal H} \partial_{\eta}\varphi
  +   a^{2} \frac{\partial V}{\partial\varphi}
  = 0.
\end{eqnarray}
The Klein-Gordon equation
(\ref{eq:background-Klein-Gordon-equation}) is also derived from the
Einstein equations
(\ref{eq:background-Einstein-equations-scalar-1}) and
(\ref{eq:background-Einstein-equations-scalar-2}).
This is a well-known fact and is just due to the Bianchi identity of
the background spacetime.
However, these types of relation are useful to check whether the
derived system of equations is consistent.


\section{Equations for the first-order cosmological perturbations}
\label{sec:Equations-for-the-first-order-cosmological-perturbations}


On the cosmological background spacetime in the last section, we
develop the perturbation theory in the gauge-invariant manner.
In this section, we summarize the first-order perturbation of the
Einstein equation and the Klein-Gordon equations.
In Sec.~\ref{sec:Gauge-invariant-metric-perturbations}, we show that
Conjecture~\ref{conjecture:decomp_conjecture_for_hab} for the
linear-order metric perturbation is correct except for the special
modes of perturbations.
In Sec.~\ref{sec:First-order-Einstein-equations}, we summarize the
first-order perturbation of the Einstein equation.
Finally, in Sec.~\ref{sec:First-order-Klein-Gordon-equations}, we show
the first-order perturbation of the Klein-Gordon equation.


\subsection{Gauge-invariant metric perturbations}
\label{sec:Gauge-invariant-metric-perturbations}


Here, we consider the first-order metric perturbation $h_{ab}$ and
show that Conjecture~\ref{conjecture:decomp_conjecture_for_hab} is
correct in the background metric (\ref{eq:background-metric}) expect
for the special modes of perturbations.
Although the outline of the proof of
Conjecture~\ref{conjecture:decomp_conjecture_for_hab} for the general
metric is given in
Appendix~\ref{sec:Outline-of-the-proof-of-the-decomposition-conjecture},
we show a specific approach of
Conjecture~\ref{conjecture:decomp_conjecture_for_hab} which is valid
only in the case of cosmological perturbations.


As the starting point of our arguments, we consider the linear metric
perturbation on the background spacetime with the metric
(\ref{eq:background-metric}):
\begin{eqnarray}
  h_{ab}
  &=&
      h_{\eta\eta}(d\eta)_{a}(d\eta)_{b}
      + 2 h_{\eta i} (d\eta)_{(a}(dx^{i})_{b)}
     + h_{ij} (dx^{i})_{a} (dx^{j})_{b}
      .
      \label{eq:cosmological-linear-perturbations-bare}
\end{eqnarray}
Furthermore, we consider the decomposition of the set of the above
component $\{h_{\eta\eta},h_{\eta i},h_{ij}\}$ as
\begin{eqnarray}
  h_{ab}
  &=&
      h_{\eta\eta}(d\eta)_{a}(d\eta)_{b}
     + 2 \left(
     D_{i}h_{(VL)} + h_{(V)i}
     \right)(d\eta)_{(a}(dx^{i})_{b)}
     \label{eq:cosmological-linear-perturbations}
  \\
  &&
     + a^{2} \left\{
     h_{(L)} \gamma_{ij}
     + \left(D_{i}D_{j} - \frac{1}{3}\gamma_{ij}\Delta\right)h_{(TL)}
     + 2 D_{(i}h_{(TV)j)} + {h_{(TT)ij}}
     \right\} (dx^{i})_{a}(dx^{j})_{b}
     ,
     \nonumber
\end{eqnarray}
where $h_{(V)i}$, $h_{(TV)j}$, and ${h_{(TT)ij}}$ satisfy the
properties
\begin{eqnarray}
  D^{i}h_{(V)i} = 0, \quad
  D^{i} h_{(TV)i} = 0, \quad
  h_{(TT)ij} = h_{(TT)ji}, \quad
  D^{i} h_{(TT)ij} = 0,
  \label{eq:bare-parturbative-property}
\end{eqnarray}
and $D_{i}$ is the covariant derivative associated with the metric
$\gamma_{ij}$ and the operator $\Delta:=D^{i}D_{i}$ is the Laplacian
which is an elliptic operator.
The decomposition (\ref{eq:cosmological-linear-perturbations}) of the
symmetric tensor with the properties
(\ref{eq:bare-parturbative-property}) is originated from
Refs.~\cite{J.W.York-Jr.-1974,S.Deser-1967} and used in many
literature.


To examine the one-to-one correspondence between the set of variables
$\{h_{\eta\eta},$ $h_{\eta i},$ $h_{ij}\}$ and the set of variables
$\{h_{\eta\eta},$ $h_{(VL)},$ $h_{(V)i},$ $h_{(L)},$ $h_{(TL)},$
$h_{(TV)i},$ $h_{(TT)ij}\}$ through the decomposition
(\ref{eq:cosmological-linear-perturbations}), we consider the inverse
relation of the variable transformation from the original components
in Eq.~(\ref{eq:cosmological-linear-perturbations-bare}) to the decomposed
components in Eq.~(\ref{eq:cosmological-linear-perturbations}), i.e.,
from the set $\{h_{\eta\eta},$ $h_{\eta i},$ $h_{ij}\}$ to the set
$\{h_{\eta\eta},$ $h_{(VL)},$ $h_{(V)i},$ $h_{(L)},$ $h_{(TL)},$
$h_{(TV)i},$ $h_{(TT)ij}\}$:
\begin{eqnarray}
  h_{\eta\eta}
  &=&
      h_{\eta\eta}
      , \quad
  h_{(VL)}
  =
      \Delta^{-1}D^{i}h_{\eta i}
      , \quad
  h_{(V)i}
  =
  \left(h_{\eta i} - D_{i}\Delta^{-1}D^{j}h_{\eta j}\right)
  ,
  \label{eq:h-decomp-inv-1}
  \\
  a^{2} h_{(L)}
  &=&
      \frac{1}{3} \gamma^{ij} h_{ij}
      ,
  \quad
  a^{2} h_{(TL)}
  =
      \frac{3}{2}
      \left[\Delta+3K\right]^{-1}
      \Delta^{-1}
      D^{k}D^{l}
      h_{(T)kl}
      ,
      \label{eq:h-decomp-inv-5}
  \\
  a^{2}h_{(TV)i}
  &=&
      \left[\Delta+2K\right]^{-1}
      \left[
      \gamma_{i}^{\;\;m}
      -
      D_{i} \Delta^{-1} D^{m}
      \right]
      D^{k} h_{(T)mk}
      ,
      \label{eq:h-decomp-inv-6}
  \\
  a^{2} {h_{(TT)ij}}
  &=&
      h_{(T)ij}
%
      -
      \frac{3}{2}
      \left(D_{i}D_{j} - \frac{1}{3}\gamma_{ij}\Delta\right)
      \left[\Delta+3K\right]^{-1}
      \Delta^{-1}
      D^{k}D^{l}
      h_{(T)kl}
      \nonumber\\
  && \quad\quad\quad
     -
     2
     \gamma_{(i}^{\;\;\;l}
     D_{j)}^{}
     \left[\Delta+2K\right]^{-1}
     \left[
     \gamma_{l}^{\;\;m}
     -
     D_{l} \Delta^{-1} D^{m}
     \right] D^{k} h_{(T)mk}
     ,
  \label{eq:h-decomp-inv-7}
\end{eqnarray}
where $h_{(T)}$ is the traceless part of the components $h_{ij}$
defined by
\begin{eqnarray}
  \label{eq:hT-def}
  h_{(T)ij} := h_{ij} - \frac{1}{3} \gamma_{ij} \gamma^{kl} h_{kl}
  =
  \left( D_{i}D_{j} - \frac{1}{3} \gamma_{ij}\Delta \right) h_{(TL)}
  +
  2 D_{(i}h_{(TV)j)}
  +
  h_{(TT)ij}
  .
\end{eqnarray}
In Eqs.~(\ref{eq:h-decomp-inv-1})--(\ref{eq:h-decomp-inv-7}),
the operator $\Delta^{-1}$, $(\Delta + 2 K)^{-1}$, and
$(\Delta + 3 K)^{-1}$ are the Green functions of the elliptic
derivative operators $\Delta$, $\Delta+2K$, and $\Delta+3K$,
respectively, and $K$ is the curvature constant of the maximally
symmetric three space.


Equations~(\ref{eq:h-decomp-inv-1})--(\ref{eq:h-decomp-inv-7})
indicate that the decomposition
(\ref{eq:cosmological-linear-perturbations}) is non-local, since its
inverse relations~(\ref{eq:h-decomp-inv-1})--(\ref{eq:h-decomp-inv-7})
requires the Green functions $\Delta^{-1}$,
$\left[\Delta+2K\right]^{-1}$, and $\left[\Delta+3K\right]^{-1}$.
More importantly, the inverse relation
(\ref{eq:h-decomp-inv-1})--(\ref{eq:h-decomp-inv-7}) indicates that
the decomposition (\ref{eq:cosmological-linear-perturbations}) does
not includes the modes which belong to the kernels of the derivative
operators $\Delta$, $\Delta+2K$, and $\Delta+3K$.
Actually, there is one-to-one correspondence between the set
$\{h_{\eta\eta},h_{\eta  i},h_{ij}\}$ and the set
$\{h_{\eta\eta},h_{(VL)},h_{(V)i},h_{(L)},h_{(TL)},h_{(TV)i},h_{(TT)ij}\}$
of metric perturbations if the Green functions $\Delta^{-1}$,
$\left[\Delta+2K\right]^{-1}$, and $\left[\Delta+3K\right]^{-1}$
exist.
This implies that the representation
(\ref{eq:cosmological-linear-perturbations}) of the metric
perturbation $h_{ab}$ does not include the kernel modes of the
operators $\Delta$, $\Delta+2K$, and $\Delta+3K$, while the
representation (\ref{eq:cosmological-linear-perturbations-bare}) may
include these kernel modes.
In this sense, we should regard that the set
$\{h_{\eta\eta},h_{(VL)},h_{(V)i},h_{(L)},h_{(TL)},h_{(TV)i},h_{(TT)ij}\}$
of the perturbative metric is a subset of the original set
$\{h_{\eta\eta},h_{\eta  i},h_{ij}\}$ due to the lack of these kernel
modes of the operators $\Delta$, $\Delta+2K$, and $\Delta+3K$.
In spite of this fact, in this review, we ignore these kernel modes,
for simplicity, keeping in our mind the importance of these kernel
modes.
The importance of these kernel modes are discussed in
Sec.~\ref{sec:summary}.


In terms of the perturbative variables
$\{h_{\eta\eta},$ $h_{(VL)},$ $h_{(V)i},$ $h_{(L)},$ $h_{(TL)},$
$h_{(TV)i},$ $h_{(TT)ij}\}$ for the metric perturbations, we consider
the construction of the gauge-invariant variables for the linear-order
metric perturbations.
From the gauge-transformation rule (\ref{eq:linear-metric-decomp})
with the generator
\begin{eqnarray}
  \label{eq:xia-components}
  \xi_{a} = \xi_{\eta}(d\eta)_{a} + \xi_{i}(d\xi^{i})_{a}
  ,
\end{eqnarray}
we can derive the gauge-transformation rules for the components
$\{h_{\eta\eta},$ $h_{\eta i},$ $h_{ij}\}$ as
\begin{eqnarray}
  {}_{{\cal Y}}\!h_{\eta\eta} - {}_{{\cal X}}\!h_{\eta\eta}
  &=&
      2 \left(\partial_{\eta} - {\cal H}\right)\xi_{\eta}
      ,
      \label{eq:hetaeta-gauge-trans}
  \\
  {}_{{\cal Y}}\!h_{\eta i} - {}_{{\cal X}}\!h_{\eta i}
  &=&
      D_{i}\xi_{\eta} + 2 \left(\partial_{\eta} - 2 {\cal H}\right) \xi_{i}
      ,
      \label{eq:hetai-gauge-trans}
  \\
  {}_{{\cal Y}}\!h_{ij} - {}_{{\cal X}}\!h_{ij}
  &=&
      2 D_{(i}\xi_{j)} - 2 {\cal H} \gamma_{ij} \xi_{\eta}
      .
      \label{eq:hij-gauge-trans}
\end{eqnarray}
From these gauge-transformation rules
(\ref{eq:hetaeta-gauge-trans})--(\ref{eq:hij-gauge-trans}), we can
derive the gauge-transformation rules for the variables
$\{h_{\eta\eta},$ $h_{(VL)},$ $h_{(V)i},$ $h_{(L)},$ $h_{(TL)},$ $h_{(TV)i},$
$h_{(TT)ij}\}$ as
\begin{eqnarray}
  &&
     {}_{{\cal Y}}\!h_{\eta\eta} - {}_{{\cal X}}\!h_{\eta\eta}
     =
     2 \left(\partial_{\eta} - {\cal H}\right)\xi_{\eta}
     ,
     \label{eq:hetaeta-gauge-trans-2}
  \\
  &&
     {}_{{\cal Y}}\!h_{(VL)} - {}_{{\cal X}}\!h_{(VL)}
     =
     \xi_{\eta}
     +
     \left(\partial_{\eta} - 2 {\cal H}\right) \xi_{(L)}
     ,
     \label{eq:hVL-gauge-trans}
  \\
  &&
     {}_{{\cal Y}}\!h_{(V)i} - {}_{{\cal X}}\!h_{(V)i}
     =
     \left(\partial_{\eta} - 2 {\cal H} \right) \xi_{(T)i}
     ,
     \label{eq:hVi-gauge-trans}
  \\
  &&
     a^{2}{}_{{\cal Y}}\!h_{(L)} - a^{2}{}_{{\cal X}}\!h_{(L)}
     =
     - 2 {\cal H} \xi_{\eta} + \frac{2}{3} \Delta \xi_{(L)}
     ,
     \label{eq:hL-gauge-trans}
  \\
  &&
     a^{2}{}_{{\cal Y}}\!h_{(TL)} - a^{2}{}_{{\cal X}}\!h_{(TL)}
     =
     2 \xi_{(L)}
     ,
     \label{eq:hTL-gauge-trans}
  \\
  &&
     a^{2}{}_{{\cal Y}}\!h_{(TV)i} - a^{2}{}_{{\cal X}}\!h_{(TV)i}
     =
     \xi_{(T)i}
     ,
     \label{eq:hTVi-gauge-trans}
  \\
  &&
     a^{2}{}_{{\cal Y}}\!h_{(TT)ij} - a^{2}{}_{{\cal X}}\!h_{(TT)ij}
     =
     0
     ,
     \label{eq:hTTij-gauge-trans}
\end{eqnarray}
where we decomposed the component $\xi_{i}$ as
\begin{eqnarray}
  \label{eq:xii-decomp}
  \xi_{i} = D_{i}\xi_{(L)} + \xi_{(V)i}, \quad D^{i}\xi_{(V)i} = 0.
\end{eqnarray}


First, we derive the definition of the gauge-variant part $X_{a}$ of the
metric perturbation in Eq.~(\ref{eq:linear-metric-decomp}).
From Eqs.~(\ref{eq:xia-components}), (\ref{eq:hVL-gauge-trans}),
(\ref{eq:hL-gauge-trans}), (\ref{eq:hTL-gauge-trans}), and
(\ref{eq:xii-decomp}), we define the variable $X_{a}$ as
\begin{eqnarray}
  X_{a}
  &:=&
       \left(
       h_{(VL)} - \frac{1}{2} a^{2}\partial_{\eta}h_{(TL)}
       \right) (d\eta)_{a}
     +
     a^{2} \left(
     h_{(TV)i}
     + \frac{1}{2} D_{i}h_{(TL)}
     \right)
     (dx^{i})_{a}
      \label{eq:gauge-variant-part-Xa-explicit}
  \\
  &=:&
       X_{\eta} (d\eta)_{a}
       +
       X_{i} (dx^{i})_{a}
       .
       \label{eq:gauge-variant-part-Xa-compare-}
\end{eqnarray}
We can easily check this vector field $X_{a}$ satisfies
Eq.~(\ref{eq:linear-metric-decomp-gauge-trans}).


Now, we derive the definition of the gauge-invariant part
${\cal H}_{ab}$.
First, we note that the gauge-transformation rule
(\ref{eq:hTTij-gauge-trans}) indicates that the $h_{(TT)ij}$ is gauge
invariant itself:
\begin{eqnarray}
  \label{eq:chiij-def}
  \stackrel{(1)}{\chi}_{ij} := h_{(TT)ij},
  \quad
  \gamma^{ij}\stackrel{(1)}{\chi}_{ij}=0 = D^{i}\stackrel{(1)}{\chi}_{ij}.
\end{eqnarray}
Second, from Eqs.~(\ref{eq:hVi-gauge-trans}) and
(\ref{eq:hTVi-gauge-trans}), we can easily check that the combination
\begin{eqnarray}
  \label{eq:nui-def}
  a^{2}\stackrel{(1)}{\nu}_{i} :=  h_{(V)i} - a^{2} \partial_{\eta}h_{(TV)i}
\end{eqnarray}
is gauge invariant.
The gauge-invariant variable $\stackrel{(1)}{\nu}_{i}$ is called a
``vector mode'' in the context of cosmological perturbations.
It satisfies the equation
\begin{eqnarray}
  \label{eq:divergenceless}
  D^{i}\stackrel{(1)}{\nu}_{i} = 0
\end{eqnarray}
from the divergenceless property of the variables $h_{(V)i}$ and
$h_{(TV)i}$.
Third, using the component $X_{\eta}$ of the gauge-variant part
$X_{a}$ given by Eq.~(\ref{eq:gauge-variant-part-Xa-explicit}), we can
define the gauge-invariant scalar variable $\stackrel{(1)}{\Phi}$ as
\begin{eqnarray}
  \label{eq:Newton-potential-def}
  - 2 a^{2}\stackrel{(1)}{\Phi}
  :=
  h_{\eta\eta} - 2 (\partial_{\eta}-{\cal H}) X_{\eta}
  .
\end{eqnarray}
This scalar variable $\stackrel{(1)}{\Phi}$ corresponds to the Newton
potential.
Finally, from the gauge-transformation rules
(\ref{eq:hL-gauge-trans}), (\ref{eq:hTL-gauge-trans}), and the
gauge-transformation rules for the component $X_{\eta}$ of the
variable defined by Eq.~(\ref{eq:gauge-variant-part-Xa-explicit}),
we can define the gauge-invariant variable $\stackrel{(1)} {\Psi}$ by
\begin{eqnarray}
  \label{eq:curvature-perturbation-potential-def}
  - 2 a^{2} \stackrel{(1)}{\Psi}
  :=
  a^{2} \left(
  h_{(L)} - \frac{1}{3} \Delta h_{(TL)}
  \right)
  +
  2 {\cal H} X_{\eta}
  .
\end{eqnarray}
This scalar variable $\stackrel{(1)}{\Psi}$ is called curvature
perturbation in the context of cosmological perturbations.
The two scalar functions $\stackrel{(1)}{\Phi}$ and
$\stackrel{(1)}{\Psi}$ are called ``scalar perturbations.''


In terms of the components of the gauge-variant variable $X_{a}$
defined by Eq.~(\ref{eq:gauge-variant-part-Xa-explicit}) and
gauge-invariant variables $\{\stackrel{(1)}{\Phi},$
$\stackrel{(1)}{\Psi},$ $\stackrel{(1)}{\nu}_{i},$
$\stackrel{(1)}{\chi}_{ij}\}$ defined by Eqs.~(\ref{eq:chiij-def}),
(\ref{eq:nui-def}), (\ref{eq:Newton-potential-def}), and
(\ref{eq:curvature-perturbation-potential-def}), the original
components $\{h_{\eta\eta},$ $h_{\eta i},$ $h_{ij}\}$ of the metric
perturbation are given by
\begin{eqnarray}
  h_{\eta\eta}
  &=&
      - 2 a^{2} \stackrel{(1)}{\Phi}
      +
      2 (\partial_{\eta} - {\cal H}) X_{\eta}
      ,
      \label{eq:hetaeta-express}
  \\
  h_{\eta i}
  &=&
      a^{2} \stackrel{(1)}{\nu}_{i}
      +
      D_{i}X_{\eta} + \partial_{\eta}X_{i}
      - 2 {\cal H} X_{i}
      ,
      \label{eq:hetai-express}
  \\
  h_{ij}
  &=&
      - 2 a^{2} \stackrel{(1)}{\Psi} \gamma_{ij}
      +
      a^{2} \stackrel{(1)}{\chi}_{ij}
     +
     2 D_{(i}X_{j)} - 2 {\cal H} \gamma_{ij} X_{\eta}
     .
     \label{eq:hij-express}
\end{eqnarray}
These expression are summarized in the covariant form
Eq.~(\ref{eq:linear-metric-decomp}) through the identification of the
gauge-invariant part ${\cal H}_{ab}$ as
\begin{eqnarray}
  {\cal H}_{ab}
  &=&
      a^{2} \left\{
      - 2 \stackrel{(1)}{\Phi} (d\eta)_{a}(d\eta)_{b}
      + 2 \stackrel{(1)}{\nu}_{i} (d\eta)_{(a}(dx^{i})_{b)}
     +
     \left( - 2 \stackrel{(1)}{\Psi} \gamma_{ij}
     + \stackrel{(1)}{\chi}_{ij} \right)
     (dx^{i})_{a}(dx^{j})_{b}
     \right\}
     .
     \label{eq:components-calHab}
\end{eqnarray}


Thus, we may say that our assumption for the decomposition
(\ref{eq:linear-metric-decomp}) in linear-order metric perturbation is
correct in the case of cosmological perturbations.
We have to emphasize that to accomplish
Eq.~(\ref{eq:linear-metric-decomp}), we implicitly assumed the
existence of the Green functions $\Delta^{-1}$, $(\Delta + 2 K)^{-1}$,
and $(\Delta + 3 K)^{-1}$.
This assumption is necessary to guarantee the one-to-one
correspondence between the variables $\{h_{\eta\eta},$ $h_{i\eta},$
$h_{ij}\}$ and $\{h_{\eta\eta},$ $h_{(VL)},$ $h_{(V)i},$ $h_{(L)},$
$h_{(TL)},$ $h_{(TV)j},$ ${h_{(TT)ij}}\}$, but excludes some
perturbative modes of the metric perturbations which belong to the
kernel of the operator $\Delta$, $(\Delta+2K)$, and $(\Delta+3 K)$
in the variable set $\{h_{\eta\eta},$ $h_{\eta i},$ $h_{ij}\}$ from
our consideration.
For example, we should regard that homogeneous modes, which belong to
the kernel of the operator $\Delta$, are not included in the
decomposition formula (\ref{eq:cosmological-linear-perturbations}).
If we have to treat these exceptional modes, the special treatments
for these modes are necessary, as mentioned above.
We call this problem of the treatments of these special mode as
{\it zero-mode problem}.


We also note the fact that the definition
(\ref{eq:linear-metric-decomp}) of the gauge-invariant variables is
not unique.
This comes from the fact that we can always construct new
gauge-invariant quantities by the combination of the gauge-invariant
variables.
For example,  using the gauge-invariant variables
$\stackrel{(1)}{\Phi}$ and $\stackrel{(1)}{\nu_{i}}$ of the
first-order metric perturbation, we can define a vector field $Z_{a}$
by
\begin{eqnarray}
  \label{eq:Za-example}
  Z_{a} := - a \stackrel{(1)}{\Phi} (d\eta)_{a} + a \stackrel{(1)}{\nu_{i}} (dx^{i})_{a}.
\end{eqnarray}
which is gauge-invariant.
We have to emphasize that the vector field (\ref{eq:Za-example}) is
just an example.
We can construct infinitely many different gauge-invariant vector
field $Z^{a}$.
Then, we can rewrite the decomposition formula
(\ref{eq:linear-metric-decomp}) for the linear-order metric
perturbation as
\begin{eqnarray}
  h_{ab}
  &=&
  {\cal H}_{ab} - {\pounds}_{Z}g_{ab}
  + {\pounds}_{Z}g_{ab} + {\pounds}_{X}g_{ab},
  \nonumber\\
  &=:&
  {\cal K}_{ab} + {\pounds}_{X+Z}g_{ab},
  \label{eq:non-uniqueness-of-gauge-invariant-metric-perturbation}
\end{eqnarray}
where we have defined new gauge-invariant variable
${\cal K}_{ab}$ by
${\cal K}_{ab}:={\cal H}_{ab}-{\pounds}_{Z}g_{ab}$.
Clearly, ${\cal K}_{ab}$ is gauge-invariant and the vector field
$X^{a}+Z^{a}$ satisfies
Eq.~(\ref{eq:linear-metric-decomp-gauge-trans}).
Therefore, we can construct infinitely many gauge-invariant variables
by changing the definition of the gauge-invariant vector field
$Z_{a}$.
In spite of this non-uniqueness of gauge-invariant variables, we
specify the components of the tensor ${\cal H}_{ab}$ as
Eq.~(\ref{eq:components-calHab}), which is the gauge-invariant part of
the linear-order metric perturbation associated with the longitudinal
gauge.


The existence of such infinitely many definitions of gauge-invariant
variables corresponds to the fact that there are infinitely many
``gauge-fixing'' method, in principle.
Due to the non-uniqueness of gauge-invariant variables, we can
consider the gauge-fixing in the first-order metric perturbation from
two different points of view.
The first point of view is that the gauge-fixing is to specify the
gauge-variant part $X^{a}$.
For example, the longitudinal gauge is realized by the gauge fixing
$X^{a}=0$.
Due to this gauge fixing $X^{a}=0$, we can regard the fact that
perturbative variables in the longitudinal gauge are the completely
gauge fixed variables.
On the other hand, we may also regard that the gauge fixing is the
specification of the gauge-invariant vector field $Z^{a}$ in Eq.~(\ref{eq:non-uniqueness-of-gauge-invariant-metric-perturbation}).
In this point of view, we do not specify the vector field $X^{a}$.
Instead, we have to specify the gauge-invariant vector $Z^{a}$ or
equivalently to specify the gauge-invariant metric perturbation
${\cal K}_{ab}$ without specifying $X^{a}$ so that the first-order
metric perturbation $h_{ab}$ coincides with the gauge-invariant
variables ${\cal K}_{ab}$ when we fix the gauge $X^{a}$ so that
$X^{a}+Z^{a}=0$.
These two different point of view of ``gauge fixing'' is equivalent
with each other due to the non-uniqueness of the definition
(\ref{eq:non-uniqueness-of-gauge-invariant-metric-perturbation})
of the gauge-invariant variables.
These two understandings of ``gauge fixing'' are explicitly discussed
through the derivation of the correspondence between the Poisson
gauge and the flat gauge in
Ref.~\cite{A.J.Christopherson-et.al-2011}.
As the result, we reach to the statement that our general formulation
is equivalent to the formulation developed by K.~A.~Malik and
Wands~\cite{K.A.Malik-D.Wands-2004}.
Recently, second-order cosmological perturbations in the
synchoronous gauge-fixing and its correspondence with the Poisson
gauge are extensively discussed in Refs.~\cite{B.Wang-Y.Zhang-2017a,B.Wang-Y.Zhang-2017b,B.Wang-Y.Zhang-2018,B.Wang-Y.Zhang-2019}.


\subsection{First-order Einstein equations}
\label{sec:First-order-Einstein-equations}


Here, we derive the linear-order Einstein equation
(\ref{eq:linear-order-Einstein-equation}).
To derive the components of the gauge-invariant part of the linearized
Einstein tensor ${}^{(1)}{\cal G}_{a}^{\;\;b}\left[{\cal H}\right]$,
which is defined by Eqs.~(\ref{eq:cal-G-def-linear}), we first derive
the components of the tensor
$H_{ab}^{\;\;\;\;c}\left[{\cal H}\right]$, which is defined in
Eq.~(\ref{eq:Habc-def-1}) with $A_{ab} = {\cal H}_{ab}$ and its
component (\ref{eq:components-calHab}).
These components are summarized in
Ref.~\cite{kouchan-cosmo-second-letter,kouchan-cosmo-second-full-paper}.


From Eq.~(\ref{eq:cal-G-def-linear}), the component of
${}^{(1)}{\cal G}_{a}^{\;\;b}\left[{\cal H}\right]$ are summarized as
\begin{eqnarray}
  a^{2} {}^{(1)}{\cal G}_{\eta}^{\;\;\eta}\left[{\cal H}\right]
  &=&
      - \left(
      - 6 {\cal H} \partial_{\eta}
      + 2 \Delta
      + 6 K
      \right) \stackrel{(1)}{\Psi}
      + 6 {\cal H}^{2} \stackrel{(1)}{\Phi}
      \label{eq:kouchan-10.120}
      ,
  \\
  a^{2} {}^{(1)}{\cal G}_{i}^{\;\;\eta}\left[{\cal H}\right]
  &=&
      - 2 \partial_{\eta} D_{i} \stackrel{(1)}{\Psi}
      - 2 {\cal H} D_{i} \stackrel{(1)}{\Phi}
      + \frac{1}{2} \left(
      \Delta
      + 2 K
      \right)
      \stackrel{(1)\;\;}{\nu_{i}}
  \label{eq:kouchan-10.121}
  ,
  \\
  a^{2} {}^{(1)}{\cal G}_{\eta}^{\;\;i}\left[{\cal H}\right]
  &=&
      2 \partial_{\eta} D^{i} \stackrel{(1)}{\Psi}
      + 2 {\cal H} D^{i} \stackrel{(1)}{\Phi}
      + \frac{1}{2} \left(
      - \Delta
      + 2 K
      + 4 {\cal H}^{2}
      - 4 \partial_{\eta}{\cal H}
      \right)
      \stackrel{(1)\;\;}{\nu^{i}}
  \label{eq:kouchan-10.122}
  ,
  \\
  a^{2} {}^{(1)}{\cal G}_{i}^{\;\;j}\left[{\cal H}\right]
  &=&
    D_{i} D^{j} \left(\stackrel{(1)}{\Psi} - \stackrel{(1)}{\Phi}\right)
    +
    \left\{
      \left(
        -   \Delta
        + 2 \partial_{\eta}^{2}
        + 4 {\cal H} \partial_{\eta}
        - 2 K
      \right)
      \stackrel{(1)}{\Psi}
      + \left(
          2 {\cal H} \partial_{\eta}
        + 4 \partial_{\eta}{\cal H}
        + 2 {\cal H}^{2}
        + \Delta
      \right)
      \stackrel{(1)}{\Phi}
    \right\}
    \gamma_{i}^{\;\;j}
  \nonumber\\
  &&
    - \frac{1}{2a^{2}} \partial_{\eta} \left\{
      a^{2} \left(
        D_{i} \stackrel{(1)\;\;}{\nu^{j}} + D^{j} \stackrel{(1)\;\;}{\nu_{i}}
      \right)
    \right\}
    + \frac{1}{2} \left(
      \partial_{\eta}^{2}
      + 2 {\cal H} \partial_{\eta}
      + 2 K
      - \Delta
    \right) \stackrel{(1)\;\;\;\;}{\chi_{i}^{\;\;j}}
     .
     \label{eq:kouchan-10.123}
\end{eqnarray}
Straightforward calculations show that these components of the
first-order gauge-invariant perturbation
${}^{(1)}{\cal G}_{a}^{\;\;b}\left[{\cal H}\right]$ of the Einstein
tensor satisfies the identity
(\ref{eq:linear-order-divergence-of-calGab}).
Although this confirmation is also possible without specification of
the tensor ${\cal H}_{ab}$, the confirmation of
Eq.~(\ref{eq:linear-order-divergence-of-calGab}) through the explicit
components (\ref{eq:kouchan-10.120})--(\ref{eq:kouchan-10.123})
implies that we have derived the components of
${}^{(1)}{\cal G}_{a}^{\;\;b}\left[{\cal H}\right]$ consistently.


Next, we summarize the first-order perturbation of the energy momentum
tensor for a scalar field.
Since, at the background level, we assume that the scalar field
$\varphi$ is homogeneous as
Eq.~(\ref{eq:background-varphi-is-homogeneous}), the components of the
gauge-invariant part of the first-order energy-momentum tensor
${}^{(1)}\!{\cal T}_{a}^{\;\;b}$ are given by
\begin{eqnarray}
  a^{2} {}^{(1)}\!{\cal T}_{\eta}^{\;\;\eta}
  &=&
      - \partial_{\eta}\varphi_{1}\partial_{\eta}\varphi
      + \stackrel{(1)}{\Phi} (\partial_{\eta}\varphi)^{2}
      - a^{2}\frac{dV}{d\varphi}\varphi_{1}
      \label{eq:kouchan-10.161}
      ,
  \\
  a^{2}{}^{(1)}\!{\cal T}_{i}^{\;\;\eta}
  &=&
      -
      D_{i}\varphi_{1}\partial_{\eta}\varphi
      ,
      \label{eq:kouchan-10.162}
  \\
  a^{2} {}^{(1)}\!{\cal T}_{\eta}^{\;\;i}
  &=&
      \partial_{\eta}\varphi D^{i}\varphi_{1}
      + (\partial_{\eta}\varphi)^{2} \stackrel{(1)\;\;}{\nu^{i}}
      ,
      \label{eq:kouchan-10.163}
  \\
  a^{2} {}^{(1)}\!{\cal T}_{i}^{\;\;j}
  &=&
      \gamma_{i}^{\;\;j}\left(
      \partial_{\eta}\varphi_{1} \partial_{\eta}\varphi
      - \stackrel{(1)}{\Phi} (\partial_{\eta}\varphi)^{2}
      - a^{2} \frac{dV}{d\varphi} \varphi_{1}
      \right)
      .
      \label{eq:kouchan-10.165}
\end{eqnarray}
The second equation in (\ref{eq:kouchan-10.163}) shows that there is
no anisotropic stress in the energy-momentum tensor of the single
scalar field.
Then, we obtain
\begin{eqnarray}
  \label{eq:absence-of-anisotropic-stress-Einstein-i-j-traceless-scalar}
  \stackrel{(1)}{\Phi} = \stackrel{(1)}{\Psi}.
\end{eqnarray}
From
Eqs.~(\ref{eq:kouchan-10.120})--(\ref{eq:kouchan-10.163}) and
(\ref{eq:absence-of-anisotropic-stress-Einstein-i-j-traceless-scalar}),
the components of scalar parts of the linearized Einstein equation
(\ref{eq:linear-order-Einstein-equation}) are given
as~\cite{Mukhanov-Feldman-Brandenberger-1992}
\begin{eqnarray}
  \left(
        \Delta
    - 3 {\cal H} \partial_{\eta}
    + 4 K
    - \partial_{\eta}{\cal H}
    - 2 {\cal H}^{2}
  \right) \stackrel{(1)}{\Phi}
  &=&
  4 \pi G \left(
    \partial_{\eta}\varphi_{1} \partial_{\eta}\varphi
    + a^{2}\frac{dV}{d\varphi}\varphi_{1}
  \right)
  \label{eq:kouchan-18.185}
  , \\
  \partial_{\eta}\stackrel{(1)}{\Phi} + {\cal H} \stackrel{(1)}{\Phi}
  &=&
  4 \pi G \varphi_{1} \partial_{\eta}\varphi
  \label{eq:kouchan-18.186}
  , \\
  \left(
        \partial_{\eta}^{2}
    + 3 {\cal H} \partial_{\eta}
    +   \partial_{\eta}{\cal H}
    + 2 {\cal H}^{2}
  \right)
  \stackrel{(1)}{\Phi}
     &=&
  4 \pi G
  \left(
    \partial_{\eta}\varphi_{1} \partial_{\eta}\varphi
    - a^{2} \frac{dV}{d\varphi} \varphi_{1}
  \right)
  \label{eq:kouchan-18.187}
  .
\end{eqnarray}
In the derivation of
Eqs.~(\ref{eq:kouchan-18.185})--(\ref{eq:kouchan-18.187}), we have
used Eq.~(\ref{eq:background-Einstein-equations-scalar-3}).
We also note that only two of these equations are independent.
Further, the vector part of the component
${}^{(1)}\!{\cal G}_{i}^{\;\;\eta}\left[{\cal H}\right]=8\pi G{}^{(1)}\!{\cal T}_{i}^{\;\;\eta}$
shows that
\begin{eqnarray}
  \label{eq:no-first-order-vector-mode-scalar-field-case}
  \stackrel{(1)}{\nu}_{i} = 0.
\end{eqnarray}
The equation for the tensor mode $\stackrel{(1)\;\;}{\chi_{ij}}$ is
given by
\begin{eqnarray}
  \label{eq:linearized-Einstein-i-j-traceless-tensor}
  \left(
        \partial_{\eta}^{2}
    + 2 {\cal H} \partial_{\eta}
    + 2 K
    - \Delta
  \right) \stackrel{(1)\;\;\;\;}{\chi_{i}^{\;\;j}}
  =
  0
  .
\end{eqnarray}


Combining Eqs.~(\ref{eq:kouchan-18.185}) and
(\ref{eq:kouchan-18.187}), we eliminate the potential term of the
scalar field and thereby obtain
\begin{eqnarray}
  \left(
        \partial_{\eta}^{2}
    +   \Delta
    + 4 K
  \right) \stackrel{(1)}{\Phi}
  =
  8 \pi G \partial_{\eta}\varphi_{1} \partial_{\eta}\varphi.
  \label{eq:scalar-linearized-Einstein-scalar-master-eq-pre}
\end{eqnarray}
Further, using Eq.~(\ref{eq:kouchan-18.186}) to express
$\partial_{\eta}\varphi_{1}$ in terms of
$\partial_{\eta}\stackrel{(1)}{\Phi}$ and $\stackrel{(1)}{\Phi}$, we
also eliminate $\partial_{\eta}\varphi_{1}$ in
Eq.~(\ref{eq:scalar-linearized-Einstein-scalar-master-eq-pre}).
Hence, we have
\begin{eqnarray}
  &&
     \left\{
     \partial_{\eta}^{2}
     + 2 \left(
     {\cal H}
     - \frac{2\partial_{\eta}^{2}\varphi}{\partial_{\eta}\varphi}
     \right) \partial_{\eta}
     - \Delta
     - 4 K
     + 2
     \left( \partial_{\eta}{\cal H}
     - \frac{{\cal H}\partial_{\eta}^{2}\varphi}{\partial_{\eta}\varphi}
     \right)
     \right\} \stackrel{(1)}{\Phi}
     =
     0.
     \label{eq:scalar-linearized-Einstein-scalar-master-eq}
\end{eqnarray}
This is the master equation for the scalar mode perturbation of the
cosmological perturbation in universe filled with a single scalar field.
It is also known that
Eq.~(\ref{eq:scalar-linearized-Einstein-scalar-master-eq}) reduces to
a simple equation through a change of
variables~\cite{Mukhanov-Feldman-Brandenberger-1992}.


\subsection{First-order Klein-Gordon equations}
\label{sec:First-order-Klein-Gordon-equations}


Next, we consider the first-order perturbation of the Klein-Gordon
equation (\ref{eq:Klein-Gordon-eq-first-gauge-inv-def}).
By the straightforward calculations using
Eqs.~(\ref{eq:background-metric}), (\ref{eq:components-calHab}),
(\ref{eq:background-varphi-is-homogeneous}),
(\ref{eq:background-Klein-Gordon-equation}), and the components
$H_{a}^{\;\;ac}$ summarized in
Ref.~\cite{kouchan-cosmo-second-letter,kouchan-cosmo-second-full-paper},
the gauge-invariant part $\stackrel{(1)}{{\cal C}_{(K)}}$ of the
first-order Klein-Gordon equation defined by
Eq.~(\ref{eq:Klein-Gordon-eq-first-gauge-inv-def}) is given by
\begin{eqnarray}
  - a^{2} \stackrel{(1)}{{\cal C}_{(K)}}
  &=&
      \partial_{\eta}^{2}\varphi_{1}
  + 2 {\cal H} \partial_{\eta}\varphi_{1}
  -   \Delta\varphi_{1}
  - \left(
        \partial_{\eta}\stackrel{(1)}{\Phi}
    + 3 \partial_{\eta}\stackrel{(1)}{\Psi}
  \right) \partial_{\eta}\varphi
  + 2 a^{2} \stackrel{(1)}{\Phi} \frac{\partial V}{\partial\bar{\varphi}}(\varphi)
  +   a^{2}\varphi_{1} \frac{\partial^{2}V}{\partial\bar{\varphi}^{2}}(\varphi)
  \nonumber\\
  &=& 0
  \label{eq:kouchan-17.806-first-explicit}
  .
\end{eqnarray}


Through the background Einstein equations
(\ref{eq:background-Einstein-equations-scalar-1}),
(\ref{eq:background-Einstein-equations-scalar-2}), and the first-order
perturbations (\ref{eq:kouchan-18.186}) and
(\ref{eq:scalar-linearized-Einstein-scalar-master-eq}) of the Einstein
equation, we can easily derive the first-order perturbation of the
Klein-Gordon equation
(\ref{eq:kouchan-17.806-first-explicit})~\cite{kouchan-second-cosmo-consistency}.
Hence, the first-order perturbation of the Klein-Gordon equation is
not independent of the background and the first-order perturbation of
the Einstein equation.
Therefore, from the viewpoint of the Cauchy problem, any information
obtained from the first-order perturbation of the Klein-Gordon
equation should also be obtained from the set of the background and
the first-order the Einstein equation, in principle.


\section{Equations for the second-order cosmological perturbations}
\label{sec:Equations-for-the-second-order-cosmological-perturbations}


Now, we develop the second-order perturbation theory on the
cosmological background spacetime in
Sec.~\ref{sec:Cosmological-Background-spacetime-equations} within the
general framework of the gauge-invariant perturbation theory reviewed
in Sec.~\ref{sec:General-framework-of-GR-GI-perturbation-theory}.
Since we have already confirm the important step of our general
framework, i.e., the assumption for the decomposition
(\ref{eq:linear-metric-decomp}) of the linear-order metric
perturbation is correct except for some special modes which we ignore
here.
Hence, the general framework reviewed in
Sec.~\ref{sec:General-framework-of-GR-GI-perturbation-theory} is
applicable.
Applying this framework, we define the second-order gauge-invariant
variables of the metric perturbation in
Sec.~\ref{sec:Second-order-gauge-invariant-metric-variables}.
In Sec.~\ref{sec:Second-order-gauge-invariant-energy-momentum}, we
summarize the explicit components of the gauge-invariant parts of the
second-order perturbation of the Einstein tensor.
In
Sec.~\ref{sec:Second-order-gauge-invariant-energy-momentum-Klein-Gordon},
we summarize the explicit components of the second-order perturbation
of the energy-momentum tensor and the Klein-Gordon equations.
Then, in Sec.~\ref{sec:Secnd-order-cosmological-Einstein-equations},
we derive the second-order Einstein equations in terms of
gauge-invariant variables.
The resulting equations have the source terms which constitute of the
quadratic terms of the linear-order perturbations.
Although these source terms have complicated forms, we give identities
which comes from the consistency of all the second-order perturbations
of the Einstein equation and the Klein-Gordon equation in
Sec.~\ref{sec:Consistency-of-the-second-order-perturbations}.


\subsection{Gauge-invariant metric perturbations}
\label{sec:Second-order-gauge-invariant-metric-variables}


First, we consider the components of the gauge-invariant variables for
the metric perturbation of second order.
The variable $\hat{L}_{ab}$ defined by Eq.~(\ref{eq:Lhatab-def}) is
transformed as Eq.~(\ref{eq:kouchan-4.67}) under the gauge
transformation and we may regard the generator $\sigma_{a}$ defined by
Eq.~(\ref{eq:sigma-def}) as an arbitrary vector field on
${\cal M}_{0}$ from the fact that the generator $\xi_{2}^{a}$ in
Eq.~(\ref{eq:sigma-def}) is arbitrary.
We can apply the procedure to find gauge-invariant variables for the
first-order metric perturbations (\ref{eq:components-calHab}) in
Sec.~\ref{sec:Gauge-invariant-metric-perturbations}.
Then, we can accomplish the decomposition
(\ref{eq:Lhatab-decomposition}).
Following to the same argument as in the linear case, we may choose
the components of the gauge-invariant variables ${\cal L}_{ab}$ in
Eq.~(\ref{eq:H-ab-in-gauge-X-def-second-1}) as
\begin{eqnarray}
  {\cal L}_{ab}
  &=&
  - 2 a^{2} \stackrel{(2)}{\Phi} (d\eta)_{a}(d\eta)_{b}
  + 2 a^{2} \stackrel{(2)\;\;}{\nu_{i}} (d\eta)_{(a}(dx^{i})_{b)}
  + a^{2}
  \left( - 2 \stackrel{(2)}{\Psi} \gamma_{ij}
    + \stackrel{(2)\;\;\;\;}{{\chi}_{ij}} \right)
  (dx^{i})_{a}(dx^{j})_{b},
  \label{eq:second-order-gauge-inv-metrc-pert-components}
\end{eqnarray}
where $\stackrel{(2)}{\nu}_{i}$ and
$\stackrel{(2)\;\;\;\;}{\chi_{ij}}$ satisfy the equations
\begin{eqnarray}
  && D^{i}\stackrel{(2)\;\;}{\nu_{i}} = 0, \quad
  \stackrel{(2)\;\;\;\;}{\chi^{i}_{\;\;i}} = 0, \quad
   D^{i}\stackrel{(2)\;\;\;\;}{\chi_{ij}} = 0.
\end{eqnarray}
The gauge-invariant variables $\stackrel{(2)}{\Phi}$ and
$\stackrel{(2)}{\Psi}$ are the scalar mode perturbations of second
order, and $\stackrel{(2)\;\;}{\nu_{i}}$ and
$\stackrel{(2)\;\;\;\;}{\chi_{ij}}$ are the second-order vector and
tensor modes of the metric perturbations, respectively.


Here, we also note the fact that the decomposition
(\ref{eq:H-ab-in-gauge-X-def-second-1}) is not unique.
This situation is similar to the case of the linear-order metric
perturbation $h_{ab}$ discussed above, but more complicated.
In the definition of the gauge-invariant variables of the second-order
metric perturbation, we may replace
\begin{eqnarray}
  \label{eq:non-uniqueness-of-gauge-invariant-2nd-order-metric-perturbation-1}
  X^{a} = X^{'a} - Z^{'a},
\end{eqnarray}
where $Z^{'a}$ is gauge invariant and $X^{'a}$ is transformed as
\begin{eqnarray}
  \label{eq:non-uniqueness-of-gauge-invariant-2nd-order-metric-perturbation-2}
  {}_{{\cal Y}}\!X^{'a} - {}_{{\cal X}}\!X^{'a} = \xi^{a}_{1}
\end{eqnarray}
under the gauge transformation ${\cal X}_{\lambda}$
$\rightarrow$ ${\cal Y}_{\lambda}$.
This $Z^{'a}$ may be different from the vector $Z^{a}$ in
Eq.~(\ref{eq:non-uniqueness-of-gauge-invariant-metric-perturbation}).
By the replacement
(\ref{eq:non-uniqueness-of-gauge-invariant-2nd-order-metric-perturbation-1}),
the second-order metric perturbation
(\ref{eq:H-ab-in-gauge-X-def-second-1}) is given in the form
\begin{eqnarray}
  l_{ab}
  &=:&
  {\cal J}_{ab}
  + 2 {\pounds}_{X'} h_{ab}
  + \left(
      {\pounds}_{Y'}
    - {\pounds}_{X'}^{2}
  \right)
  g_{ab},
  \label{eq:non-uniqueness-of-gauge-invariant-2nd-order-metric-perturbation-4}
\end{eqnarray}
where we defined
\begin{eqnarray}
  {\cal J}_{ab}
  &:=&
  {\cal L}_{ab}
  -   {\pounds}_{W}g_{ab}
  - 2 {\pounds}_{Z'}{\cal K}_{ab}
  - 2 {\pounds}_{Z'}{\pounds}_{Z}g_{ab}
  +   {\pounds}_{Z'}^{2}g_{ab}
  \label{eq:calJ-def}
  , \\
  Y^{'a} &:=& Y^{a}+W^{a}+[X',Z']^{a}
  .
  \label{eq:Y-prime-transform}
\end{eqnarray}
Here, the vector field $W^{a}$ in Eq.~(\ref{eq:Y-prime-transform})
constitute of some components of gauge-invariant second-order metric
perturbation ${\cal L}_{ab}$ like $Z^{a}$ in
Eq.~(\ref{eq:non-uniqueness-of-gauge-invariant-metric-perturbation}).
The tensor field ${\cal J}_{ab}$ is manifestly gauge invariant.
The gauge transformation rule of the new gauge-variant part $Y^{'a}$
of the second-order metric perturbation is given by
\begin{eqnarray}
  {}_{{\cal Y}}\!Y^{'a}
  -
  {}_{{\cal X}}\!Y^{'a}
  &=&
  \xi_{(2)}^{a} + [\xi_{(1)},X']^{a}
  .
\end{eqnarray}
Although
Eq.~(\ref{eq:non-uniqueness-of-gauge-invariant-2nd-order-metric-perturbation-4})
is similar to Eq.~(\ref{eq:H-ab-in-gauge-X-def-second-1}),
the tensor fields ${\cal L}_{ab}$ and ${\cal J}_{ab}$ are different
from each other.
Thus, the definition of the gauge-invariant variables for the
second-order metric perturbation is not unique in a more complicated
way than the linear order.
This non-uniqueness of gauge-invariant variables for the metric
perturbations propagates to the definition
(\ref{eq:matter-gauge-inv-def-1.0}) and
(\ref{eq:matter-gauge-inv-def-2.0}) of the gauge-invariant variables
for matter fields.


In spite of the existence of infinitely many definitions of the
gauge-invariant variables, in this  paper, we consider the components
of ${\cal L}_{ab}$ given by
Eq.~(\ref{eq:second-order-gauge-inv-metrc-pert-components}).
Eq.~(\ref{eq:second-order-gauge-inv-metrc-pert-components})
corresponds to the second-order extension of the longitudinal gauge,
which is called Poisson gauge $X^{a}=Y^{a}=0$.


\subsection{Einstein tensor}
\label{sec:Second-order-gauge-invariant-energy-momentum}


Here, we evaluate the second-order perturbation of the Einstein tensor
(\ref{eq:second-Einstein-2,0-0,2}) with the cosmological background
(\ref{eq:background-metric}).
We evaluate the term
${}^{(1)}\!{\cal G}_{a}^{\;\;b}\left[{\cal L}\right]$ and
${}^{(2)}\!{\cal G}_{a}^{\;\;b}\left[{\cal H}, {\cal H}\right]$ in the
Einstein equation (\ref{eq:second-order-Einstein-equation}).


First, we evaluate the term
${}^{(1)}\!{\cal G}_{a}^{\;\;b}\left[{\cal L}\right]$ in the Einstein
equation (\ref{eq:second-order-Einstein-equation}).
Because the components
(\ref{eq:second-order-gauge-inv-metrc-pert-components}) of
${\cal L}_{ab}$ are obtained through the replacements
\begin{eqnarray}
  \label{eq:replacement-from-calHab-to-calLab}
  \stackrel{(1)}{\Phi} \rightarrow \stackrel{(2)}{\Phi}, \quad
  \stackrel{(1)\;\;}{\nu_{i}} \rightarrow \stackrel{(2)\;\;}{\nu}_{i}, \quad
  \stackrel{(1)}{\Psi} \rightarrow \stackrel{(2)}{\Psi}, \quad
  \stackrel{(1)\;\;\;\;}{\chi_{ij}} \rightarrow
  \stackrel{(2)\;\;\;\;}{\chi_{ij}}
\end{eqnarray}
in the components (\ref{eq:components-calHab}) of ${\cal H}_{ab}$, we
easily obtain the components of
${}^{(1)}{\cal G}_{a}^{\;\;b}\left[{\cal L}\right]$ through the
replacements (\ref{eq:replacement-from-calHab-to-calLab}) in Eqs.~(\ref{eq:kouchan-10.120})--(\ref{eq:kouchan-10.123}).


From Eq.~(\ref{eq:components-calHab}), we can derive the components of
${}^{(2)}\!{\cal G}_{a}^{\;\;b}={}^{(2)}\!{\cal G}_{a}^{\;\;b}[{\cal H},{\cal H}]$
defined by Eqs.~(\ref{eq:cal-G-def-second})--(\ref{eq:Habc-def-2}) in
a straightforward manner.
Here, we use the results
(\ref{eq:absence-of-anisotropic-stress-Einstein-i-j-traceless-scalar})
and (\ref{eq:no-first-order-vector-mode-scalar-field-case}) of the
first-order Einstein equations, for simplicity.
Then the explicit components
${}^{(2)}\!{\cal G}_{a}^{\;\;b}={}^{(2)}\!{\cal G}_{a}^{\;\;b}[{\cal H},{\cal H}]$
are summarized as
\begin{eqnarray}
  \frac{a^{2}}{2}
  {}^{(2)}\!{\cal G}_{\eta}^{\;\;\eta}
  &=&
      -  3 D_{k}\stackrel{(1)}{\Phi} D^{k}\stackrel{(1)}{\Phi}
      -  8 \stackrel{(1)}{\Phi} \Delta\stackrel{(1)}{\Phi}
      -  3 \left(\partial_{\eta}\stackrel{(1)}{\Phi}\right)^{2}
      - 12 \left({\cal H}^{2} + K\right) \left(\stackrel{(1)}{\Phi}\right)^{2}
      +             D_{l}D_{k}\stackrel{(1)}{\Phi} \stackrel{(1)}{\chi^{lk}}
      \nonumber\\
  &&
     + \frac{1}{8} \partial_{\eta}\stackrel{(1)}{\chi^{kl}}
     \left(
     \partial_{\eta}
     + 8 {\cal H}
     \right) \stackrel{(1)}{\chi_{kl}}
     + \frac{1}{2} D_{k}\stackrel{(1)}{\chi_{lm}} D^{[l}\stackrel{(1)}{\chi^{k]m}}
     - \frac{1}{8} D_{k}\stackrel{(1)}{\chi_{lm}} D^{k}\stackrel{(1)}{\chi^{ml}}
     - \frac{1}{2} \stackrel{(1)}{\chi^{lm}} \left(
     \Delta - K
     \right) \stackrel{(1)}{\chi_{lm}}
     ,
     \label{eq:generic-2-calG-eta-eta}
  \\
  \frac{a^{2}}{2}
  {}^{(2)}\!{\cal G}_{\eta}^{\;\;i}
  &=&
      8  \stackrel{(1)}{\Phi} \partial_{\eta}D^{i}\stackrel{(1)}{\Phi}
      -  D_{j}\stackrel{(1)}{\Phi} \partial_{\eta}\stackrel{(1)}{\chi^{ji}}
      -  \left(
      \partial_{\eta}D_{j}\stackrel{(1)}{\Phi}
      + 2 {\cal H} D_{j}\stackrel{(1)}{\Phi}
      \right) \stackrel{(1)}{\chi^{ij}}
      + \frac{1}{4} \partial_{\eta}\stackrel{(1)}{\chi_{jk}} D^{i}\stackrel{(1)}{\chi^{kj}}
      + \stackrel{(1)}{\chi_{kl}} \partial_{\eta}D^{[i}\stackrel{(1)}{\chi^{k]l}}
      ,
      \label{eq:generic-2-calG-eta-i}
  \\
  \frac{a^{2}}{2}
  {}^{(2)}\!{\cal G}_{i}^{\;\;\eta}
  &=&
      8 {\cal H} \stackrel{(1)}{\Phi} D_{i}\stackrel{(1)}{\Phi}
      - 2 D_{i}\stackrel{(1)}{\Phi} \partial_{\eta}\stackrel{(1)}{\Phi}
      +   D^{j}\stackrel{(1)}{\Phi} \partial_{\eta}\stackrel{(1)}{\chi_{ij}}
      -   \partial_{\eta}D^{j}\stackrel{(1)}{\Phi} \stackrel{(1)}{\chi_{ij}}
      -   \frac{1}{4} \partial_{\eta}\stackrel{(1)}{\chi^{kj}} D_{i}\stackrel{(1)}{\chi_{kj}}
      +  \stackrel{(1)}{\chi^{kj}} \partial_{\eta}D_{[j}\stackrel{(1)}{\chi_{i]k}}
      ,
      \label{eq:generic-2-calG-i-eta}
  \\
  \frac{a^{2}}{2}
  {}^{(2)}\!{\cal G}_{i}^{\;\;j}
  &=&
      \left\{
      -  3  D_{k}\stackrel{(1)}{\Phi} D^{k}\stackrel{(1)}{\Phi}
      -  4 \stackrel{(1)}{\Phi} \left( \Delta + K \right) \stackrel{(1)}{\Phi}
      -  \partial_{\eta}\stackrel{(1)}{\Phi} \partial_{\eta}\stackrel{(1)}{\Phi}
      -  8  {\cal H} \stackrel{(1)}{\Phi} \partial_{\eta}\stackrel{(1)}{\Phi}
      -  4 \left( 2 \partial_{\eta}{\cal H} + {\cal H}^{2} \right) \left(\stackrel{(1)}{\Phi}\right)^{2}
      \right\} \gamma_{i}^{\;\;j}
  \nonumber\\
  &&
     + 2 D_{i}\stackrel{(1)}{\Phi} D^{j}\stackrel{(1)}{\Phi}
     + 4 \stackrel{(1)}{\Phi} D_{i}D^{j}\stackrel{(1)}{\Phi}
     + \stackrel{(1)}{\chi_{i}^{\;\;j}} \left( \partial_{\eta}^{2} + 2 {\cal H} \partial_{\eta} \right) \stackrel{(1)}{\Phi}
     + D_{k}\stackrel{(1)}{\Phi} \left( D_{i}\stackrel{(1)}{\chi^{jk}} + D^{j}\stackrel{(1)}{\chi_{ik}} \right)
  \nonumber\\
  &&
    - 2 D^{k}\stackrel{(1)}{\Phi} D_{k}\stackrel{(1)}{\chi_{i}^{\;\;j}}
    - 2 \stackrel{(1)}{\Phi} \left( \Delta - 2 K \right) \stackrel{(1)}{\chi_{i}^{\;\;j}}
    -             \Delta \stackrel{(1)}{\Phi} \stackrel{(1)}{\chi_{i}^{\;\;j}}
    +             D_{k}D_{i}\stackrel{(1)}{\Phi} \stackrel{(1)}{\chi^{jk}}
    +             D^{m}D^{j}\stackrel{(1)}{\Phi} \stackrel{(1)}{\chi_{im}}
    -             D_{l}D_{k}\stackrel{(1)}{\Phi} \stackrel{(1)}{\chi^{lk}} \gamma_{i}^{\;\;j}
  \nonumber\\
  &&
    - \frac{1}{2} \partial_{\eta}\stackrel{(1)}{\chi_{ik}} \partial_{\eta}\stackrel{(1)}{\chi^{kj}}
    + D_{k}\stackrel{(1)}{\chi_{il}} D^{[k}\stackrel{(1)}{\chi^{l]j}}
    + \frac{1}{4} D^{j}\stackrel{(1)}{\chi_{lk}} D_{i}\stackrel{(1)}{\chi^{lk}}
    + \frac{1}{2} \stackrel{(1)}{\chi_{lm}} D_{i}D^{j}\stackrel{(1)}{\chi^{ml}}
    - \frac{1}{2} \stackrel{(1)}{\chi_{lm}} D^{l}D_{i}\stackrel{(1)}{\chi^{mj}}
  \nonumber\\
  &&
     -  \frac{1}{2} \stackrel{(1)}{\chi^{lm}} D_{l}D^{j}\stackrel{(1)}{\chi_{mi}}
     + \frac{1}{2} \stackrel{(1)}{\chi^{lm}} D_{m}D_{l}\stackrel{(1)}{\chi_{i}^{\;\;j}}
     -  \frac{1}{2} \stackrel{(1)}{\chi^{jk}} \left(
     \partial_{\eta}^{2}  + 2 {\cal H} \partial_{\eta} - \Delta  + 2 K
     \right) \stackrel{(1)}{\chi_{ik}}
  \nonumber\\
  &&
    + \frac{1}{2} \left\{
        \frac{3}{4} \partial_{\eta}\stackrel{(1)}{\chi_{lk}} \partial_{\eta}\stackrel{(1)}{\chi^{kl}}
      +   \stackrel{(1)}{\chi_{kl}} \left(
            \partial_{\eta}^{2}
        + 2 {\cal H} \partial_{\eta}
        -   \Delta
        +   K
      \right) \stackrel{(1)}{\chi^{lk}}
      - \frac{1}{4} D_{k}\stackrel{(1)}{\chi_{lm}} D^{k}\stackrel{(1)}{\chi^{ml}}
      +             D_{k}\stackrel{(1)}{\chi_{lm}} D^{[l}\stackrel{(1)}{\chi^{k]m}}
    \right\} \gamma_{i}^{\;\;j}
     .
      \nonumber\\
     \label{eq:generic-2-calG-i-j}
\end{eqnarray}
We have checked the identity (\ref{eq:second-div-of-calGab-1,1})
through
Eqs.~(\ref{eq:generic-2-calG-eta-eta})--(\ref{eq:generic-2-calG-i-j}).
Then, we may say that the expressions
(\ref{eq:generic-2-calG-eta-eta})--(\ref{eq:generic-2-calG-i-j}) are
self-consistent.


\subsection{Energy-momentum tensor and Klein-Gordon equation}
\label{sec:Second-order-gauge-invariant-energy-momentum-Klein-Gordon}


Here, we summarize the explicit components of the gauge-invariant
parts (\ref{eq:second-order-energy-momentum-scalar-gauge-inv}) of the
second-order perturbation of energy momentum tensor for a single
scalar field in terms of gauge-invariant variables.
Through Eqs.~(\ref{eq:background-varphi-is-homogeneous}),
(\ref{eq:components-calHab}),
(\ref{eq:second-order-gauge-inv-metrc-pert-components}), the
components of
Eq.~(\ref{eq:second-order-energy-momentum-scalar-gauge-inv}) are
derived by the straightforward calculations.
In this paper, we just summarize the components of
${}^{(2)}\!{\cal T}_{a}^{\;\;b}$ in the situation where the first-order
Einstein equations
(\ref{eq:absence-of-anisotropic-stress-Einstein-i-j-traceless-scalar})
and (\ref{eq:no-first-order-vector-mode-scalar-field-case}) are
satisfied:
\begin{eqnarray}
  a^{2} {}^{(2)}\!{\cal T}_{\eta}^{\;\;\eta}
  &=&
      -  \partial_{\eta}\varphi \partial_{\eta}\varphi_{2}
      +  (\partial_{\eta}\varphi)^{2} \stackrel{(2)}{\Phi}
      -  a^{2} \varphi_{2}\frac{\partial V}{\partial\varphi}
      + 4 \partial_{\eta}\varphi \stackrel{(1)}{\Phi} \partial_{\eta}\varphi_{1}
      -  4 (\partial_{\eta}\varphi)^{2} \left(\stackrel{(1)}{\Phi}\right)^{2}
      \nonumber\\
  &&
     -  (\partial_{\eta}\varphi_{1})^{2}
     -  D_{i}\varphi_{1} D^{i}\varphi_{1}
     -  a^{2} (\varphi_{1})^{2} \frac{\partial^{2}V}{\partial\varphi^{2}}
     \label{eq:kouchan-19.209}
     ,
  \\
  a^{2} {}^{(2)}\!{\cal T}_{i}^{\;\;\eta}
  &=&
      -  \partial_{\eta}\varphi D_{i}\varphi_{2}
      + 4 \partial_{\eta}\varphi D_{i}\varphi_{1} \stackrel{(1)}{\Phi}
      -  2 D_{i}\varphi_{1} \partial_{\eta}\varphi_{1}
      \label{eq:kouchan-19.211}
      ,
  \\
  a^{2} {}^{(2)}\!{\cal T}_{\eta}^{\;\;i}
  &=&
            \partial_{\eta}\varphi D^{i}\varphi_{2}
      + 2 \partial_{\eta}\varphi_{1} D^{i}\varphi_{1}
      + 4 \partial_{\eta}\varphi \stackrel{(1)}{\Phi} D^{i}\varphi_{1}
      -  2 \partial_{\eta}\varphi \stackrel{(1)}{\chi^{il}} D_{l}\varphi_{1}
      \label{eq:kouchan-19.210}
      ,
  \\
  a^{2} {}^{(2)}\!{\cal T}_{i}^{\;\;j}
  &=&
      D_{i}\varphi_{1} D^{j}\varphi_{1}
      \nonumber\\
  &&
     + \frac{1}{2} \gamma_{i}^{\;\;j}
     \left\{
        \partial_{\eta}\varphi \partial_{\eta}\varphi_{2}
     -  4 \partial_{\eta}\varphi \stackrel{(1)}{\Phi} \partial_{\eta}\varphi_{1}
     + 4 (\partial_{\eta}\varphi)^{2} \left(\stackrel{(1)}{\Phi}\right)^{2}
     -  (\partial_{\eta}\varphi)^{2} \stackrel{(2)}{\Phi}
     + (\partial_{\eta}\varphi_{1})^{2}
     -  D_{l}\varphi_{1} D^{l}\varphi_{1}
     \right.
     \nonumber\\
  && \quad\quad\quad\quad
     \left.
     -  a^{2} \varphi_{2} \frac{\partial V}{\partial\varphi}
     -  a^{2} (\varphi_{1})^{2} \frac{\partial^{2}V}{\partial\varphi^{2}}
     \right\}
     .
     \label{eq:kouchan-19.212}
\end{eqnarray}
More generic formulae for the components of
${}^{(2)}\!{\cal T}_{a}^{b}$ are given in
Ref.~\cite{kouchan-second-cosmo-matter}.


Next, we show the gauge-invariant second-order the Klein-Gordon
equation.
We only consider the simple situation where
Eqs.~(\ref{eq:absence-of-anisotropic-stress-Einstein-i-j-traceless-scalar})
and (\ref{eq:no-first-order-vector-mode-scalar-field-case}) are
satisfied.
The formulae for more generic situation is given in
Ref.~\cite{kouchan-second-cosmo-matter}.
Through Eqs.~(\ref{eq:components-calHab}),
(\ref{eq:second-order-gauge-inv-metrc-pert-components}),
(\ref{eq:background-varphi-is-homogeneous}), the second-order
perturbation of the Klein-Gordon equation
(\ref{eq:Klein-Gordon-eq-second-gauge-inv-def}) is given by
\begin{eqnarray}
  - a^{2} \stackrel{(2)}{{\cal C}_{(K)}}
  &=&
       \partial_{\eta}^{2}\varphi_{2}
  +  2 {\cal H} \partial_{\eta}\varphi_{2}
  -    \Delta\varphi_{2}
  -    \left(
    \partial_{\eta}\stackrel{(2)}{\Phi}
    +  3 \partial_{\eta}\stackrel{(2)}{\Psi}
  \right) \partial_{\eta}\varphi
  +  2 a^{2} \stackrel{(2)}{\Phi} \frac{\partial V}{\partial\bar{\varphi}}(\varphi)
  +    a^{2}\varphi_{2}\frac{\partial^{2}V}{\partial\bar{\varphi}^{2}}(\varphi)
  - \Xi_{(K)}
  \nonumber\\
  &=& 0
  ,
  \label{eq:Klein-Gordon-eq-second-gauge-inv-explicit}
\end{eqnarray}
where we defined
\begin{eqnarray}
  \Xi_{(K)}
  &:=&
     8 \partial_{\eta}\stackrel{(1)}{\Phi} \partial_{\eta}\varphi_{1}
  +  8 \stackrel{(1)}{\Phi} \Delta\varphi_{1}
  -  4 a^{2} \stackrel{(1)}{\Phi} \varphi_{1} \frac{\partial^{2}V}{\partial\bar{\varphi}^{2}}(\varphi)
  -    a^{2} (\varphi_{1})^{2}\frac{\partial^{3}V}{\partial\bar{\varphi}^{3}}(\varphi)
  +  8 \stackrel{(1)}{\Phi} \partial_{\eta}\stackrel{(1)}{\Phi} \partial_{\eta}\varphi
  \nonumber\\
  &&
  -  2 \stackrel{(1)}{\chi^{ij}} D_{j}D_{i}\varphi_{1}
  +   \partial_{\eta}\varphi \stackrel{(1)}{\chi^{ij}} \partial_{\eta}\stackrel{(1)}{\chi_{ij}}
  .
  \label{eq:Klein-Gordon-eq-second-gauge-inv-explicit-source}
\end{eqnarray}


In Eq.~(\ref{eq:Klein-Gordon-eq-second-gauge-inv-explicit}),
$\Xi_{(K)}$ is the source term which is the collection of the
quadratic terms of the linear-order perturbations in the second-order
perturbation of the Klein-Gordon equation. If we ignore this source
term, Eq.~(\ref{eq:Klein-Gordon-eq-second-gauge-inv-explicit})
coincide with the first-order perturbation of the Klein-Gordon
equation.
From this source term
(\ref{eq:Klein-Gordon-eq-second-gauge-inv-explicit-source}) of the
Klein-Gordon equation, we can see that the mode-mode coupling due to
the non-linear effects appear in the second-order Klein-Gordon
equation.


We cannot discuss solutions to
Eq.~(\ref{eq:Klein-Gordon-eq-second-gauge-inv-explicit}) only through
this equation, since this equation includes metric perturbations.
To determine the behavior of the metric perturbations, we have to
treat the Einstein equations simultaneously.
The second-order Einstein equation is shown in
Sec.~\ref{sec:Secnd-order-cosmological-Einstein-equations}.


\subsection{Einstein equations}
\label{sec:Secnd-order-cosmological-Einstein-equations}


Here, we show the all components of the second-order Einstein equation
(\ref{eq:second-order-Einstein-equation}).
All components of Eq.~(\ref{eq:second-order-Einstein-equation}) are
summarized as
\begin{eqnarray}
  &&
     \left(
     - 3 {\cal H} \partial_{\eta}
     +   \Delta
     + 3 K
     \right) \stackrel{(2)}{\Psi}
     +
     \left(
     -   \partial_{\eta}{\cal H}
     - 2 {\cal H}^{2}
     +   K
     \right)
     \stackrel{(2)}{\Phi}
     - 4 \pi G
     \left(
     \partial_{\eta}\varphi \partial_{\eta}\varphi_{2}
     +   a^{2} \varphi_{2} \frac{\partial V}{\partial\varphi}
     \right)
     =
     \Gamma_{0}
     ,
     \label{eq:kouchan-18.218}
     \\
  &&
     2 \partial_{\eta} D_{i} \stackrel{(2)}{\Psi}
     + 2 {\cal H} D_{i} \stackrel{(2)}{\Phi}
     - \frac{1}{2} \left(
     \Delta
     + 2 K
     \right)
     \stackrel{(2)}{\nu_{i}}
     -
     8\pi G D_{i}\varphi_{2} \partial_{\eta}\varphi
     =
     \Gamma_{i}
     ,
     \label{eq:kouchan-18.199}
  \\
  &&
     D_{i} D_{j} \left( \stackrel{(2)}{\Psi} - \stackrel{(2)}{\Phi} \right)
     +
     \left\{
     \left(
     -   \Delta
     + 2 \partial_{\eta}^{2}
     + 4 {\cal H} \partial_{\eta}
     - 2 K
     \right)
     \stackrel{(2)}{\Psi}
     + \left(
        2 {\cal H} \partial_{\eta}
     + 2 \partial_{\eta}{\cal H}
     + 4 {\cal H}^{2}
     + \Delta
     + 2 K
     \right)
     \stackrel{(2)}{\Phi}
     \right\}
     \gamma_{ij}
     \nonumber\\
  &&
  - \frac{1}{a^{2}} \partial_{\eta} \left(
    a^{2} D_{(i} \stackrel{(2)}{\nu_{j)}}
  \right)
  + \frac{1}{2} \left(
    \partial_{\eta}^{2}
    + 2 {\cal H} \partial_{\eta}
    + 2 K
    - \Delta
  \right) \stackrel{(2)}{\chi}_{ij}
  - 8 \pi G \left(
    \partial_{\eta}\varphi\partial_{\eta}\varphi_{2}
    - a^{2} \varphi_{2}\frac{\partial V}{\partial\varphi}(\varphi)
  \right) \gamma_{ij} = \Gamma_{ij}
  \label{eq:kouchan-18.207}
  ,
\end{eqnarray}
where $\Gamma_{0}$, $\Gamma_{i}$ $\Gamma_{ij}$ are the collection of
the quadratic term of the first-order perturbations as follows:
\begin{eqnarray}
  \Gamma_{0}
  &:=&
       4 \pi G \left(
       (\partial_{\eta}\varphi_{1})^{2}
       +   D_{i}\varphi_{1} D^{i}\varphi_{1}
       +   a^{2} (\varphi_{1})^{2} \frac{\partial^{2}V}{\partial\varphi^{2}}
       \right)
       -          4  \partial_{\eta}{\cal H} \left(\stackrel{(1)}{\Phi}\right)^{2}
       -          2  \stackrel{(1)}{\Phi} \partial_{\eta}^{2}\stackrel{(1)}{\Phi}
       -          3  D_{k}\stackrel{(1)}{\Phi} D^{k}\stackrel{(1)}{\Phi}
       -         10  \stackrel{(1)}{\Phi} \Delta\stackrel{(1)}{\Phi}
       \nonumber\\
  &&
       -          3  \left(\partial_{\eta}\stackrel{(1)}{\Phi}\right)^{2}
       -         16  K \left(\stackrel{(1)}{\Phi}\right)^{2}
       -          8  {\cal H}^{2} \left(\stackrel{(1)}{\Phi}\right)^{2}
       +             D_{l}D_{k}\stackrel{(1)}{\Phi} \stackrel{(1)}{\chi^{lk}}
       + \frac{1}{8} \partial_{\eta}\stackrel{(1)}{\chi_{lk}} \partial_{\eta}\stackrel{(1)}{\chi^{kl}}
       +             {\cal H} \stackrel{(1)}{\chi_{kl}} \partial_{\eta}\stackrel{(1)}{\chi^{lk}}
       \nonumber\\
  &&
       - \frac{3}{8} D_{k}\stackrel{(1)}{\chi_{lm}} D^{k}\stackrel{(1)}{\chi^{ml}}
       + \frac{1}{4} D_{k}\stackrel{(1)}{\chi_{lm}} D^{l}\stackrel{(1)}{\chi^{mk}}
       - \frac{1}{2} \stackrel{(1)}{\chi^{lm}} \Delta\stackrel{(1)}{\chi_{lm}}
       + \frac{1}{2} K \stackrel{(1)}{\chi_{lm}} \stackrel{(1)}{\chi^{lm}}
       ;
  \label{eq:kouchan-19.337}
  \\
  \Gamma_{i}
  &:=&
       16  \pi G \partial_{\eta}\varphi_{1} D_{i}\varphi_{1}
       -          4  \partial_{\eta}\stackrel{(1)}{\Phi} D_{i}\stackrel{(1)}{\Phi}
       +          8  {\cal H} \stackrel{(1)}{\Phi} D_{i}\stackrel{(1)}{\Phi}
       -          8  \stackrel{(1)}{\Phi} \partial_{\eta}D_{i}\stackrel{(1)}{\Phi}
       +          2  D^{j}\stackrel{(1)}{\Phi} \partial_{\eta}\stackrel{(1)}{\chi_{ji}}
       -          2  \partial_{\eta}D^{j}\stackrel{(1)}{\Phi} \stackrel{(1)}{\chi_{ij}}
       \nonumber\\
  &&
     - \frac{1}{2} \partial_{\eta}\stackrel{(1)}{\chi_{jk}} D_{i}\stackrel{(1)}{\chi^{kj}}
     -             \stackrel{(1)}{\chi_{kl}} \partial_{\eta}D_{i}\stackrel{(1)}{\chi^{lk}}
     +             \stackrel{(1)}{\chi^{kl}} \partial_{\eta}D_{k}\stackrel{(1)}{\chi_{il}}
     \label{eq:kouchan-19.338}
     ;
  \\
  \Gamma_{ij}
  &:=&
       16 \pi G D_{i}\varphi_{1} D_{j}\varphi_{1}
       +  8 \pi G \left\{
       (\partial_{\eta}\varphi_{1})^{2}
       - D_{l}\varphi_{1} D^{l}\varphi_{1}
       - a^{2} (\varphi_{1})^{2} \frac{\partial^{2}V}{\partial\varphi^{2}}
       \right\} \gamma_{ij}
       -          4  D_{i}\stackrel{(1)}{\Phi} D_{j}\stackrel{(1)}{\Phi}
       -          8  \stackrel{(1)}{\Phi} D_{i}D_{j}\stackrel{(1)}{\Phi}
       \nonumber\\
  &&
     + \left(
     6  D_{k}\stackrel{(1)}{\Phi} D^{k}\stackrel{(1)}{\Phi}
     +          4  \stackrel{(1)}{\Phi} \Delta\stackrel{(1)}{\Phi}
     +          2  \left(\partial_{\eta}\stackrel{(1)}{\Phi}\right)^{2}
     +          8  \partial_{\eta}{\cal H} \left(\stackrel{(1)}{\Phi}\right)^{2}
     +         16  {\cal H}^{2} \left(\stackrel{(1)}{\Phi}\right)^{2}
     +         16  {\cal H} \stackrel{(1)}{\Phi} \partial_{\eta}\stackrel{(1)}{\Phi}
     -          4  \stackrel{(1)}{\Phi} \partial_{\eta}^{2}\stackrel{(1)}{\Phi}
     \right) \gamma_{ij}
  \nonumber\\
  &&
  -          4  {\cal H} \partial_{\eta}\stackrel{(1)}{\Phi} \stackrel{(1)}{\chi_{ij}}
  -          2  \partial_{\eta}^{2}\stackrel{(1)}{\Phi} \stackrel{(1)}{\chi_{ij}}
  -          4  D^{k}\stackrel{(1)}{\Phi} D_{(i}\stackrel{(1)}{\chi_{j)k}}
  +          4  D^{k}\stackrel{(1)}{\Phi} D_{k}\stackrel{(1)}{\chi_{ij}}
  -          8  K \stackrel{(1)}{\Phi} \stackrel{(1)}{\chi_{ij}}
  +          4  \stackrel{(1)}{\Phi} \Delta\stackrel{(1)}{\chi_{ij}}
  -          4  D^{k}D_{(i}\stackrel{(1)}{\Phi} \stackrel{(1)}{\chi_{j)k}}
  \nonumber\\
  &&
  +          2  \Delta \stackrel{(1)}{\Phi} \stackrel{(1)}{\chi_{ij}}
  +          2  D_{l}D_{k}\stackrel{(1)}{\Phi} \stackrel{(1)}{\chi^{lk}} \gamma_{ij}
  +             \partial_{\eta}\stackrel{(1)}{\chi_{ik}} \partial_{\eta}\stackrel{(1)}{\chi_{j}^{\;\;k}}
  -             D^{k}\stackrel{(1)}{\chi_{il}} D_{k}\stackrel{(1)}{\chi_{j}^{\;\;l}}
  +             D^{k}\stackrel{(1)}{\chi_{il}} D^{l}\stackrel{(1)}{\chi_{jk}}
  - \frac{1}{2} D_{i}\stackrel{(1)}{\chi^{lk}} D_{j}\stackrel{(1)}{\chi_{lk}}
  \nonumber\\
  &&
  -             \stackrel{(1)}{\chi_{lm}} D_{i}D_{j}\stackrel{(1)}{\chi^{ml}}
  +          2  \stackrel{(1)}{\chi^{lm}} D_{l}D_{(i}\stackrel{(1)}{\chi_{j)m}}
  -             \stackrel{(1)}{\chi^{lm}} D_{m}D_{l}\stackrel{(1)}{\chi_{ij}}
  \nonumber\\
  &&
  - \frac{1}{4} \left(
      3 \partial_{\eta}\stackrel{(1)}{\chi_{lk}} \partial_{\eta}\stackrel{(1)}{\chi^{kl}}
    - 3 D_{k}\stackrel{(1)}{\chi_{lm}} D^{k}\stackrel{(1)}{\chi^{ml}}
    + 2 D_{k}\stackrel{(1)}{\chi_{lm}} D^{l}\stackrel{(1)}{\chi^{mk}}
    - 4 K \stackrel{(1)}{\chi_{lm}} \stackrel{(1)}{\chi^{lm}}
  \right) \gamma_{ij}
  \label{eq:kouchan-19.339}
  .
\end{eqnarray}
Here, we used Eqs.~(\ref{eq:background-Einstein-equations-scalar-3}),
(\ref{eq:absence-of-anisotropic-stress-Einstein-i-j-traceless-scalar}),
(\ref{eq:kouchan-18.186}),
(\ref{eq:no-first-order-vector-mode-scalar-field-case}) and
(\ref{eq:scalar-linearized-Einstein-scalar-master-eq-pre}).


The tensor part of Eq.~(\ref{eq:kouchan-18.207}) is given by
\begin{eqnarray}
  \left(
    \partial_{\eta}^{2} + 2 {\cal H} \partial_{\eta} + 2 K  - \Delta
  \right)
  \stackrel{(2)\;\;\;\;}{\chi_{ij}}
  &=&
  2 \Gamma_{ij}
  - \frac{2}{3} \gamma_{ij} \Gamma_{k}^{\;\;k}
  - 3
  \left(
    D_{i}D_{j} - \frac{1}{3} \gamma_{ij} \Delta
  \right)
  \left( \Delta + 3 K \right)^{-1}
  \left(
    \Delta^{-1} D^{k}D_{l}\Gamma_{k}^{\;\;l}
    - \frac{1}{3} \Gamma_{k}^{\;\;k}
  \right)
  \nonumber\\
  &&
  + 4
  \left\{
      D_{(i} (\Delta+2K)^{-1} D_{j)}\Delta^{-1}D^{l}D_{k}\Gamma_{l}^{\;\;k}
    - D_{(i}(\Delta+2K)^{-1}D^{k}\Gamma_{j)k}
  \right\}
  .
  \label{eq:kouchan-18.215}
\end{eqnarray}
This tensor mode is also called the second-order gravitational waves.


Further, the vector part of Eqs.~(\ref{eq:kouchan-18.199}) and
(\ref{eq:kouchan-18.207}) yields the initial value constraint and the
evolution equation of the vector mode $\stackrel{(2)\;\;}{\nu_{j}}$:
\begin{eqnarray}
  &&
  \stackrel{(2)}{\nu_{i}}
  =
  \frac{2}{\Delta + 2 K}
  \left\{
    D_{i} \Delta^{-1} D^{k} \Gamma_{k}
    - \Gamma_{i}
  \right\}
  ,
  \quad
  \partial_{\eta}
  \left(
    a^{2} \stackrel{(2)}{\nu_{i}}
  \right)
  =
  \frac{2 a^{2}}{\Delta + 2 K}
  \left\{
    D_{i}\Delta^{-1} D^{k}D_{l}\Gamma_{k}^{\;\;l}
    - D_{k}\Gamma_{i}^{\;\;k}
  \right\}
  .
  \label{eq:kouchan-18.214}
\end{eqnarray}


Finally, scalar part of
Eqs.~(\ref{eq:kouchan-18.218})--(\ref{eq:kouchan-18.207}) are
summarized as
\begin{eqnarray}
  &&
     2 \partial_{\eta} \stackrel{(2)}{\Psi}
     + 2 {\cal H} \stackrel{(2)}{\Phi}
     -
     8\pi G \varphi_{2} \partial_{\eta}\varphi
     =
     \Delta^{-1} D^{k} \Gamma_{k}
     ,
  \label{eq:kouchan-18.199-2}
  , \\
  &&
  \stackrel{(2)}{\Psi} - \stackrel{(2)}{\Phi}
  =
  \frac{3}{2}
  (\Delta + 3 K)^{-1}
  \left\{
    \Delta^{-1} D^{i}D_{j}\Gamma_{i}^{\;\;j} - \frac{1}{3} \Gamma_{k}^{\;\;k}
  \right\}
  \label{eq:kouchan-18.213}
  , \\
  &&
  \left(
    -   \partial_{\eta}^{2}
    - 5 {\cal H} \partial_{\eta}
    + \frac{4}{3} \Delta
    + 4 K
  \right) \stackrel{(2)}{\Psi}
  -
  \left(
      2 \partial_{\eta}{\cal H}
    +   {\cal H} \partial_{\eta}
    + 4 {\cal H}^{2}
    +   \frac{1}{3} \Delta
  \right)
  \stackrel{(2)}{\Phi}
  - 8 \pi G a^{2} \varphi_{2} \frac{\partial V}{\partial\varphi}
  =
  \Gamma_{0} - \frac{1}{6} \Gamma_{k}^{\;\;k}
  ,
  \label{eq:kouchan-18.228-2}
  \\
  &&
  \left\{
    \partial_{\eta}^{2}
    + 2 \left(
      {\cal H}
      - \frac{\partial_{\eta}^{2}\varphi}{\partial_{\eta}\varphi}
    \right)
    \partial_{\eta}
    -             \Delta
    -          4  K
    + 2 \left(
      \partial_{\eta}{\cal H}
      - \frac{\partial_{\eta}^{2}\varphi}{\partial_{\eta}\varphi} {\cal H}
    \right)
  \right\}
  \stackrel{(2)}{\Phi}
     \nonumber\\
  &=&
  - \Gamma_{0}
  - \frac{1}{2} \Gamma_{k}^{\;\;k}
  +
  \Delta^{-1} D^{i}D_{j}\Gamma_{i}^{\;\;j}
      \nonumber\\
  &&
  + \left(
    \partial_{\eta}
    - \frac{\partial_{\eta}^{2}\varphi}{\partial_{\eta}\varphi}
  \right)
  \Delta^{-1}D^{k}\Gamma_{k}
  -
  \frac{3}{2}
  \left\{
    \partial_{\eta}^{2}
    - \left(
      \frac{2\partial_{\eta}^{2}\varphi}{\partial_{\eta}\varphi} - {\cal H}
    \right)
    \partial_{\eta}
  \right\}
  (\Delta + 3 K)^{-1}
  \left\{
    \Delta^{-1} D^{i}D_{j}\Gamma_{i}^{\;\;j} - \frac{1}{3} \Gamma_{k}^{\;\;k}
  \right\}.
  \label{eq:kouchan-18.233}
\end{eqnarray}
where $\Gamma_{i}^{\;\;j} := \gamma^{kj}\Gamma_{ik}$ and
$\Gamma_{k}^{\;\;k} = \gamma^{ij}\Gamma_{ij}$.
Eq.~(\ref{eq:kouchan-18.233}) is the second-order extension of
Eq.~(\ref{eq:scalar-linearized-Einstein-scalar-master-eq}), which is
the master equation of scalar mode of the second-order cosmological
perturbation in a universe filled with a single scalar field.


Thus, we have a set of ten equations for the second-order
perturbations of a universe filled with a single scalar field,
Eqs.~(\ref{eq:kouchan-18.215})--(\ref{eq:kouchan-18.233}).
To solve this system of equations of the second-order Einstein
equation, first of all, we have to solve the linear-order system.
This is accomplished by solving
Eq.~(\ref{eq:scalar-linearized-Einstein-scalar-master-eq}) to
obtain the potential $\stackrel{(1)}{\Phi}$, $\varphi_{1}$ is
given through (\ref{eq:kouchan-18.186}), and the tensor mode
$\stackrel{(1)}{\chi}_{ij}$ is given by solving
Eq.~(\ref{eq:linearized-Einstein-i-j-traceless-tensor}).
Next, we evaluate the quadratic terms, $\Gamma_{0}$, $\Gamma_{i}$ and
$\Gamma_{ij}$ of the linear-order perturbations, which are defined by Eqs.~(\ref{eq:kouchan-19.337})--(\ref{eq:kouchan-19.339}).
Then, using the information of
Eqs.~(\ref{eq:kouchan-19.337})--(\ref{eq:kouchan-19.339}), we estimate
the source term in Eq.~(\ref{eq:kouchan-18.233}).
If we know the two independent solutions to the linear-order master
equation (\ref{eq:scalar-linearized-Einstein-scalar-master-eq}), we
can solve Eq.~(\ref{eq:kouchan-18.233}) through the method using the
Green functions.
After constructing the solution $\stackrel{(2)}{\Phi}$ to
Eq.~(\ref{eq:kouchan-18.233}), we can obtain the second-order metric
perturbation $\stackrel{(2)}{\Psi}$ through
Eq.~(\ref{eq:kouchan-18.213}).
Then, we have obtained the second-order gauge-invariant perturbation
$\varphi_{2}$ of the scalar field through
Eq.~(\ref{eq:kouchan-18.199-2}).
Thus, the all scalar modes $\stackrel{(2)}{\Phi}$,
$\stackrel{(2)}{\Psi}$, $\varphi_{2}$ are obtained.
Equation (\ref{eq:kouchan-18.228-2}) is then used to check the
consistency of the second-order perturbation of the Klein Gordon
equation (\ref{eq:Klein-Gordon-eq-second-gauge-inv-explicit}) as in
Sec.~\ref{sec:Consistency-of-the-second-order-perturbations}.


For the vector-mode, $\stackrel{(1)\;\;}{\nu_{i}}$ of the first-order
identically vanishes due to the momentum constraint
(\ref{eq:no-first-order-vector-mode-scalar-field-case}) for the
linear-order metric perturbations.
On the other hand, in the second-order, we have evolution equation
(\ref{eq:kouchan-18.214}) of the vector mode
$\stackrel{(2)\;\;}{\nu_{i}}$ with the initial value constraint.
This evolution equation of the second-order vector mode should be
consistent with the initial value constraint, which is confirmed in
Sec.~\ref{sec:Consistency-of-the-second-order-perturbations}.
Equations (\ref{eq:kouchan-18.214}) also imply that the second-order
vector-mode perturbation may be generated by the mode couplings of the
linear order perturbations.
As the simple situations, the generation of the second-order vector
mode due to the scalar-scalar mode coupling is discussed in
Refs.~\cite{Mena:2007ve,Lu:2007cj,Lu:2008ju,Christopherson:2009bt}.


The second-order tensor mode is also generated by the mode-coupling of
the linear-order perturbations through the source term in
Eq.~(\ref{eq:kouchan-18.215}).
Note that Eq.~(\ref{eq:kouchan-18.215}) is almost same as
Eq.~(\ref{eq:linearized-Einstein-i-j-traceless-tensor}) for the
linear-order tensor mode, except for the existence of the source term
in Eq.~(\ref{eq:kouchan-18.215}).
If we know the solution to the linear-order Einstein equations
(\ref{eq:linearized-Einstein-i-j-traceless-tensor}) and
(\ref{eq:scalar-linearized-Einstein-scalar-master-eq}), we can
evaluate the source term in Eq.~(\ref{eq:kouchan-18.215}).
Further, we can solve Eq.~(\ref{eq:kouchan-18.215}) through the Green
function method.
This leads the generation of the gravitational wave of the second
order.
Actually, in the simple situation where the first-order tensor mode
neglected, the generation of the second-order gravitational waves
discussed in some
literature\cite{S.Mollerach-D.Harari-S.Matarrese-2004,Ananda:2006af,Osano:2006ew,Baumann:2007zm,Bartolo:2007vp,Martineau:2007dj,Saito:2008jc,Arroja:2009sh,H.Assadullahi-D.Wands-2009,H.Assadullahi-D.Wands-2010,K.Jedamzik-M.Lemoine-J.Martin-2010,L.Alabidi-K.Kohri-M.Sasaki-Y.Sendouda-2013,S.Saga-K.Ichiki-N.Sugiyama-2015,R.G.Cai-S.Pi-M.Sasaki-2019}.


\subsection{Consistency of equations for second-order perturbations}
\label{sec:Consistency-of-the-second-order-perturbations}


Now, we consider the consistency of the second-order perturbations of
the Einstein equations
(\ref{eq:kouchan-18.199-2})--(\ref{eq:kouchan-18.233}) for the scalar
modes, Eqs.~(\ref{eq:kouchan-18.214}) for vector mode, and the
Klein-Gordon equation
(\ref{eq:Klein-Gordon-eq-second-gauge-inv-explicit}).
The consistency check of these equations are necessary to guarantee
that the derived equations are correct, since the second-order
Einstein equations have complicated forms owing to the quadratic terms
of the linear-order perturbations that arise from the nonlinear
effects of the Einstein equations.


Since the first equation in Eqs.~(\ref{eq:kouchan-18.214}) is the
initial value constraint for the vector mode $\stackrel{(2)}{\nu_{i}}$
and it should be consistent with the evolution equation, i.e., the
second equation of Eqs.~(\ref{eq:kouchan-18.214}). these equations
should be consistent with each other from the general arguments of the
Einstein equation.
Explicitly, these equations are consistent with each other if the equation
\begin{eqnarray}
  \partial_{\eta}\Gamma_{k}
  + 2 {\cal H} \Gamma_{k}
  - D^{l}\Gamma_{lk} = 0
  \label{eq:kouchan-19.358}
\end{eqnarray}
is satisfied.
Actually, through the first-order perturbative Einstein equations
(\ref{eq:kouchan-18.186}),
(\ref{eq:scalar-linearized-Einstein-scalar-master-eq}),
(\ref{eq:linearized-Einstein-i-j-traceless-tensor}), we can confirm
the equation (\ref{eq:kouchan-19.358}).
This is a trivial result from a general viewpoint, because the
Einstein equation is the first class constrained system.
However, this trivial result implies that we have derived the source
terms $\Gamma_{i}$ and $\Gamma_{ij}$ of the second-order Einstein
equations consistently.


Next, we consider Eq.~(\ref{eq:kouchan-18.228-2}).
Through the second-order Einstein equations
(\ref{eq:kouchan-18.199-2}), (\ref{eq:kouchan-18.213}),
(\ref{eq:kouchan-18.233}), and the background Klein-Gordon equation
(\ref{eq:background-Klein-Gordon-equation}),
we can confirm that Eq.~(\ref{eq:kouchan-18.228-2}) is consistent with
the set of the background, first-order and other second-order Einstein
equation if the equation
\begin{eqnarray}
  \left(
                  \partial_{\eta}
    +          2  {\cal H}
  \right) D^{k}\Gamma_{k}
  - D^{j}D^{i}\Gamma_{ij}
  =
  0
  \label{eq:kouchan-19.345}
\end{eqnarray}
is satisfied under the background and first-order Einstein equations.
Actually, we have already seen that Eq.~(\ref{eq:kouchan-19.358}) is
satisfied under the background and first-order Einstein equations.
Taking the divergence of Eq.~(\ref{eq:kouchan-19.358}), we can
immediately confirm Eq.~(\ref{eq:kouchan-19.345}).
Then, Eq.~(\ref{eq:kouchan-18.228-2}) gives no information.


Thus, we have seen that the derived Einstein equations of the second
order (\ref{eq:kouchan-18.214})--(\ref{eq:kouchan-18.233}) are
consistent with each other through Eq.~(\ref{eq:kouchan-19.358}).
This fact implies that the derived source terms $\Gamma_{i}$ and
$\Gamma_{ij}$ of the second-order perturbations of the Einstein
equations, which are defined by Eqs.~(\ref{eq:kouchan-19.338}) and
(\ref{eq:kouchan-19.339}), are correct source terms of the
second-order Einstein equations.
On the other hand, for $\Gamma_{0}$, we have to consider the
consistency between the perturbative Einstein equations and the
perturbative Klein-Gordon equation as seen below.


Now, we consider the consistency of the second-order perturbation of
the Klein-Gordon equation and the Einstein equations.
The second-order perturbation of the Klein-Gordon equation is given by
Eq.~(\ref{eq:Klein-Gordon-eq-second-gauge-inv-explicit}) with the
source term
(\ref{eq:Klein-Gordon-eq-second-gauge-inv-explicit-source}).
Since the vector mode $\stackrel{(2)}{\nu_{i}}$ and tensor mode
$\stackrel{(2)}{\chi}_{ij}$ of the second-order do not appear in the
expressions (\ref{eq:Klein-Gordon-eq-second-gauge-inv-explicit}) of
the second-order perturbation of the Klein-Gordon equation, we may
concentrate on the Einstein equations for scalar mode of the second
order, i.e., Eqs.~(\ref{eq:kouchan-18.199-2}),
(\ref{eq:kouchan-18.213}), and (\ref{eq:kouchan-18.233}) with the
definitions (\ref{eq:kouchan-19.337})--(\ref{eq:kouchan-19.339}) of
the source terms.
As in the linear case, the second-order perturbation of the
Klein-Gordon equation should also be derived from the set of equations
consisting of the second-order perturbations of the Einstein equations
(\ref{eq:kouchan-18.199-2}), (\ref{eq:kouchan-18.213}),
(\ref{eq:kouchan-18.233}), the first-order perturbations of the
Einstein equations
(\ref{eq:absence-of-anisotropic-stress-Einstein-i-j-traceless-scalar}),
(\ref{eq:kouchan-18.186}),
(\ref{eq:scalar-linearized-Einstein-scalar-master-eq}), and the
background Einstein equations
(\ref{eq:background-Einstein-equations-scalar-1}) and
(\ref{eq:background-Einstein-equations-scalar-2}).
Actually, from these equation, we can show that the second-order
perturbation of the Klein-Gordon equation is consistent with the
background and the second-order Einstein equations if the equation
\begin{eqnarray}
  2 \left(
    \partial_{\eta} + {\cal H}
  \right) \Gamma_{0}
  -    D^{k}\Gamma_{k}
  +    {\cal H} \Gamma_{k}^{\;\;k}
  + 8 \pi G \partial_{\eta}\varphi \Xi_{(K)}
  = 0
  \label{eq:kouchan-19.374}
\end{eqnarray}
is satisfied under the background and the first-order Einstein
equations.
Further, we can also confirm Eq.~(\ref{eq:kouchan-19.374}) through the
background Einstein equations
(\ref{eq:background-Einstein-equations-scalar-1}) and
(\ref{eq:background-Einstein-equations-scalar-2}), the scalar part of
the first-order perturbation of the momentum constraint
(\ref{eq:kouchan-18.186}), the evolution equations
(\ref{eq:scalar-linearized-Einstein-scalar-master-eq}) and
(\ref{eq:linearized-Einstein-i-j-traceless-tensor}) for the first
order scalar and tensor modes in the Einstein equation.


As shown in Ref.~\cite{kouchan-second-cosmo-consistency}, the
first-order perturbation of the Klein-Gordon equation is derived from
the background and the first-order perturbations of the Einstein
equation.
In the case of the second-order perturbation, the Klein-Gordon
equation (\ref{eq:Klein-Gordon-eq-second-gauge-inv-explicit}) can be
also derived from the background, the first-order, and the
second-order Einstein equations.
The second-order perturbations of the Einstein equation and the
Klein-Gordon equation include the source terms $\Gamma_{0}$,
$\Gamma_{i}$, $\Gamma_{ij}$, and $\Xi_{(K)}$ due to the mode-coupling
of the linear-order perturbations.
The second-order perturbation of the Klein-Gordon equation gives the
relation (\ref{eq:kouchan-19.374}) between the source terms
$\Gamma_{0}$, $\Gamma_{i}$, $\Gamma_{ij}$, $\Xi_{(K)}$ and we have
also confirmed that Eq.~(\ref{eq:kouchan-19.374}) is satisfied due to
the background, the first-order perturbation of the Einstein
equations, and the Klein-Gordon equation.
Thus, the second-order perturbation of the Klein-Gordon equation is
not independent of the set of the background, the first-order, and the
second-order Einstein equations if we impose on the Einstein equation
at any conformal time $\eta$.
This also implies that the derived formulae of the source terms
$\Gamma_{0}$, $\Gamma_{i}$, $\Gamma_{ij}$, and $\Xi_{(K)}$ are
consistent with each other.
In this sense, we may say that the formulae
(\ref{eq:kouchan-19.337})--(\ref{eq:kouchan-19.339}) and (\ref{eq:Klein-Gordon-eq-second-gauge-inv-explicit-source}) for
these source terms are correct.


\section{Summary and discussions}
\label{sec:summary}


In this review, we summarized the current status of our formulation of
the gauge-invariant second-order cosmological perturbations.
Although the presentation in this article is restricted to the case of
the universe filled with a single scalar field, the essence of our
general framework of the gauge-invariant perturbation theory is
transparent through this simple case.
Our general framework of the general relativistic higher-order
gauge-invariant perturbation theory can be separated into three
parts.
First one is the general formulation to derive the
gauge-transformation rules (\ref{eq:Bruni-47-one}) and
(\ref{eq:Bruni-49-one}).
Second one is the construction of the gauge-invariant variables for
the perturbations on the generic background spacetime inspecting
gauge-transformation rules (\ref{eq:Bruni-47-one}) and
(\ref{eq:Bruni-49-one}) and the decomposition formula
(\ref{eq:matter-gauge-inv-decomp-1.0}) and
(\ref{eq:matter-gauge-inv-decomp-2.0}) for perturbations of any tensor
field.
Third one is the application of the above general framework of the
gauge-invariant perturbation theory to the cosmological situations.


To derive the gauge-transformation rules (\ref{eq:Bruni-47-one}) and
(\ref{eq:Bruni-49-one}), we considered the general arguments on the
Taylor expansion of an arbitrary tensor field on a manifold, the
general class of the diffeomorphism which is wider than the well-known
exponential map, and the general formulation of the perturbation
theory.
This general class of diffeomorphism is represented in terms of the
Taylor expansion (\ref{eq:Taylor-expansion-of-f}) of its pull-back.
The generality of the representation of the Taylor expansion
(\ref{eq:Taylor-expansion-of-f}) can be seen in its derivation shown
in Appendix~\ref{sec:derivation-of-Taylor-expansion}.
We note that the derivation in shown in
Appendix~\ref{sec:derivation-of-Taylor-expansion} does not require any
information of the connection, the metric, nor the special coordinate
systems on the manifold.
Therefore, the formula for the Taylor expansion
(\ref{eq:Taylor-expansion-of-f}) is quite general.


As commented in
Sec.~\ref{sec:Taylor-expansion-of-tensors-on-a-manifold}, this general
class of diffeomorphism does not form a one-parameter group of
diffeomorphism as shown through
Eq.~(\ref{eq:Phi-is-not-one-parameter-group-of-diffeomorphism}).
However, the properties
(\ref{eq:Phi-is-not-one-parameter-group-of-diffeomorphism}) do not
directly mean that this general class of diffeomorphism does not form
a group, as emphasized in
Sec.~\ref{sec:Taylor-expansion-of-tensors-on-a-manifold}.
One of the key points of the properties of this diffeomorphism is the
non-commutativity of generators $\xi_{1}^{a}$ and $\xi_{2}^{a}$ of
each order.
The expression of the $n$-th order Taylor expansion of the pull-back
of this general class is discussed in
Ref.~\cite{M.Bruni-S.Sonego-CQG1999}.
When we consider the situation of the $n$-th order perturbation, this
non-commutativity becomes important\cite{kouchan-gauge-inv}.
Therefore, to clarify the properties of this general class of
diffeomorphism, we have to take care of this non-commutativity of
generators.
Thus, there is a room to clarify the properties of this general class
of diffeomorphism.


Further, in Sec.~\ref{sec:Formulation-of-perturbation-theory}, we
introduced a gauge choice ${\cal X}_{\lambda}$ as an exponential map,
for simplicity.
On the other hand, we have the concept of the general class of
diffeomorphism which is wider than the class of the exponential map.
Therefore, we may introduce a gauge choice as one of the element of
this general class of diffeomorphism.
However, the gauge-transformation rules (\ref{eq:Bruni-47-one}) and
(\ref{eq:Bruni-49-one}) will not be changed even if we generalize the
definition of a each gauge choice as emphasized in
Sec.~\ref{sec:Formulation-of-perturbation-theory}.
Although there is a room to sophisticate in logical arguments to
derive the gauge-transformation rules (\ref{eq:Bruni-47-one}) and
(\ref{eq:Bruni-49-one}), these are harmless to the development of the
general framework of the gauge-invariant perturbation theory shown in
Secs.~\ref{sec:Formulation-of-perturbation-theory},
\ref{sec:gauge-invariant-variables},
\ref{sec:Perturbation-of-the-field-equations}, and their application
to cosmological perturbations in
Sec.~\ref{sec:Cosmological-Background-spacetime-equations}.


On the other hand, as emphasize in
Sec.~\ref{sec:gauge-invariant-variables}, our starting point to
construct gauge invariant variables is
Conjecture~\ref{conjecture:decomp_conjecture_for_hab} in
Sec.~\ref{sec:gauge-invariant-variables}.
Conjecture~\ref{conjecture:decomp_conjecture_for_hab} on a generic
background spacetime is highly nontrivial.
The procedure to accomplish the decomposition
(\ref{eq:linear-metric-decomp}) completely depends on the details of
the background spacetime.
Although we propose a scenario of the proof of
Conjecture~\ref{conjecture:decomp_conjecture_for_hab} in
Refs.~\cite{K.Nakamura-CQG-Letter-2011,K.Nakamura-Progress-Construction-2013},
this scenario is still incomplete due to the non-local properties in
the statement of
Conjecture~\ref{conjecture:decomp_conjecture_for_hab}.
This situation is briefly explained in
Appendix~\ref{sec:Outline-of-the-proof-of-the-decomposition-conjecture}.
Even in the case of the cosmological perturbations in
Sec.~\ref{sec:Gauge-invariant-metric-perturbations}, we assume the
existence of some Green functions for the elliptic differential
operators $\Delta$, $\Delta+2K$, $\Delta+3K$, for simplicity.
This assumption on the existence of Green functions is an appearance
of the non-local nature of the statement of
Conjecture~\ref{conjecture:decomp_conjecture_for_hab} and corresponds
to ignoring the kernel modes of the elliptic differential operators
$\Delta$, $\Delta+2K$, $\Delta+3K$.
We call these kernel modes as {\it zero mode}.
To includes these kernel modes even in the case of cosmological
perturbations, separate treatments of perturbative modes are
required.
We call the problem to develop the treatments of these zero mode as
{\it zero mode problem}.
For example, homogeneous modes of perturbations are excluded in our
current arguments of the cosmological perturbation theory.
These homogeneous modes is physically important because these are
necessary to discuss the comparison with the arguments based on the
long-wavelength approximation.
On the other hand, we can also say that if we resolve this zero mode
problem, we can complete the proof of the
Conjecture~\ref{conjecture:decomp_conjecture_for_hab} at least in the
case of cosmological perturbations.
Therefore, we have to say that there is a room to clarify even in the
cosmological perturbation theory.


It is shown that the non-locality in
Conjecture~\ref{conjecture:decomp_conjecture_for_hab} appears even in
the scenario of its proof for a generic background spacetime shown in
Appendix~\ref{sec:Outline-of-the-proof-of-the-decomposition-conjecture}.
Therefore, we easily expect that {\it zero mode problem} essentially
exists in perturbations on generic background spacetime.
In this sense, we have to say that the scenario of the proof of
Conjecture~\ref{conjecture:decomp_conjecture_for_hab} in
Appendix~\ref{sec:Outline-of-the-proof-of-the-decomposition-conjecture}
and in
Refs.~\cite{K.Nakamura-CQG-Letter-2011,K.Nakamura-Progress-Construction-2013}
is still incomplete.
In spite of this incompleteness, the
Conjecture~\ref{conjecture:decomp_conjecture_for_hab} is almost correct in
some background
spacetime\cite{kouchan-paper-string-I-2000,kouchan-paper-string-II-2001,kouchan-paper-string-initial-2002,kouchan-paper-cylindrical-domain-wall-2002,kouchan-paper-string-comparison-2003}
in the sense of Sec.~\ref{sec:Gauge-invariant-metric-perturbations}.
Furthermore, once we accept
Conjecture~\ref{conjecture:decomp_conjecture_for_hab}, we can develop
the higher-order perturbation theory in an independent manner of the
details of the background spacetime.
We also expect that our general framework of the gauge-invariant
perturbation theory is extensible to an arbitrary-order perturbation
theory on an arbitrary background spacetime.
Actually, the recursive structure in the construction of gauge-invariant
variables for any order perturbations on arbitrary background
spacetime was found in Ref.~\cite{K.Nakamura-CQG-Recursive-2014} and
we can define the gauge-invariant variables on a generic background
spacetime to arbitrary order, although the
Conjecture~\ref{conjecture:decomp_conjecture_for_hab} is still
incomplete and the other algebraic conjecture (Conjecture 4.1 in
Ref.~\cite{K.Nakamura-CQG-Recursive-2014}) should be proved.
This situation indicates that the zero-mode problem for the
perturbations on a generic background spacetime, which is similar
to that of cosmological perturbations, is physically essential problem
not only of linear-order perturbations but also of non-linear
perturbations.
Rather, in higher-order perturbations, this zero-mode problem is a
serious problem and zero modes should also be included in higher-order
perturbations, because
Conjecture~\ref{conjecture:decomp_conjecture_for_hab} is used in the
construction of gauge-invariant variables for second-order
perturbations shown in Sec.~\ref{sec:gauge-invariant-variables}.
This situation is also same in the extension to any order
perturbations~\cite{K.Nakamura-CQG-Recursive-2014}.
Thus, we may say that the most important nontrivial part of our
general framework of higher-order gauge-invariant perturbation theory
is in this zero-mode problem.


Even if Conjecture~\ref{conjecture:decomp_conjecture_for_hab} is
correct on any background spacetime, the other problem exists in the
interpretations of the gauge-invariant variables.
We have commented on the non-uniqueness in the definitions of the
gauge-invariant variables through
Eqs.~(\ref{eq:non-uniqueness-of-gauge-invariant-metric-perturbation})
and
(\ref{eq:non-uniqueness-of-gauge-invariant-2nd-order-metric-perturbation-4}).
Although this non-uniqueness corresponds to the fact that there are
infinitely many ``gauge-fixing'' method, in principle, this
non-uniqueness also leads some ambiguities in the interpretations of
gauge-invariant variables.
On the other hand, as emphasize in
Sec.~\ref{sec:Formulation-of-perturbation-theory}, any observations
and experiments are carried out only on the physical spacetime through
the physical processes within the physical spacetime.
For this reason, any direct observables in any observations or
experiments should be independent of the gauge choice, i.e., gauge
invariant.
However, it is not trivial {\it which gauge-invariant variable corresponds
  to the direct observable in a specific observation or experiment.}
This non-triviality also comes from the non-uniqueness in the definitions
the gauge-invariant variables expressed by
Eqs.~(\ref{eq:non-uniqueness-of-gauge-invariant-metric-perturbation})
and
(\ref{eq:non-uniqueness-of-gauge-invariant-2nd-order-metric-perturbation-4})
that have the same form as the decomposition formulae
(\ref{eq:matter-gauge-inv-decomp-1.0}) and
(\ref{eq:matter-gauge-inv-decomp-2.0}).
If we can specify the variable which is the direct observable in an
experiment or observation, this variable should be automatically
gauge invariant.
Furthermore, non-uniqueness of gauge-invariant variables will be no longer
serious problem, since the terms that bring the non-uniqueness of
gauge-invariant variables have the same form as its gauge-variant parts in
Eqs.~(\ref{eq:matter-gauge-inv-decomp-1.0}) and
(\ref{eq:matter-gauge-inv-decomp-2.0}).
These will be confirmed by the clarification of the relations between
gauge-invariant variables and direct observables in experiments or
observations.
To accomplish this, we have to specify the concrete process of
experiments, to clarify the problem what are the direct observables in
the experiments or observations, and to derive the relations between
the gauge-invariant variables and direct observables in a specific
experiment.
If these arguments are completed, we will be able to show that the
gauge degree of freedom is just unphysical degree of freedom and the
non-uniqueness of the gauge-invariant variables is not essential to
the direct observables in the concrete observation or experiment,
simultaneously.
In addition, these considerations will give the precise physical
interpretations of the gauge-invariant variables.


This problem of the interpretation of gauge-invariant variables is
closely related to ``the gauge-dependence of second-order
gravitational waves generated by the mode-coupling of the
first-order perturbations'' which is recently pointed out by
J.~c-.~Hwang et al. in Ref.~\cite{J.c-.Hwang-D.Jeong-H.Noh-2017}.
Usually, so called $\Omega_{GW}$ is estimated the amplitude of
gravitational waves in many literature.
This $\Omega_{GW}$ is justified by the arguments on the
pseudo-energy-momentum tensor of gravitational
field in many text books (for example,
see~\cite{Wald-book,L.D.Landau-E.M.Lifshitz-1962}).
However, we have to emphasize that we are proposing a different
formulation of higher-order perturbation theories of gravity from
those in some text books (for exam. in
Ref.~\cite{Wald-book,L.D.Landau-E.M.Lifshitz-1962}).
In spite of this difference, $\Omega_{GW}$ is used in many literature.
In this sense, the appearance of gauge-dependence in
$\Omega_{GW}$ is not that surprising, because the theoretical
context is different.
From the arguments in this paper, we can simply say that the
gauge-dependence of $\Omega_{GW}$ for higher-order perturbations
indicates that $\Omega_{GW}$ is no longer direct observable in any
experiment nor any observations within our perturbation theory, though
$\Omega_{GW}$ for higher-order perturbations might be one of
indicators to estimate the amplitude of gravitational waves in some
sense.


As another example in cosmology, in case of the CMB physics, we can
easily see that the linear-order perturbation of the CMB temperature is
automatically gauge-invariant from
Eq.~(\ref{eq:matter-gauge-inv-decomp-1.0}), because the background
temperature of CMB is isotropic Planck distribution.
On the other hand, the decomposition formula
(\ref{eq:matter-gauge-inv-decomp-2.0}) yields that the theoretical prediction
of the second-order perturbation of the CMB temperature may depend on gauge
choice, since we do know the existence of the first-order fluctuations as the
temperature anisotropy in CMB.
However, as emphasized above, the direct observables in observations
should be gauge-invariant and the gauge-variant term in
Eq.~(\ref{eq:matter-gauge-inv-decomp-2.0}) should be disappear in the
direct observables.
Therefore, we have to clarify the how gauge-invariant variables are
related to the directly observed temperature fluctuations and have to
confirm the disappearance of the gauge-variant terms in the direct
observable.
This will be an important problem for our higher-order cosmological
perturbation theory.


Although there are some rooms to accomplish the complete formulation
of the second-order cosmological perturbation theory as mentioned above,
we derived all the components of the second-order perturbation of the
Einstein equation without ignoring any types modes (scalar-, vector-,
tensor-types) of perturbations in the case of a scalar field system.
In our formulation, any gauge fixing is not necessary and we can
obtain all equations in the gauge-invariant form, which are equivalent
to the complete gauge fixing.
In other words, our formulation gives complete gauge-fixed equations
without any gauge fixing.
In this sense, the equations shown here are irreducible.
This is one of the advantages of the gauge-invariant perturbation
theory.
Our second-order gauge-invariant cosmological perturbation theory
reviewed here is also extensively discussed by Uggla and
Wainwright in their series of papers~\cite{C.Uggla-J.Wainwright-2011,C.Uggla-J.Wainwright-2012,C.Uggla-J.Wainwright-2013a,C.Uggla-J.Wainwright-2013b,C.Uggla-J.Wainwright-2014a,C.Uggla-J.Wainwright-2014b,C.Uggla-J.Wainwright-2018,C.Uggla-J.Wainwright-2019a,C.Uggla-J.Wainwright-2019b,C.Uggla-J.Wainwright-2019c}.
As discussed in these papers, we may obtain more simple equations for
second-order cosmological perturbations due to the restriction of the
physical situations and the classification of the physical effects
such as ``super horizon effects'', ``Newtonian effects'', and
``post-Newtonian effects.''
Furthermore, we may also obtain more simple equations by the inclusion
of some parts of the source terms in second-order Einstein equations
to the gauge-invariant variables for second-order perturbations as in
the case of the conventional post-Newtonian expansion
theory~\cite{C.M.Will-1993}.


The explicit Einstein equations of the second order show that any type
of mode-coupling appears as the quadratic terms of the linear-order
perturbations due to the nonlinear effect of the Einstein equations,
in principle.
Perturbations in cosmological situations are classified into three
types: scalar, vector, and tensor.
In the second-order perturbations, we also have these three types of
perturbations as in the case of the first-order perturbations.
Furthermore, in the equations for the second-order perturbations,
there are many quadratic terms of linear-order perturbations due to
the nonlinear effects of the system.
Owing to these nonlinear effects, the above three types of
perturbations couple with each other.
In the scalar field system shown in this paper, the first-order vector
mode does not appear due to the momentum constraint of the first-order
perturbation of the Einstein equation.
Therefore, we have seen that three types of mode-coupling appear in
the second-order Einstein equations, i.e., scalar-scalar,
scalar-tensor, and tensor-tensor type of mode coupling.
In general, all types of mode-coupling may appear in the second-order
Einstein equations.
Actually, in Ref.~\cite{kouchan-second-cosmo-consistency}, we also
derived the all components of the Einstein equations for a perfect
fluid system and we can see all types of mode-coupling, i.e.,
scalar-scalar, scalar-vector, scalar-tensor, vector-vector,
vector-tensor, tensor-tensor types mode-coupling, appear in the
second-order Einstein equation, in general.
Of course, in the some realistic situations of cosmology, we may
neglect some modes and some mode-coupling terms.
However, even in this case, we should keep in mind the fact that all
types of mode-couplings may appear in principle when we discuss the
realistic situations of cosmology.
We cannot deny the possibility that the mode-couplings of any type
produces observable effects when the quite high accuracy of
observations is accomplished.


Even in the case of the single scalar field discussed in this paper,
the source terms of the second-order Einstein equation show the
mode-coupling of scalar-scalar, scalar-tensor, and the  tensor-tensor
types as mentioned above.
Since the tensor mode of the linear order is also generated due to
quantum fluctuations during the inflationary phase, the mode-couplings
of the scalar-tensor and tensor-tensor types may appear in the
inflation.
If these mode-couplings occur during the inflationary phase, these
effects will depend on the scalar-tensor ratio $r$.
If so, there is a possibility that the accurate observations of the
second-order effects in the fluctuations of the scalar type in our
universe also restrict the scalar-tensor ratio $r$ or give some
consistency relations between the other observations of primordial
gravitational waves such as the measurements of the B-mode of the
polarization of CMB.
This will be a new effect that gives some information on the
scalar-tensor ratio $r$.


Furthermore, we have also checked the consistency between the
second-order perturbations of the equations of motion of matter field
and the Einstein equations.
In the case of a scalar field, we checked the consistency between the
second-order perturbations of the Klein-Gordon equation and the
Einstein equations.
Due to this consistency check, we have obtained the consistency
relations between the source terms in these equations $\Gamma_{0}$,
$\Gamma_{i}$, $\Gamma_{ij}$, and $\Xi_{(K)}$, which are given by
Eqs.~(\ref{eq:kouchan-19.358}) and (\ref{eq:kouchan-19.374}).
We note that the relation (\ref{eq:kouchan-19.358}) comes from the
consistency in the Einstein equations of the second order by itself,
while the relation (\ref{eq:kouchan-19.374}) comes from the
consistency between the second-order perturbation of the Klein-Gordon
equation and the Einstein equation.
We also showed that these relations between the source terms are
satisfied through the background and the first-order perturbation of
the Einstein equations in
Ref.~\cite{kouchan-second-cosmo-consistency}.
This implies that the set of all equations are self-consistent and the
derived source terms $\Gamma_{0}$, $\Gamma_{i}$, $\Gamma_{ij}$, and
$\Xi_{(K)}$ are correct.
We also note that these relations are independent of the details of
the potential of the scalar field.


Thus, we have derived the self-consistent set of equations of the
second-order perturbation of the Einstein equations and the evolution
equations of matter fields in terms of gauge-invariant variables.
As the current status of the second-order gauge-invariant cosmological
perturbation theory, we may say that the curvature terms in the
second-order Einstein tensor
(\ref{eq:second-order-Einstein-equation}), i.e., the second-order
perturbations of the Einstein tensor, are almost completely derived,
although we have the ``zero-mode problem'' as an remaining problem, as
mentioned above.
After resolving this zero-mode problem, we have to clarify the
physical behaviors of the second-order cosmological perturbation
in the single scalar field system in the context of the inflationary
scenario.
This will be a preliminary step to clarify the quantum behaviors of
second-order perturbations in the inflationary universe.
Further, we also have to carry out the comparison with the result by
long-wavelength approximations.
If these issues are completed, we may say that we have completely
understood the properties of the second-order perturbation of the
Einstein tensor.
The next task is to clarify the nature of the second-order
perturbation of the energy-momentum tensor through the extension to
multi-fluid or multi-field systems.
Further, we also have to extend our arguments to the Einstein
Boltzmann system to discuss CMB physics, since we have to treat photon
and neutrinos through the Boltzmann distribution functions.
This issue is also discussed in some
literature\cite{N.Bartolo-S.Matarrese-A.Riotto-2004c,N.Bartolo-S.Matarrese-A.Riotto-2004d,N.Bartolo-E.Komatsu-S.Matarrese-A.Riotto-2004,N.Bartolo-S.Matarrese-A.Riotto-2006,Bartolo:2006cu,Bartolo:2006fj,Nitta:2009jp,Pitrou:2008ak,Senatore:2008vi,Pitro-2007,Pitro-2009}.
If we accomplish these extension, we will be able to clarify the
non-linear effects in CMB physics.


Finally, readers might think that the ingredients of this paper is too
mathematical as Astronomy.
However, we have to emphasize that a high degree of the theoretical
sophistication leads unambiguous theoretical predictions in many
case.
As in the case of the linear-order cosmological perturbation theory,
the developments in observations are also supported by the theoretical
sophistication and the theoretical sophistication are accomplished
motivated by observations.
In this sense, now, we have an opportunity to develop the general
relativistic second-order perturbation theory to a high degree of
sophistication which is motivated by observations.
We also expect that this theoretical sophistication will be also
useful to discuss the theoretical predictions of non-Gaussianity in
CMB and comparison with observations.
Therefore, I think that this opportunity is opened not only for
observational cosmologists but also for theoretical and mathematical
physicists.


\section*{Acknowledgments}


The author thanks participants in the GCOE/YITP workshop YITP-W-0901
on ``Non-linear cosmological perturbations'' which was held at YITP in
Kyoto, Japan in April, 2009, for valuable discussions, in particular,
Prof. M.~Bruni, Prof. R.~Maartens, Prof. M.~Sasaki, Prof. T.~Tanaka,
and Prof. K.~Tomita.
This review is an extension of the contribution to this workshop by
the author.


\appendix
\section{Derivation of the generic representation of the Taylor
  expansion of tensors on a manifold}
\label{sec:derivation-of-Taylor-expansion}


In this Appendix, we derive the representation of the coefficients of
the formal Taylor expansion (\ref{eq:Taylor-expansion-of-f}) of the
pull-back of a diffeomorphism in terms of the suitable derivative
operators.
The guide principle of our arguments is the following theorem\cite{Bruni-Gualtieri-Sopuerta-2003,Kobayashi-Nomizu-I-1996}.


\begin{theorem}
  \label{theorem:Bruni-Gualtieri-Sopuerta-2003-Appendix}
  Let ${\cal D}$ be a derivative operator acting on the set of all the
  tensor fields defined on a differentiable manifold ${\cal M}$ and
  satisfying the following conditions: (i) it is linear and satisfies
  the Leibniz rule; (ii) it is tensor-type preserving; (iii) it
  commutes with every contraction of a tensor field; and (iv) it
  commutes with the exterior differentiation $d$.
  Then, ${\cal D}$ is equivalent to the Lie derivative operator with
  respect to some vector field $\xi$, i.e.,
  ${\cal D}={\pounds}_{\xi}$.
\end{theorem}


The prove of the assertion of Theorem
\ref{theorem:Bruni-Gualtieri-Sopuerta-2003-Appendix} is given in
Ref.~\cite{Bruni-Gualtieri-Sopuerta-2003} as follows.
When acting on functions, the derivative operator ${\cal D}$ defines a
vector field $\xi$ through the relation
\begin{eqnarray}
  \label{eq:Bruni-Gualtieri-Sopuerta-2003-A.1}
  {\cal D}f =: \xi(f) = {\pounds}_{\xi}f,
  \quad \forall f\in{\cal F}(M)
\end{eqnarray}
where ${\cal F}(M)$ denotes the algebra of $C^{\infty}$ functions on
${\cal M}$.
The assertion of the Theorem for an arbitrary tensor field is hold if
and only if the assertions for an arbitrary scalar function and for an
arbitrary vector field $V$ are hold.
To do this, we consider the scalar function $V(f)$ and we obtain
\begin{eqnarray}
  {\cal D}(V(f)) = \xi(V(f))
\end{eqnarray}
through Eq.~(\ref{eq:Bruni-Gualtieri-Sopuerta-2003-A.1}).
Through the conditions (i)-(iv) of ${\cal D}$, ${\cal D}(V(f))$ is
also given by
\begin{eqnarray}
  {\cal D}(V(f))
  &=& {\cal D}(df(V))
  = {\cal D}\left\{
    {\cal C}(df\otimes V)
  \right\}
  \nonumber\\
  &=&
  {\cal C}\left\{
    {\cal D}(df\otimes V)
  \right\}
  \nonumber\\
  &=&
  {\cal C}\left\{
    {\cal D}(df)\otimes V
    +
    df \otimes{\cal D}V
  \right\}
  \nonumber\\
  &=&
  {\cal C}\left\{
    d({\cal D}f)\otimes V
    +
    df \otimes{\cal D}V
  \right\}
  \nonumber\\
  &=&
  d({\cal D}f)(V) + df({\cal D}V)
  \nonumber\\
  &=&
  V({\cal D}f) + ({\cal D}V)(f)
\end{eqnarray}
Then we obtain
\begin{eqnarray}
  ({\cal D}V)(f) &=& \xi(V(f)) - V(\xi(f))
  = \left[\xi, V\right](f)
  \nonumber\\
  &=& ({\pounds}_{\xi}V)(f)
\end{eqnarray}
for an arbitrary $f$, i.e.,
\begin{eqnarray}
  \label{eq:Bruni-Gualtieri-Sopuerta-2003-A.4}
  {\cal D}V = {\pounds}_{\xi}V.
\end{eqnarray}
Through Eqs.~(\ref{eq:Bruni-Gualtieri-Sopuerta-2003-A.1}) and
(\ref{eq:Bruni-Gualtieri-Sopuerta-2003-A.4}), we can recursively
show
\begin{eqnarray}
  \label{eq:Bruni-Gualtieri-Sopuerta-2003-A.2}
  {\cal D}Q = {\pounds}_{\xi}Q
\end{eqnarray}
for an arbitrary tensor field $Q$\cite{Kobayashi-Nomizu-I-1996}.


Now, we consider the derivation of the Taylor expansion
(\ref{eq:symbolic-Taylor-expansion-of-f}).
As in the main text, we first consider the representation of the
Taylor expansion of $\Phi^{*}_{\lambda}f$ for an arbitrary scalar
function $f\in{\cal F}(M)$:
\begin{eqnarray}
  (\Phi^{*}_{\lambda}f)(p)
  &=&
  f(p)
  +
  \lambda
  \left\{\frac{\partial}{\partial\lambda}(\Phi^{*}_{\lambda}f)\right\}_{\lambda=0}
  +
  \frac{1}{2} \lambda^{2}
  \left\{\frac{\partial^{2}}{\partial\lambda^{2}}(\Phi^{*}_{\lambda}f)\right\}_{\lambda=0}
  + O(\lambda^{3}).
  \label{eq:symbolic-Taylor-expansion-of-f-appendix}
\end{eqnarray}
Although the operator $\partial/\partial\lambda$ in the bracket
$\{*\}_{\lambda=0}$ of
Eq.~(\ref{eq:symbolic-Taylor-expansion-of-f-appendix}) are simply
symbolic notation, we stipulate the properties
\begin{eqnarray}
  \left\{
    \frac{\partial^{2}}{\partial\lambda^{2}}(\Phi^{*}_{\lambda}f)
  \right\}_{\lambda=0}
  &=&
  \left\{
    \frac{\partial}{\partial\lambda}\left(
      \frac{\partial}{\partial\lambda}(\Phi^{*}_{\lambda}f)
    \right)
  \right\}_{\lambda=0}
  \label{eq:properties-of-partial-over-partial-lambda-1}
  ,\\
  \left\{
    \frac{\partial}{\partial\lambda}(\Phi^{*}_{\lambda}f)^{2}
  \right\}_{\lambda=0}
  &=&
  \left\{
    2 \Phi^{*}_{\lambda}f \frac{\partial}{\partial\lambda}(\Phi^{*}_{\lambda}f)
  \right\}_{\lambda=0}
  \label{eq:properties-of-partial-over-partial-lambda-2}
  .
\end{eqnarray}
for $\forall f\in{\cal F}({\cal M})$.
These properties imply that the operator $\partial/\partial\lambda$ is
in fact not simply symbolic notation but indeed the usual partial
differential operator on $\RF$.
We note that the property
(\ref{eq:properties-of-partial-over-partial-lambda-2}) is the Leibniz
rule, which plays important roles when we derive the representation of
the Taylor expansion
(\ref{eq:symbolic-Taylor-expansion-of-f-appendix}) in terms of
suitable Lie derivatives.


Together with the property
(\ref{eq:properties-of-partial-over-partial-lambda-2}), Theorem
\ref{theorem:Bruni-Gualtieri-Sopuerta-2003-Appendix} yields that there
exists a vector field $\xi_{1}$ so that
\begin{eqnarray}
  \left\{
    \frac{\partial}{\partial\lambda}(\Phi^{*}_{\lambda}f)
  \right\}_{\lambda=0}
  &=:&
  {\pounds}_{\xi_{1}} f
  .
  \label{eq:def-of-calL1}
\end{eqnarray}
Actually, the conditions (ii)-(iv) in Theorem
\ref{theorem:Bruni-Gualtieri-Sopuerta-2003-Appendix} are satisfied
from the fact that $\Phi_{\lambda}^{*}$ is the pull-back of a
diffeomorphism $\Phi_{\lambda}$ and (i) is satisfied due to the
property (\ref{eq:properties-of-partial-over-partial-lambda-2}).


Next, we consider the second-order term in
Eq.~(\ref{eq:symbolic-Taylor-expansion-of-f-appendix}).
Since we easily expect that the second-order term in
Eq.~(\ref{eq:symbolic-Taylor-expansion-of-f-appendix}) may includes
${\pounds}_{\xi_{1}}^{2}$, we define the derivative operator
${\cal L}_{2}$ by
\begin{eqnarray}
  \left\{
    \frac{\partial^{2}}{\partial\lambda^{2}}(\Phi^{*}_{\lambda}f)
  \right\}_{\lambda=0}
  &=:&
  \left({\cal L}_{2} + a {\pounds}_{\xi_{1}}^{2}\right) f
  \label{eq:def-of-calL2}
  ,
\end{eqnarray}
where $a$ is determined so that ${\cal L}_{2}$ satisfy the conditions
of Theorem~\ref{theorem:Bruni-Gualtieri-Sopuerta-2003-Appendix}.
The conditions (ii)-(iv) in
Theorem~\ref{theorem:Bruni-Gualtieri-Sopuerta-2003-Appendix} for
${\cal L}_{2}$ are satisfied from the fact that $\Phi_{\lambda}^{*}$
is the pull-back of a diffeomorphism $\Phi_{\lambda}$.
Further, ${\cal L}_{2}$ is obviously linear but we have to check
${\cal L}_{2}$ satisfy the Leibniz rule, i.e.,
\begin{eqnarray}
  \label{eq:Leibnitz-rule-of-calL2}
  {\cal L}_{2}\left(f^{2}\right) = 2 f {\cal L}_{2} f
\end{eqnarray}
for $\forall f \in {\cal F}({\cal M})$.
To do this, we use the properties
(\ref{eq:properties-of-partial-over-partial-lambda-1}) and
(\ref{eq:properties-of-partial-over-partial-lambda-2}), then we can
easily see that the Leibniz rule (\ref{eq:Leibnitz-rule-of-calL2}) is
satisfied iff $a=1$ and we may regard ${\cal L}_{2}$ as the Lie
derivative with respect to some vector field.
Then, when and only when $a=1$, there exists a vector field $\xi_{2}$
such that
\begin{eqnarray}
  {\cal L}_{2}f = {\pounds}_{\xi_{2}}f
  \label{eq:def-of-calL2-is-Lie}
\end{eqnarray}
and
\begin{eqnarray}
  \left\{
    \frac{\partial^{2}}{\partial\lambda^{2}}(\Phi^{*}_{\lambda}f)
  \right\}_{\lambda=0}
  &=:&
  \left({\pounds}_{\xi_{2}} + {\pounds}_{\xi_{1}}^{2}\right) f
  \label{eq:second-order-term-of-Taylor-expansion-is-Lie}
  .
\end{eqnarray}
Thus, we have seen that the Taylor expansion
(\ref{eq:symbolic-Taylor-expansion-of-f-appendix}) for an arbitrary
scalar function $f$ is given by Eq.~(\ref{eq:Taylor-expansion-of-f}).


Although the formula (\ref{eq:Taylor-expansion-of-f}) of the Taylor
expansion is for an arbitrary scalar function, we can easily extend
this formula to that for an arbitrary tensor field $Q$ as the
assertion of
Theorem~\ref{theorem:Bruni-Gualtieri-Sopuerta-2003-Appendix}.
The proof of the extension of the formula
(\ref{eq:Taylor-expansion-of-f}) to an arbitrary tensor field $Q$ is
completely parallel to the proof of the formula
(\ref{eq:Taylor-expansion-of-f}) for an arbitrary scalar function if
we stipulate the properties
\begin{eqnarray}
  \left\{
    \frac{\partial^{2}}{\partial\lambda^{2}}(\Phi^{*}_{\lambda}Q)
  \right\}_{\lambda=0}
  &=&
  \left\{
    \frac{\partial}{\partial\lambda}\left(
      \frac{\partial}{\partial\lambda}(\Phi^{*}_{\lambda}Q)
    \right)
  \right\}_{\lambda=0}
  \label{eq:properties-of-partial-over-partial-lambda-1-tensor}
  ,\\
  \left\{
    \frac{\partial}{\partial\lambda}(\Phi^{*}_{\lambda}Q)^{2}
  \right\}_{\lambda=0}
  &=&
  \left\{
    2 \Phi^{*}_{\lambda}Q \frac{\partial}{\partial\lambda}(\Phi^{*}_{\lambda}Q)
  \right\}_{\lambda=0}
  \label{eq:properties-of-partial-over-partial-lambda-2-tensor}
\end{eqnarray}
instead of
Eqs.~(\ref{eq:properties-of-partial-over-partial-lambda-1}) and
(\ref{eq:properties-of-partial-over-partial-lambda-2}).
As the result, we obtain the representation of the Taylor
expansion for an arbitrary tensor field $Q$.


\section{Derivation of the perturbative Einstein tensors}
\label{sec:derivation-of-pert-Einstein-tensors}


Following the outline of the calculations explained in
Sec.~\ref{sec:Perturbation-of-the-Einstein-tensor}, we first calculate
the perturbative expansion of the inverse metric.
The perturbative expansion of the inverse metric can be easily derived
from Eq.~(\ref{eq:metric-expansion}) and the definition of the inverse
metric
\begin{eqnarray}
  \label{eq:inverse-metric-def}
  \bar{g}^{ab}\bar{g}_{bc} = \delta^{a}_{c}.
\end{eqnarray}
We also expand the inverse metric $\bar{g}^{ab}$ in the form
\begin{eqnarray}
  \label{eq:inverse-metric-expansion}
  \bar{g}^{ab} = g^{ab} + \lambda {}^{(1)}\!\bar{g}^{ab} +
  \frac{1}{2} \lambda^{2} {}^{(2)}\!\bar{g}^{ab}.
\end{eqnarray}
Then, each term of the expansion of the inverse metric is given by
\begin{eqnarray}
  \label{eq:inverse-metric-each-order}
  {}^{(1)}\!\bar{g}^{ab} = - h^{ab}, \quad
  {}^{(2)}\!\bar{g}^{ab} = 2 h^{ac} h_{c}^{\;\;b} - l^{ab}.
\end{eqnarray}


To derive the formulae for the perturbative expansion of the Riemann
curvature, we have to derive the formulae for the perturbative
expansion of the tensor $C^{c}_{\;\;ab}$ given by
Eq.~(\ref{eq:c-connection}).
The tensor $C^{c}_{\;\;ab}$ is also expanded in the same form as
Eq.~(\ref{eq:Bruni-39-one}).
The first-order perturbations of $C^{c}_{\;\;ab}$ have the well-known
form\cite{Wald-book}
\begin{eqnarray}
  {}^{(1)}\!C^{c}_{\;\;ab}
  =
  \nabla_{(a}h_{b)}^{\;\;c} - \frac{1}{2} \nabla^{c}h_{ab}
  =:
  H_{ab}^{\;\;\;\;c}\left[h\right],
  \label{eq:KN2005-3.12}
\end{eqnarray}
where $H_{ab}^{\;\;\;\;c}\left[A\right]$ is defined by
Eq.~(\ref{eq:Habc-def-1}) for an arbitrary tensor field $A_{ab}$
defined on the background spacetime ${\cal M}_{0}$.
In terms of the tensor field $H_{ab}^{\;\;\;\;c}$ defined by
(\ref{eq:Habc-def-1}) the second-order perturbation
${}^{(2)}\!C^{c}_{\;\;ab}$ of the tensor field $C^{c}_{\;\;ab}$ is
given by
\begin{eqnarray}
  {}^{(2)}\!C^{c}_{\;\;ab}
  =
  H_{ab}^{\;\;\;\;c}\left[l\right] - 2 h^{cd} H_{abd}\left[h\right].
  \label{eq:KN2005-3.13}
\end{eqnarray}
The Riemann curvature (\ref{eq:phys-riemann-back-riemann-rel}) on the
physical spacetime ${\cal M}_{\lambda}$ is also expanded in the form
(\ref{eq:Bruni-39-one}):
\begin{eqnarray}
  \bar{R}_{abc}^{\;\;\;\;\;\;d}
  &=:&
  R_{abc}^{\;\;\;\;\;\;d}
  +
  \lambda {}^{(1)}\!R_{abc}^{\;\;\;\;\;\;d}
  +
  +
  \frac{1}{2} \lambda^{2} {}^{(2)}\!R_{abc}^{\;\;\;\;\;\;d}
  + O(\lambda^{3}).
\end{eqnarray}
The first- and the second-order perturbation of the Riemann curvature
are given by
\begin{eqnarray}
  {}^{(1)}\!R_{abc}^{\;\;\;\;\;\;d}
  &=&
  - 2 \nabla_{[a}^{} {}^{(1)}\!C^{d}_{\;\;b]c},
  \label{eq:KN2005-3.15}
  \\
  {}^{(2)}\!R_{abc}^{\;\;\;\;\;\;d}
  &=&
  - 2 \nabla_{[a}^{} {}^{(2)}\!C^{d}_{\;\;b]c}
  + 4 {}^{(1)}\!C^{e}_{\;\;c[a} {}^{(1)}\!C^{d}_{\;\;b]e}
  \label{eq:KN2005-3.16}
\end{eqnarray}
Substituting Eqs.~(\ref{eq:KN2005-3.12}) and (\ref{eq:KN2005-3.13})
into Eqs.~(\ref{eq:KN2005-3.15}) and (\ref{eq:KN2005-3.16}), we obtain
the perturbative form of the Riemann curvature in terms of the
variables defined by Eq.~(\ref{eq:Habc-def-1}) and
(\ref{eq:Habc-def-2}):
\begin{eqnarray}
  {}^{(1)}\!R_{abc}^{\;\;\;\;\;\;d}
  &=&
  - 2 \nabla_{[a} H_{b]c}^{\;\;\;\;\;d}\left[h\right],
  \label{eq:KN2005-3.15-2}
  \\
  {}^{(2)}\!R_{abc}^{\;\;\;\;\;\;d}
  &=&
  - 2 \nabla_{[a} H_{b]c}^{\;\;\;\;\;d}\left[l\right]
  + 4 H_{[a}^{\;\;\;de}\left[h\right] H_{b]ce}\left[h\right]
  + 4 h^{de} \nabla_{[a} H_{b]ce}\left[h\right].
  \label{eq:KN2005-3.16-2}
\end{eqnarray}


To write down the perturbative curvatures (\ref{eq:KN2005-3.15-2}) and
(\ref{eq:KN2005-3.16-2}) in terms of the gauge invariant and variant
variables defined by Eqs.~(\ref{eq:linear-metric-decomp}) and
(\ref{eq:H-ab-in-gauge-X-def-second-1}), we first derive an expression
for the tensor field $H_{abc}[h]$ in terms of the gauge invariant
variables, and then, we derive a perturbative expression for the
Riemann curvature.


First, we consider the linear-order perturbation
(\ref{eq:KN2005-3.15-2}) of the Riemann curvature.
Using the decomposition (\ref{eq:linear-metric-decomp}) and the
identity $R_{[abc]}^{\;\;\;\;\;\;\;\;d}=0$, we can easily derive the
relation
\begin{eqnarray}
  \label{eq:KN2005-3.20}
  H_{abc}\left[h\right]
  =
  H_{abc}\left[{\cal H}\right]
  +
  \nabla_{a}\nabla_{b}X_{c}
  +
  R_{bca}^{\;\;\;\;\;\;d} X_{d}
  ,
\end{eqnarray}
where the variable $H_{abc}\left[{\cal H}\right]$ is defined by
Eqs.~(\ref{eq:Habc-def-1}) and (\ref{eq:Habc-def-2}) with
$A_{ab}={\cal H}_{ab}$.
Clearly, the variable $H_{ab}^{\;\;\;\;c}\left[{\cal H}\right]$ is
gauge invariant.
Taking the derivative and using the Bianchi identity
$\nabla_{[a}R_{bc]de}=0$, we obtain
\begin{eqnarray}
  {}^{(1)}\!R_{abc}^{\;\;\;\;\;\;d}
  &=&
  - 2 \nabla_{[a}^{} H_{b]c}^{\;\;\;\;\;d}\left[{\cal H}\right]
  +
  {\pounds}_{X}R_{abc}^{\;\;\;\;\;\;d}
  \label{eq:KN2005-3.23}
  .
\end{eqnarray}
Similar but some cumbersome calculations yield
\begin{eqnarray}
  {}^{(2)}\!R_{abc}^{\;\;\;\;\;d}
  &=&
  - 2 \nabla_{[a}^{} H_{b]c}^{\;\;\;\;d}\left[{\cal L}\right]
  +
  4 H_{[a}^{\;\;\;de}\left[{\cal H}\right]
  H_{b]ce}^{}\left[{\cal H}\right]
  +
  4 {\cal H}_{e}^{\;\;d}
  \nabla_{[a}^{}H_{b]c}^{\;\;\;\;e}\left[{\cal H}\right]
  + 2 {\pounds}_{X}{}^{(1)}\!R_{abc}^{\;\;\;\;\;d}
  + \left(
    {\pounds}_{Y} - {\pounds}_{X}^{2}
  \right)R_{abc}^{\;\;\;\;\;d}
  \label{eq:KN2005-3.32}
  .
\end{eqnarray}
Equations (\ref{eq:KN2005-3.23}) and (\ref{eq:KN2005-3.32}) have the
same for as the decomposition formulae
(\ref{eq:matter-gauge-inv-decomp-1.0}) and
(\ref{eq:matter-gauge-inv-decomp-2.0}), respectively, as the result.


Contracting the indices $b$ and $d$ in Eqs.~(\ref{eq:KN2005-3.23}) and
(\ref{eq:KN2005-3.32}) of the perturbative Riemann curvature, we can
directly derive the formulae for the perturbative expansion of the
Ricci curvature:
expanding the Ricci curvature
\begin{eqnarray}
  \bar{R}_{ab}
  =:
  R_{ab}
  +
  \lambda {}^{(1)}\!R_{ab}
  +
  +
  \frac{1}{2} \lambda^{2} {}^{(2)}\!R_{ab}
  + O(\lambda^{3}),
\end{eqnarray}
we obtain the first-order Ricci curvature as
\begin{eqnarray}
   {}^{(1)}\!R_{ab}
   &=&
   - 2 \nabla_{[a}^{} H_{c]b}^{\;\;\;\;\;c}\left[{\cal H}\right]
   +
   {\pounds}_{X}R_{ab}
  \label{eq:KN2005-3.45}
  .
\end{eqnarray}
and we also obtain the second-order Ricci curvature as
\begin{eqnarray}
  {}^{(2)}\!R_{ab}
  &=&
  - 2 \nabla_{[a}H_{c]b}^{\;\;\;\;\;c}\left[{\cal L}\right]
  +
   4 H_{[a}^{\;\;\;cd}\left[{\cal H}\right] H_{c]bd}\left[{\cal H}\right]
  +
  4 {\cal H}_{d}^{\;\;c} \nabla_{[a}H_{b]c}^{\;\;\;\;\;d}\left[{\cal H}\right]
  +
  2 {\pounds}_{X}{}^{(1)}\!R_{ab}
  + \left(
    {\pounds}_{Y} - {\pounds}_{X}^{2}
  \right)R_{ab}
  \label{eq:KN2005-3.46}
  .
\end{eqnarray}


The scalar curvature on the physical spacetime ${\cal M}$ is given by
$\bar{R} = \bar{g}^{ab}\bar{R}_{ab}$.
To obtain the perturbative form of the scalar curvature, we expand the
$\bar{R}$ in the form (\ref{eq:Bruni-39-one}), i.e.,
\begin{eqnarray}
  \bar{R} =: R + \lambda {}^{(1)}\!R
  + \frac{1}{2}\lambda^{2} {}^{(2)}\!R + O(\lambda^{3})
\end{eqnarray}
and $\bar{g}^{ab}\bar{R}_{ab}$ is expanded through the Leibniz rule.
Then, the perturbative formula for the scalar curvature at each order
is derived from perturbative form of the inverse metric
(\ref{eq:inverse-metric-each-order}) and the Ricci curvature
(\ref{eq:KN2005-3.45}) and (\ref{eq:KN2005-3.46}).
Straightforward calculations lead to the expansion of the scalar
curvature as
\begin{eqnarray}
  {}^{(1)}\!R
  &=&
  - 2 \nabla_{[a} H_{b]}^{\;\;\;ab}\left[{\cal H}\right]
  - R_{ab} {\cal H}^{ab}
  + {\pounds}_{X}R
  \label{eq:KN2005-3.52}
  , \\
  {}^{(2)}\!R
  &=&
  - 2 \nabla_{[a}^{} H_{b]}^{\;\;\;ab}\left[{\cal L}\right]
  + R^{ab} \left(
    2 {\cal H}_{ca} {\cal H}_{b}^{\;\;c}
    - {\cal L}_{ab}
  \right)
  + 4 H_{[a}^{\;\;\;cd}\left[{\cal H}\right]
  H_{c]\;\;d}^{\;\;\;a}\left[{\cal H}\right]
  + 4 {\cal H}_{c}^{\;\;b} \nabla_{[a}H_{b]}^{\;\;\;ac}\left[{\cal H}\right]
  + 4 {\cal H}^{ab} \nabla_{[a}^{} H_{d]b}^{\;\;\;\;\;d}\left[{\cal H}\right]
  \nonumber\\
  &&
  + 2 {\pounds}_{X}{}^{(1)}\!R
  + \left(
    {\pounds}_{Y} - {\pounds}_{X}^{2}
  \right)R
  \label{eq:KN2005-3.54}
  .
\end{eqnarray}
We also note that the expansion formulae (\ref{eq:KN2005-3.52}) and
(\ref{eq:KN2005-3.54}) have the same for as the decomposition formulae
(\ref{eq:matter-gauge-inv-decomp-1.0}) and
(\ref{eq:matter-gauge-inv-decomp-2.0}), respectively, as the result.


Next, we consider the perturbative form of the Einstein tensor
$\bar{G}_{ab}:=\bar{R}_{ab}-\frac{1}{2}\bar{g}_{ab}\bar{R}$ and we
expand $\bar{G}_{ab}$ as in the form (\ref{eq:Bruni-39-one}):
\begin{eqnarray}
  \bar{G}_{ab} =: G_{ab} + \lambda {}^{(1)}\!\left(G_{ab}\right)
  + \frac{1}{2}\lambda^{2} {}^{(2)}\!\left(G_{ab}\right) + O(\lambda^{3})
  .
\end{eqnarray}
As in the case of the scalar curvature, straightforward calculations
lead
\begin{eqnarray}
  {}^{(1)}\!\left(G_{ab}\right)
  &=&
  - 2 \nabla_{[a}^{} H_{d]b}^{\;\;\;\;\;d}\left[{\cal H}\right]
  + g_{ab} \nabla_{[c}^{} H_{d]}^{\;\;\;cd}\left[{\cal H}\right]
  - \frac{1}{2} R {\cal H}_{ab}
  + \frac{1}{2} g_{ab} R_{cd} {\cal H}^{cd}
  + {\pounds}_{X}G_{ab}
  \label{eq:KN2005-3.59}
  ,
  \\
  {}^{(2)}\!\left(G_{ab}\right)
  &=&
  - 2 \nabla_{[a}^{} H_{c]b}^{\;\;\;\;\;c}\left[{\cal L}\right]
  + 4 H_{[a}^{\;\;\;cd}\left[{\cal H}\right] H_{c]bd}\left[{\cal H}\right]
  + 4 {\cal H}_{c}^{\;\;d} \nabla_{[a} H_{d]b}^{\;\;\;\;\;c}\left[{\cal H}\right]
  + 2 {\cal H}_{ab} \nabla_{[c}H_{d]}^{\;\;\;cd}\left[{\cal H}\right]
  \nonumber\\
  &&
  - \frac{1}{2} g_{ab} \left(
    - 2 \nabla_{[c}^{} H_{d]}^{\;\;\;cd}\left[{\cal L}\right]
    + 2 R_{de}{\cal H}_{c}^{\;\;d}{\cal H}^{ec}
    - R_{de}{\cal L}^{de}
    + 4 H_{[c}^{\;\;\;de}\left[{\cal H}\right] H_{d]\;\;e}^{\;\;\;c}\left[{\cal H}\right]
  \right.
  \nonumber\\
  && \quad\quad\quad\quad
  \left.
    + 4 {\cal H}_{e}^{\;\;d} \nabla_{[c}^{} H_{d]}^{\;\;\;ce}\left[{\cal H}\right]
    + 4 {\cal H}^{ce} \nabla_{[c}^{} H_{d]e}^{\;\;\;\;\;d}\left[{\cal H}\right]
  \right)
  + {\cal H}_{ab} {\cal H}^{cd} R_{cd}
  - \frac{1}{2} R {\cal L}_{ab}
  \nonumber\\
  &&
  + 2 {\pounds}_{X}{}^{(1)}\!\left(G_{ab}\right)
  + \left({\pounds}_{Y} - {\pounds}_{X}^{2}\right)G_{ab}
  \label{eq:KN2005-3.61}
  .
\end{eqnarray}
We note again that Eqs.~(\ref{eq:KN2005-3.59}) and
(\ref{eq:KN2005-3.61}) have the same form as the decomposition
formulae (\ref{eq:matter-gauge-inv-decomp-1.0}) and
(\ref{eq:matter-gauge-inv-decomp-2.0}), respectively.


The perturbative formulae for the perturbation of the Einstein tensor
\begin{eqnarray}
  \bar{G}_{a}^{\;\;b}=\bar{g}^{bc}\bar{G}_{ac}
\end{eqnarray}
is derived by the similar manner to the case of the perturbations of
the scalar curvature.
Through these formulae summarized above, straightforward calculations
leads Eqs.~(\ref{eq:linear-Einstein})--(\ref{eq:(2)Sigma-def-second}).
We have to note that to derive the formulae
(\ref{eq:cal-G-def-second}) with Eq.~(\ref{eq:(2)Sigma-def-second}),
we have to consider the general relativistic gauge-invariant
perturbation theory with two infinitesimal parameters which is
developed in Refs.~\cite{kouchan-gauge-inv,kouchan-second}, as
commented in the main text.


\section{A Scenario of the proof of Conjecture~\ref{conjecture:decomp_conjecture_for_hab}}
\label{sec:Outline-of-the-proof-of-the-decomposition-conjecture}


In this Appendix, we give a scenario of a proof of
Conjecture~\ref{conjecture:decomp_conjecture_for_hab} in
Sec.~\ref{sec:gauge-invariant-variables} for an arbitrary background
spacetime.
To do this, we assume that the background spacetime admits ADM
decomposition.
Therefore, the background spacetime ${\cal M}_{0}$ (at least the
portion of ${\cal M}_{0}$ that we are addressing) considered here is
$n-1+1$-dimensional spacetime, which is described by the direct
product $\RF\times\Sigma$. Here, $\RF$ is a time direction and
$\Sigma$ is the spacelike hypersurface ($\dim\Sigma=n-1$) embedded in
${\cal M}_{0}$.
This means that ${\cal M}_{0}$ is foliated by the one-parameter family
of spacelike hypersurface $\Sigma(t)$, where $t\in\RF$ is a time
function.
In this setup, the metric on ${\cal M}_{0}$ is described by
\begin{eqnarray}
  g_{ab}
  &=&
      - \alpha^{2} (dt)_{a} (dt)_{b}
      + q_{ij} (dx^{i}+\beta^{i}dt)_{a} (dx^{j}+\beta^{j}dt)_{b},
  \label{eq:background-ADM-decomp}
\end{eqnarray}
where $\alpha$ is the lapse function, $\beta^{i}$ is the shift vector,
and $q_{ab}=q_{ij}(dx^{i})_{a}(dx^{j})_{b}$ is the metric on
$\Sigma(t)$.


Since the ADM decomposition (\ref{eq:background-ADM-decomp}) of the
metric is a local decomposition, we may regard the arguments in this
paper as being restricted to that for a single patch in
${\cal M}_{0}$, which is covered by the metric
(\ref{eq:background-ADM-decomp}).
Further, we may change the region that is covered by the metric
(\ref{eq:background-ADM-decomp}) through the choice of the lapse
function $\alpha$ and the shift vector $\beta^{i}$.
The choice of $\alpha$ and $\beta^{i}$ is regarded as the first kind
of gauge choice explained in Sec.~\ref{sec:first-kind-gauge}, which
has nothing to do with the second kind of gauge as emphasized in
Sec.~\ref{sec:second-kind-gauge}.
Since we may regard the representation
(\ref{eq:background-ADM-decomp}) of the background metric as being
that on a single patch in ${\cal M}_{0}$, in a general situation, each
$\Sigma$ may have its boundary $\partial\Sigma$.
For example, in asymptotically flat spacetime, $\partial\Sigma$
includes asymptotically flat regions~\cite{Wald-book}.
Furthermore, if necessary, we may regard $\Sigma(t)$ as a portion of
the spacelike hypersurface in ${\cal M}_{0}$ and add disjoint
components to the boundary $\partial\Sigma$.
For example, when the formation of black holes occurs, we may exclude
the region inside the black holes from $\Sigma$.
In any case, when we consider the spacelike hypersurface $\Sigma$ with
boundary $\partial\Sigma$, we have to impose appropriate boundary
conditions at the boundary $\partial\Sigma$.


To consider the decomposition (\ref{eq:hab-calHab+LieXgab}) of the
first-order metric perturbation $h_{ab}$, we first consider the
components of the metric $h_{ab}$ as
\begin{eqnarray}
  h_{ab}
  &=&
      h_{tt} (dt)_{a} (dt)_{b} + 2 h_{ti} (dt)_{(a} (dx^{i})_{b)}
     + h_{ij} (dx^{i})_{a} (dx^{j})_{b}.
     \label{eq:hab-ADM-decomposition}
\end{eqnarray}
The components $h_{tt}$, $h_{ti}$, and $h_{ij}$ are regarded as a
scalar function, components of a vector field, and the components of a
symmetric tensor field on the spacelike hypersurface $\Sigma$,
respectively.
Under the gauge-transformation rule (\ref{eq:hab-gauge-trans}) the
components $\{h_{tt},h_{ti},h_{ij}\}$ are transformed as
\begin{eqnarray}
  {}_{\;{\cal Y}}\!h_{tt} - {}_{\;{\cal X}}\!h_{tt}
  &=&
      2 \partial_{t}\xi_{t}
      -
      \frac{2}{\alpha} \left(
      \partial_{t}\alpha + \beta^{i} D_{i}\alpha
      - \beta^{i}\beta^{j} K_{ij}
      \right)\xi_{t}
      \nonumber\\
  &&
      - \frac{2}{\alpha} \left(
      \beta^{i}\beta^{k}\beta^{j}K_{kj}
      - \beta^{i}\partial_{t}\alpha
      + \alpha q^{ij} \partial_{t}\beta_{j}
      + \alpha^{2} D^{i}\alpha
     - \alpha\beta^{k} D^{i}\beta_{k}
      - \beta^{i}\beta^{j} D_{j}\alpha
      \right) \xi_{i}
      ,
      \label{eq:ADM-htt-gauge-trans}
  \\
  {}_{\;{\cal Y}}\!h_{ti} - {}_{\;{\cal X}}\!h_{ti}
  &=&
      \partial_{t}\xi_{i} + D_{i}\xi_{t}
      - \frac{2}{\alpha} \left( D_{i}\alpha - \beta^{j} K_{ij}\right) \xi_{t}
      - \frac{2}{\alpha} M_{i}^{\;\;j} \xi_{j}
      ,
      \label{eq:ADM-hti-gauge-trans}
  \\
  {}_{\;{\cal Y}}\!h_{ij} - {}_{\;{\cal X}}\!h_{ij}
  &=&
      2 D_{(i}\xi_{j)} + \frac{2}{\alpha} K_{ij} \xi_{t}
      - \frac{2}{\alpha} \beta^{k} K_{ij} \xi_{k}
      ,
      \label{eq:ADM-hij-gauge-trans}
\end{eqnarray}
where $M_{i}^{\;\;j}$ is defined by
\begin{eqnarray}
  \label{eq:ADM-hti-termMij-def}
  M_{i}^{\;\;J}
  :=
  - \alpha^{2} K_{i}^{\;\;j}
  + \beta^{j} \beta^{k} K_{ki}
  - \beta^{j}D_{i}\alpha
  + \alpha D_{i}\beta^{j}
  .
\end{eqnarray}
Here, $K_{ij}$ are the components of the extrinsic curvature of
$\Sigma$ in ${\cal M}_{0}$ and $D_{i}$ is the covariant derivative
associated with the metric $q_{ij}$ ($D_{i}q_{jk}=0$).
The extrinsic curvature $K_{ij}$ and its trace $K$ are related to the
time derivative of the metric $q_{ij}$ by
\begin{eqnarray}
  \label{eq:ADM-extrinsic-def}
  K_{ij}
  =
  - \frac{1}{2\alpha} \left[
  \frac{\partial}{\partial t} q_{ij}
  -
  2
  D_{(i}\beta_{j)}
  \right]
  ,
  \quad
  K := q^{ij}K_{ij}.
\end{eqnarray}


We also note that the gauge-transformation rules
(\ref{eq:ADM-htt-gauge-trans})--(\ref{eq:ADM-hij-gauge-trans})
represent a gauge-transformation of the second kind, which has nothing
to do with the gauge degree of freedom of the first kind as explained
in Sec.~\ref{sec:Gauge-degree-of-freedom-in-general-relativity}.


To exclude the gauge degree of freedom of the second kind, we define
the variables $h_{(VL)}$, $h_{(V)i}$, $h_{(L)}$, $h_{(TV)i}$,
$h_{(TT)ij}$ by the following decomposition formulae for the
components $h_{ti}$ and $h_{ij}$:
\begin{eqnarray}
  h_{ti}
  &=:&
       D_{i}h_{(VL)}
       +
       h_{(V)i}
       -
       \frac{2}{\alpha} \left(
       D_{i}\alpha - \beta^{k}K_{ik}
       \right)
       \left(
       h_{(VL)} - \Delta^{-1} D^{k}\partial_{t}h_{(TV)k}
       \right)
       -
       \frac{2}{\alpha} M_{i}^{\;\;k} h_{(TV)k}
       ,
     \label{eq:hVL-hVi-hL-hTVi-hTTij-def-1}
  \\
  h_{ij}
  &=:&
       \frac{1}{n-1} q_{ij} h_{(L)}
       + (Lh_{(TV)})_{ij}
       + h_{(TT)ij}
     + \frac{2}{\alpha} \left(
     h_{(VL)} - \Delta^{-1} D^{k}\partial_{t}h_{(TV)k}
     \right)
     - \frac{2}{\alpha} K_{ij} \beta^{k} h_{(TV)k}
     ,
     \label{eq:hVL-hVi-hL-hTVi-hTTij-def-2}
  \\
  &&
     D^{i}h_{(V)i} = 0, \quad
     q^{ij} h_{(TT)ij} = 0 = D^{i} h_{(TT)ij}
     ,
     \label{eq:hVL-hVi-hL-hTVi-hTTij-def-3}
\end{eqnarray}
where
\begin{eqnarray}
  \label{eq:York-decomp-operator}
  (Lh_{(TV)})_{ij}
  &:=&
       D_{i}h_{(TV)j}
       +
       D_{j}h_{(TV)i}
       -
       \frac{2}{n} q_{ij} D^{l} h_{(TV)l}
       ,
\end{eqnarray}
and $\Delta^{-1}$ is the Green function of the Laplacian
$\Delta:=D^{i}D_{i}$.
We note that equations~(\ref{eq:hVL-hVi-hL-hTVi-hTTij-def-1}) and
(\ref{eq:hVL-hVi-hL-hTVi-hTTij-def-2}) have the non-trivial form.
The detailed explanations of the issue how to reach to these
expression (\ref{eq:hVL-hVi-hL-hTVi-hTTij-def-1}) and
(\ref{eq:hVL-hVi-hL-hTVi-hTTij-def-2}) are described in
Refs.~\cite{K.Nakamura-CQG-Letter-2011,K.Nakamura-Progress-Construction-2013}.
Here, we just accept the expressions of
Eqs.~(\ref{eq:hVL-hVi-hL-hTVi-hTTij-def-1}) and
(\ref{eq:hVL-hVi-hL-hTVi-hTTij-def-2}) as the definitions of the
variables $h_{(VL)}$, $h_{(V)i}$, $h_{(L)}$, $h_{(TV)i}$, and
$h_{(TT)ij}$.


\subsection{Inverse relation of
  Eqs.~(\ref{eq:hVL-hVi-hL-hTVi-hTTij-def-1}) and
  (\ref{eq:hVL-hVi-hL-hTVi-hTTij-def-2})}
\label{sec:inverse-relation-Appendix-C}


Here, we check that the definitions
(\ref{eq:hVL-hVi-hL-hTVi-hTTij-def-1}) and
(\ref{eq:hVL-hVi-hL-hTVi-hTTij-def-2}) are invertible.
We note that this check is essential to our discussion.
If the expression (\ref{eq:hVL-hVi-hL-hTVi-hTTij-def-1}) and
(\ref{eq:hVL-hVi-hL-hTVi-hTTij-def-2}) are not invertible, one-to-one
correspondence with the set $\{h_{ti},h_{ij}\}$ of the original components is
not guaranteed.


To derive the inverse relation of
Eqs.~(\ref{eq:hVL-hVi-hL-hTVi-hTTij-def-1})--(\ref{eq:hVL-hVi-hL-hTVi-hTTij-def-3}),
we first consider Eq.~(\ref{eq:hVL-hVi-hL-hTVi-hTTij-def-1}).
Assuming the existence of the Green function ${\cal F}^{-1}$ for the
elliptic derivative operator
\begin{eqnarray}
  {\cal F}
  &:=&
       \Delta
       - \frac{2}{\alpha} \left(
       D_{i}\alpha - \beta^{j} K_{ij}
       \right) D^{i}
     - 2 D^{i} \left\{
     \frac{1}{\alpha} \left(D_{i}\alpha - \beta^{j}K_{ij}\right)
     \right\}
     ,
     \label{eq:calF-operator-def}
\end{eqnarray}
we obtain the relations
\begin{widetext}
\begin{eqnarray}
  h_{(VL)}
  &=&
      {\cal F}^{-1} \left[
      D^{k}h_{tk} - D^{k}\partial_{t}h_{(TV)k}
      +
      D^{k}\left(
      \frac{2}{\alpha} M_{k}^{\;\;l} h_{(TV)l}
      \right)
      \right]
      +
      \Delta^{-1} D^{k}\partial_{t}h_{(TV)k}
      ,
      \label{eq:hVL-by-hTVi-hti}
  \\
  h_{(V)i}
  &=&
      h_{ti} - D_{i}\Delta^{-1}D^{k}\partial_{t}h_{(TV)k}
      + \frac{2}{\alpha} M_{i}^{\;\;k} h_{(TV)k}
      \nonumber\\
  &&
      +
      \left[
      D_{i}
      - \frac{2}{\alpha} \left(D_{i}\alpha-\beta^{j}K_{ij}\right)
      \right]
      {\cal F}^{-1}
      \left[
      - D^{k}h_{tk}
      + D^{k}\partial_{t} h_{(TV)k}
      - D^{k} \left(
      \frac{2}{\alpha} M_{k}^{\;\;l} h_{(TV)l}
      \right)
      \right]
      .
      \label{eq:hV-by-hTVi-hti}
\end{eqnarray}
Equations (\ref{eq:hVL-by-hTVi-hti}) and (\ref{eq:hV-by-hTVi-hti})
imply that we can obtain the relations between $\{h_{(VL)},h_{(V)i}\}$
and $\{h_{ti},h_{ij}\}$ if the relation between $h_{(TV)i}$ and
$\{h_{ti},h_{ij}\}$ is specified.
On the other hand, the trace- and the traceless-part of
Eq.~(\ref{eq:hVL-hVi-hL-hTVi-hTTij-def-2}) are given by
\begin{eqnarray}
  &&
     h_{(L)}
     =
     q^{ij}h_{ij}
     +
     \frac{2}{\alpha} K \beta^{k} h_{(TV)k}
     - \frac{2}{\alpha} K \left(
     {\cal F}^{-1} \left[
     D^{k}h_{tk} - D^{k} \partial_{t}h_{(TV)k}
     + D^{k} \left(
     \frac{2}{\alpha} M_{k}^{\;\;l} h_{(TV)l}
     \right)
     \right]
     \right)
     ,
     \label{eq:hL-by-hTVk-hij}
  \\
  &&
     h_{ij}
     -
     \frac{1}{n-1} q_{ij} q^{kl} h_{kl}
     =
     (Lh_{(TV)})_{ij}
     +
     h_{(TT)ij}
     -
     \frac{2}{\alpha} \tilde{K}_{ij} \beta^{k} h_{(TV)k}
      \nonumber\\
  && \quad\quad\quad\quad\quad\quad\quad\quad\quad
     +
     \frac{2}{\alpha} \tilde{K}_{ij} {\cal F}^{-1}\left[
     D^{k}h_{tk} - D^{k} \partial_{t}h_{(TV)k}
     + D^{k} \left(
     \frac{2}{\alpha} M_{k}^{\;\;l} h_{(TV)l}
     \right)
     \right]
     ,
     \label{eq:hij-traceless-by-hTVk-htk}
\end{eqnarray}
where we have used Eq.~(\ref{eq:hVL-by-hTVi-hti}) and defined the
traceless part $\tilde{K}_{ij}$ of the extrinsic curvature $K_{ij}$ by
$\tilde{K}_{ij}$ $:=$ $K_{ij}$ $-$ $\frac{1}{n-1} q_{ij} K$.
Taking the divergence of Eq.~(\ref{eq:hij-traceless-by-hTVk-htk}), we
obtain the single integro-differential equation for $h_{(TV)k}$:
\begin{eqnarray}
  &&
     {\cal D}_{j}^{\;\;k}h_{(TV)k}
     +
     D^{m} \left[
     \frac{2}{\alpha} \tilde{K}_{mj} \left\{
     {\cal F}^{-1} D^{k}\left(
     \frac{2}{\alpha} M_{k}^{\;\;l} h_{(TV)l} - \partial_{t}h_{(TV)k}
     \right)
     - \beta^{k} h_{(TV)k}
     \right\}
     \right]
     \nonumber\\
  && \quad\quad\quad
     =
     D^{m}\left[
     h_{mj} - \frac{1}{n-1} q_{mj} q^{lk} h_{kl}
     - \frac{2}{\alpha} \tilde{K}_{mj} {\cal F}^{-1} D^{k}h_{tk}
     \right]
     ,
     \label{eq:integro-differential-equation-for-hTVk}
\end{eqnarray}
\end{widetext}
where
\begin{eqnarray}
  {\cal D}^{ij}
  =
  q^{ij} \Delta + \left(1-\frac{2}{n-1}\right)D^{i}D^{j}
  + R^{ij}.
  \label{eq:calDij-def}
\end{eqnarray}


The existence and the uniqueness of the solution to this
integro-differential equation is highly nontrivial.
However, we assume the existence and the uniqueness of the solution
$h_{(TV)k}=h_{(TV)k}[h_{tm},h_{mn}]$ to this integro-differential
equation (\ref{eq:integro-differential-equation-for-hTVk}) here.
This solution describes the expression of the variable $h_{(TV)i}$ in
terms of the original components $\{h_{ti},$ $h_{ij}\}$ of the metric
perturbation $h_{ab}$.
Substituting the solution $h_{(TV)k}=h_{(TV)k}[h_{tm},h_{mn}]$ to
Eq.~(\ref{eq:integro-differential-equation-for-hTVk}) into
Eqs.~(\ref{eq:hVL-by-hTVi-hti})--(\ref{eq:hL-by-hTVk-hij}), we can
obtain the representation of the variables $\{h_{(VL)}, h_{(V)i},
h_{(L)}\}$ in terms of the original components $h_{ti}$ and $h_{ij}$
of $h_{ab}$.
Furthermore, the representation of the variable $h_{(TT)ij}$ in terms
of $h_{ti}$ and $h_{ij}$ are derived from
Eq.~(\ref{eq:hij-traceless-by-hTVk-htk}) through the substitution of
the solution $h_{(TV)k} = h_{(TV)k}[h_{tm},h_{mn}]$ to
Eq.~(\ref{eq:integro-differential-equation-for-hTVk}).


Thus, the decomposition formulae
(\ref{eq:hVL-hVi-hL-hTVi-hTTij-def-1})--(\ref{eq:hVL-hVi-hL-hTVi-hTTij-def-3})
are invertible if the Green functions $\Delta^{-1}$, ${\cal F}^{-1}$
exist and the solution to the integro-differential equation
(\ref{eq:integro-differential-equation-for-hTVk}) exists and is
unique.


\subsection{Gauge-transformation rules}
\label{sec:gauge-trans-appendix-C}


Through similar calculations to those in
Sec.~\ref{sec:inverse-relation-Appendix-C}, we can derive the
gauge-transformation rules for the variables $h_{(VL)}$, $h_{(V)i}$,
$h_{(L)}$, $h_{(TV)i}$, and $h_{(TT)ij}$.
From Eqs.~(\ref{eq:hVL-by-hTVi-hti}) and (\ref{eq:hV-by-hTVi-hti}),
the gauge-transformation rules (\ref{eq:ADM-hti-gauge-trans}) for the
component $h_{ti}$, we obtain the gauge-transformation rule for the
variables $h_{(VL)}$ and $h_{(V)i}$:
\begin{widetext}
\begin{eqnarray}
  {}_{\;{\cal Y}}\!h_{(VL)} - {}_{\;{\cal X}}\!h_{(VL)}
  &=&
      \xi_{t}
      +
      \Delta^{-1} D^{k}\partial_{t}\xi_{k}
      +
      {\cal F}^{-1} D^{k}\left[
      - \partial_{t}A_{k}
      + \frac{2}{\alpha} M_{k}^{\;\;l} A_{l}
      \right]
      +
      \Delta^{-1} D^{k}\partial_{t}A_{k}
      ,
      \label{eq:gauge-trans-hVL-by-xit-xik-Ak}
  \\
  {}_{\;{\cal Y}}\!h_{(V)i} - {}_{\;{\cal X}}\!h_{(V)i}
  &=&
      \partial_{t}\xi_{i}
      -
      D_{i}\Delta^{-1} D^{k}\partial_{t}\xi_{k}
      + \left[
      D_{i} - \frac{2}{\alpha} \left(D_{i}\alpha - \beta^{j}K_{ij}\right)
      \right]
      {\cal F}^{-1} D^{k}\left[
      \partial_{t}A_{k} - \frac{2}{\alpha} M_{k}^{\;\;l} A_{l}
      \right]
      \nonumber\\
  &&
      - D_{i}\Delta^{-1}D^{k}\partial_{t}A_{k}
      +
      \frac{2}{\alpha} M_{i}^{\;\;k} A_{k}
     ,
      \label{eq:gauge-trans-hVi-by-xit-xik-Ak}
\end{eqnarray}
where $A_{i}$ $:=$ ${}_{\;{\cal Y}}\!h_{(TV)i}$ $-$
${}_{\;{\cal X}}\!h_{(TV)i}$ $-$ $\xi_{i}$.
As in the case of the relations (\ref{eq:hVL-by-hTVi-hti}) and
(\ref{eq:hV-by-hTVi-hti}), these gauge-transformation rules
(\ref{eq:gauge-trans-hVL-by-xit-xik-Ak}) and
(\ref{eq:gauge-trans-hVi-by-xit-xik-Ak}) imply that we can obtain the
gauge-transformation rules for the variables $h_{(VL)}$ and $h_{(V)i}$
if the gauge-transformation rule for the variable $h_{(TV)i}$ is
specified.


From Eq.~(\ref{eq:hL-by-hTVk-hij}) and the gauge-transformation rule
(\ref{eq:ADM-hij-gauge-trans}), we can derive the gauge-transformation
rule for the variable $h_{(L)}$:
\begin{eqnarray}
  {}_{\;{\cal Y}}\!h_{(L)} - {}_{\;{\cal X}}\!h_{(L)}
  &=&
      2 D^{l}\xi_{l}
      +
      \frac{2}{\alpha} K \beta^{k} A_{k}
      +
      \frac{2}{\alpha} \left(
      {\cal F}^{-1} D^{k}\left[
      \partial_{t}A_{k} - \frac{2}{\alpha} M_{k}^{\;\;l}A_{l}
      \right]
      \right)
      .
  \label{eq:gauge-trans-hL-xil-Ak}
\end{eqnarray}
As in the case of the gauge-transformation rules
(\ref{eq:gauge-trans-hVL-by-xit-xik-Ak}) and
(\ref{eq:gauge-trans-hVL-by-xit-xik-Ak}), the gauge-transformation
rule (\ref{eq:gauge-trans-hL-xil-Ak}) also implies that we can obtain
the gauge-transformation rule for the variable $h_{(L)}$ if the
gauge-transformation rule for the variable $h_{(TV)i}$ is specified.
On the other hand, from the gauge-transformation rule for the
traceless part (\ref{eq:hij-traceless-by-hTVk-htk}) of $h_{ij}$, we
obtain the equation
\begin{eqnarray}
  \left(LA\right)_{ij}
  +
  {}_{\;{\cal Y}}\!h_{(TT)ij} - {}_{\;{\cal X}}\!h_{(TT)ij}
  -
  \frac{2}{\alpha} \tilde{K}_{ij} \beta^{k}A_{k}
  -
  \frac{2}{\alpha} \tilde{K}_{ij} {\cal F}^{-1}D^{k}\left[
  \partial_{t}A_{k} - \frac{2}{\alpha} M_{k}^{\;\;l} A_{k}
  \right]
  =
  0
  ,
  \label{eq:gauge-trans-Ak-equation-before}
\end{eqnarray}
where we have used Eqs.~(\ref{eq:ADM-hti-gauge-trans}) and
(\ref{eq:ADM-hij-gauge-trans}).
The divergence of Eq.~(\ref{eq:gauge-trans-Ak-equation-before}) yields
\begin{eqnarray}
  {\cal D}_{j}^{\;\;l}A_{l}
  -
  D^{l} \left[
  \frac{2}{\alpha} \tilde{K}_{ij} \left\{
  {\cal F}^{-1} D^{k} \left(
  \partial_{t}A_{k} - \frac{2}{\alpha} M_{k}^{\;\;l} A_{l}
  \right)
  +
  \beta^{k}A_{k}
  \right\}
  \right]
  =
  0
  .
  \label{eq:Ak-gauge-trans-eq}
\end{eqnarray}
\end{widetext}


Here, we note that we have assumed the existence and the uniqueness of
the solution to Eq.~(\ref{eq:integro-differential-equation-for-hTVk}).
Since Eq.~(\ref{eq:Ak-gauge-trans-eq}) is the homogeneous version of
Eq.~(\ref{eq:integro-differential-equation-for-hTVk}), this assumption
shows that we have the unique solution $A_{k}=0$ to
Eq.~(\ref{eq:Ak-gauge-trans-eq}), i.e.,
\begin{eqnarray}
  \label{eq:hTVi-gauge-trans-result}
  {}_{\;{\cal Y}}\!h_{(TV)i} - {}_{\;{\cal X}}\!h_{(TV)i} = \xi_{i}.
\end{eqnarray}
Thus, we have specified the gauge-transformation rule for the variable
$h_{(TV)i}$.


Substituting Eq.~(\ref{eq:hTVi-gauge-trans-result}) into
Eqs.~(\ref{eq:gauge-trans-hVL-by-xit-xik-Ak})--(\ref{eq:gauge-trans-hL-xil-Ak}),
we obtain the gauge-transformation rules for the variables $h_{(VL)}$,
$h_{(V)i}$, $h_{(L)}$, and $h_{(TT)ij}$:
\begin{eqnarray}
  &&
     {}_{\;{\cal Y}}\!h_{(VL)} - {}_{\;{\cal X}}\!h_{(VL)}
     =
     \xi_{t} + \Delta^{-1} D^{k}\partial_{t}\xi_{k}
     ,
     \label{eq:gauge-trans-hVL-by-xit-xik-final}
  \\
  &&
     {}_{\;{\cal Y}}\!h_{(V)i} - {}_{\;{\cal X}}\!h_{(V)i}
     =
     \partial_{t}\xi_{i}
     -
     D_{i}\Delta^{-1} D^{k}\partial_{t}\xi_{k}
     ,
     \label{eq:gauge-trans-hVi-by-xit-xik-final}
  \\
  &&
     {}_{\;{\cal Y}}\!h_{(L)} - {}_{\;{\cal X}}\!h_{(L)}
     =
     2 D^{l}\xi_{l}
     ,
     \label{eq:gauge-trans-hL-xil-final}
  \\
  &&
     {}_{\;{\cal Y}}\!h_{(TT)ij} - {}_{\;{\cal X}}\!h_{(TT)ij}
     =
     0
     .
     \label{eq:gauge-trans-Ak-equation-final}
\end{eqnarray}


\subsection{Gauge-invariant variables}
\label{sec:gauge-invariant-variables-appendix-C}


Inspecting gauge-transformation rules
(\ref{eq:hTVi-gauge-trans-result})--(\ref{eq:gauge-trans-Ak-equation-final}),
we define the gauge-invariant variables.
First, Eq.~(\ref{eq:gauge-trans-Ak-equation-final}) yields that the
variable $h_{(TT)ij}$ is manifestly gauge invariant and we define the
transverse-traceless gauge-invariant variable $\chi_{ij}$ as
\begin{eqnarray}
  \label{eq:TT-gauge-invariant}
  \chi_{ij} := h_{(TT)ij}.
\end{eqnarray}


To construct the other gauge-invariant variable, we consider the
gauge-variant part of the metric perturbation whose
gauge-transformation rule is given by the second equation in
Eqs.~(\ref{eq:gauge-trans-calHab-Xa}).
Since the gauge-transformation rule (\ref{eq:hTVi-gauge-trans-result})
coincides with the gauge-transformation rule for the spatial component
$X_{i}$ of gauge-variant part $X_{a}$, we may identify the variable
$X_{i}$ with $h_{(TV)i}$:
\begin{eqnarray}
  \label{eq:Xi-general-def}
  X_{i} := h_{(TV)i},
  \quad
  {}_{\;{\cal Y}}\!X_{i} - {}_{\;{\cal X}}\!X_{i} = \xi_{i}.
\end{eqnarray}
Inspecting the gauge-transformation rules
(\ref{eq:hTVi-gauge-trans-result}) and
(\ref{eq:gauge-trans-hVL-by-xit-xik-final}), we find the definition of
$X_{t}$ to be
\begin{eqnarray}
  &&
     X_{t}
     :=
     h_{(VL)} - \Delta^{-1} D^{k} \partial_{t} h_{(TV)k},
     \nonumber\\
  &&
     {}_{\;{\cal Y}}\!X_{t} - {}_{\;{\cal X}}\!X_{t} = \xi_{t}.
     \label{eq:Xt-general-def}
\end{eqnarray}
Actually, the gauge-transformation rule for $X_{t}$ defined by
Eq.~(\ref{eq:Xt-general-def}) is given by the temporal component
$X_{t}$ of the gauge-variant part $X_{a}$ in the second equation in
Eqs.~(\ref{eq:gauge-trans-calHab-Xa}).
Thus, we have constructed the gauge-variant part $X_{a}$ of the metric
perturbation as
\begin{eqnarray}
  \label{eq:Xa-general-def}
  X_{a} := X_{t} (dt)_{a} + X_{i} (dx^{i})_{a}.
\end{eqnarray}


Inspecting the gauge-transformation rules
(\ref{eq:gauge-trans-hVi-by-xit-xik-final}),
(\ref{eq:Xi-general-def}), and (\ref{eq:Xt-general-def}), we define a
gauge-invariant vector mode $\nu_{i}$ by
\begin{eqnarray}
  \label{eq:transverse-vector-nui-def}
  \nu_{i} := h_{(V)i} - \partial_{t}h_{(TV)i} + D_{i}\Delta^{-1}D^{k}\partial_{t}h_{(TV)k}.
\end{eqnarray}
Actually we can easily confirm that the variable $\nu_{i}$ is
gauge-invariant, i.e., ${}_{\;{\cal Y}}\!\nu_{i}$ $-$
${}_{\;{\cal X}}\!\nu_{i}$ $=$ $0$.
Through the divergenceless property of the variable $h_{(V)i}$, we
easily see the property $D^{i}\nu_{i}=0$.
Inspecting the gauge-transformation rule
(\ref{eq:gauge-trans-hL-xil-final}) and
(\ref{eq:hTVi-gauge-trans-result}), we define the gauge-invariant
scalar variable $\Psi$ by
\begin{eqnarray}
  \label{eq:curvature-perturbation-scalar-def}
  - 2 (n-1) \Psi := h_{(L)} - 2 D^{i}X_{i}.
\end{eqnarray}
Finally, inspecting gauge-transformation rule
(\ref{eq:ADM-htt-gauge-trans}), (\ref{eq:Xi-general-def}), and
(\ref{eq:Xt-general-def}), we can define the gauge-invariant Newton
potential $\Phi$ as
\begin{eqnarray}
  - 2 \Phi
  &:=&
       h_{tt}
       -
       2 \partial_{t}X_{t}
     +
     \frac{2}{\alpha} \left(
     \partial_{t}\alpha + \beta^{i} D_{i}\alpha
     - \beta^{i}\beta^{j} K_{ij}
     \right)X_{t}
       \nonumber\\
  &&
     + \frac{2}{\alpha} \left(
     \beta^{i}\beta^{k}\beta^{j}K_{kj}
     - \beta^{i}\partial_{t}\alpha
     + \alpha q^{ij} \partial_{t}\beta_{j}
     + \alpha^{2} D^{i}\alpha
     - \alpha\beta^{k} D^{i}\beta_{k}
     - \beta^{i}\beta^{j} D_{j}\alpha
     \right) X_{i}
     .
     \label{eq:Newton-potentail-scalar-def}
\end{eqnarray}
We can easily confirm the gauge-invariance of the variables $\Phi$ and
$\Psi$ through the definitions and gauge-transformation rules
(\ref{eq:curvature-perturbation-scalar-def}),
(\ref{eq:Newton-potentail-scalar-def}),
(\ref{eq:ADM-htt-gauge-trans}), (\ref{eq:Xi-general-def}), and
(\ref{eq:Xt-general-def}).
Here, we have chosen the factor of $\Psi$ in the definition
(\ref{eq:curvature-perturbation-scalar-def}) so that we may regard
$\Phi=\Psi$ as Newton's gravitational potential in the
four-dimensional Newton limit.


In terms of the above gauge-invariant variables $\Phi$, $\Psi$,
$\nu_{i}$, and $\chi_{ij}$, and the gauge-variant variables $X_{t}$
and $X_{i}$, the original components $\{h_{tt},h_{ti},h_{ij}\}$ of the
metric perturbation $h_{ab}$ are given by
\begin{eqnarray}
  h_{tt}
  &=&
      - 2 \Phi
      +
      2 \partial_{t}X_{t}
      -
      \frac{2}{\alpha} \left(
      \partial_{t}\alpha + \beta^{i}D_{i}\alpha - \beta^{j}\beta^{i}K_{ij}
      \right) X_{t}
      \nonumber\\
  &&
      -
      \frac{2}{\alpha} \left(
      \beta^{i}\beta^{k}\beta^{j} K_{kj}
      -
      \beta^{i} \partial_{t}\alpha
      +
      \alpha q^{ij} \partial_{t}\beta_{j}
      +
      \alpha^{2}D^{i}\alpha
      -
      \alpha \beta^{k}D^{i}\beta_{k}
      -
      \beta^{i}\beta^{j}D_{j}\alpha
      \right) X_{i}
      ,
      \label{eq:htt-Phi-Xt-Xi}
  \\
  h_{ti}
  &=&
      \nu
      +
      D_{i}X_{t}
      +
      \partial_{t}X_{i}
      -
      \frac{2}{\alpha} \left(
      D_{i}\alpha
      - \beta^{j}K_{ij}
      \right) X_{t}
      -
      \frac{2}{\alpha} M_{i}^{\;\;j} X_{j}
      ,
      \label{eq:hti-nui-Xt-Xi}
  \\
  h_{ij}
  &=&
      - 2 \Psi q_{ij}
      +
      \chi_{ij}
      +
      D_{i}X_{j}
      +
      D_{j}X_{i}
      +
      \frac{2}{\alpha} K_{ij} X_{t}
      -
      \frac{2}{\alpha} \beta^{k} K_{ij} X_{k}
      .
      \label{eq:hij-Psi-chiij-Xt-Xi}
\end{eqnarray}
Equations (\ref{eq:htt-Phi-Xt-Xi})--(\ref{eq:hij-Psi-chiij-Xt-Xi})
imply that we may identify the components of the gauge-invariant
variables ${\cal H}_{ab}$ as
\begin{eqnarray}
  \label{eq:calH-components-def}
  {\cal H}_{tt} := - 2 \Phi, \quad {\cal H}_{ti} := \nu_{i}, \quad
  {\cal H}_{ij} := - 2 \Psi q_{ij} + \chi_{ij}.
\end{eqnarray}
From Eqs.~(\ref{eq:Xa-general-def}),
(\ref{eq:htt-Phi-Xt-Xi})--(\ref{eq:hij-Psi-chiij-Xt-Xi}), we reach to
the decomposition formula (\ref{eq:hab-calHab+LieXgab}).



\end{document}